\documentstyle[12pt,epsf,epsfig]{article}
\newcount\timecount
\newcount\hours \newcount\minutes  \newcount\temp \newcount\pmhours

\hours = \time
\divide\hours by 60
\temp = \hours
\multiply\temp by 60
\minutes = \time
\advance\minutes by -\temp
\def\hour{\the\hours}
\def\minute{\ifnum\minutes<10 0\the\minutes
            \else\the\minutes\fi}
\def\clock{
\ifnum\hours=0 12:\minute\ AM
\else\ifnum\hours<12 \hour:\minute\ AM
      \else\ifnum\hours=12 12:\minute\ PM
            \else\ifnum\hours>12
                 \pmhours=\hours
                 \advance\pmhours by -12
                 \the\pmhours:\minute\ PM
                 \fi
            \fi
      \fi
\fi
}

\def\monthname{\relax\ifcase\month 0/\or January\or February\or
   March\or April\or May\or June\or July\or August\or September\or
   October\or November\or December\else\number\month/\fi}

\def\bold#1{\setbox0=\hbox{$#1$}%
     \kern-.025em\copy0\kern-\wd0
     \kern.05em\copy0\kern-\wd0
     \kern-.025em\raise.0433em\box0 }

\def\ga{\mathrel{\raise.3ex\hbox{$>$\kern-.75em\lower1ex\hbox{$\sim$}}}}
\def\la{\mathrel{\raise.3ex\hbox{$<$\kern-.75em\lower1ex\hbox{$\sim$}}}}
\def\gev{{\rm \, Ge\kern-0.125em V}}
\def\tev{{\rm \, Te\kern-0.125em V}}
\def\beq{\begin{equation}}
\def\eeq{\end{equation}}

\def\mchi{m_{\chi}}

\def\ohsq{\Omega_{\chi} h^2}

\def\m12{m_{1\!/2}}

\def\tb{\tan\beta}

\def\bsg{{{\mathrm B\!\to\!X_s}\gamma}}
\def\Bsg{{{\cal B}_{s\gamma}}}
\def\Bsgth{{{\cal B}^{theor}_{s\gamma}}}
\def\Bsgme{{{\cal B}^{meas}_{s\gamma}}}
\def\Bsgmo{{{\cal B}^{model}_{s\gamma}}}

\begin{document}
\begin{titlepage}
\pagestyle{empty}
\baselineskip=21pt
\rightline{hep-ph/0004169}
\rightline{CERN--TH/2000-106}
\rightline{MADPH-00-1166}
\rightline{MPI-PhE/2000-04}
\rightline{UMN--TH--1848/00}
\rightline{TPI--MINN--00/14}
\vskip 0.05in
\begin{center}
{\large{\bf
Supersymmetric Dark Matter in the Light of LEP and the Tevatron Collider}}
\end{center}
\begin{center}
\vskip 0.05in
{{\bf John Ellis}$^1$, {\bf Toby Falk}$^2$, {\bf Gerardo Ganis}$^3$ and
{\bf Keith A.~Olive}$^4$}\\
\vskip 0.05in
{\it
$^1${TH Division, CERN, Geneva, Switzerland}\\
$^2${Department of Physics, University of Wisconsin, Madison, WI~53706,
Wisconsin}\\
$^3${Max-Planck-Institut f\"ur Physik, Munich, Germany}\\
$^4${Theoretical Physics Institute, School of Physics and Astronomy,
University of Minnesota, Minneapolis, MN 55455, USA}\\
}
\vskip 0.05in
{\bf Abstract}
\end{center}
\baselineskip=18pt \noindent

We analyze the accelerator constraints on the parameter space of the
Minimal Supersymmetric extension of the Standard Model, comparing
those now available from LEP II and anticipating the likely
sensitivity of Tevatron Run II. The most important limits are those
from searches for charginos $\chi^{\pm}$, neutralinos $\chi_i$ and
Higgs bosons at LEP, and searches for stop squarks, charginos and
neutralinos at the Tevatron Collider.  We also incorporate the
constraints derived from $b \rightarrow s \gamma$ decay, and discuss
the relevance of charge- and colour-breaking minima in the effective
potential.  We combine and compare the different constraints on the
Higgs-mixing parameter $\mu$, the gaugino-mass parameter $m_{1/2}$ and
the scalar-mass parameter $m_0$, incorporating radiative corrections
to the physical particle masses.  We focus on the resulting
limitations on supersymmetric dark matter, assumed to be the lightest
neutralino $\chi$, incorporating coannihilation effects in the
calculation of the relic abundance.  We find that $m_\chi > 51$~GeV
and $\tan \beta > 2.2$ if all soft supersymmetry-breaking scalar
masses are universal, including those of the Higgs bosons, and that
these limits weaken to $m_\chi > 46$~GeV and $\tan \beta > 1.9$ if
non-universal scalar masses are allowed.  Light neutralino dark matter
cannot be primarily Higgsino in composition.

\end{titlepage}
\baselineskip=18pt
\section{Introduction}

The search for experimental evidence for supersymmetry
is currently approaching a transition. For several years now, many of the
most incisive experimental searches have been those at LEP~\cite{lepsusy},
whose
constraints on the parameter space of the minimal supersymmetric extension
of the Standard Model (MSSM) have grown ever more restrictive, as the
centre-of-mass energy of LEP~II has been increased in successive steps.
In parallel, improved analyses of data from Run~I of the Tevatron Collider
have been providing important complementary constraints~\cite{sugrarept}.
The
transition is marked by the termination of the
LEP~II experimental programme in late 2000 and the anticipated start of
Run~II of the Tevatron Collider in 2001.

The results of experimental searches for different MSSM particles
can usefully be compared and combined using the conventional
parameterization of the model in terms of
supersymmetry-breaking scalar and gaugino masses
$m_0, m_{1/2}$, the higgsino mixing parameter $\mu$, the ratio of
Higgs vacuum expectation values (vev's) $\tan \beta$ and a universal
trilinear supersymmetry-breaking parameter $A$. 
We work in the framework of gravity-mediated models of
supersymmetry breaking, in which it is commonly assumed
that the scalar masses $m_0$ and the gaugino masses $m_{1/2}$
are universal at some supersymmetric GUT scale. The
assumptions that these
supersymmetry-breaking parameters are universal should be
questioned, particularly for scalar masses and especially  those of the
Higgs supermultiplets, but provide a convenient way of
benchmarking comparisons and combinations of different
experimental searches. In this paper, we make such comparisons
and combinations in variants of the MSSM in which the scalar-mass
universality assumption is extended to Higgs fields (UHM, also commonly
referred to as mSUGRA or the constrained MSSM or CMSSM), and also without
this supplementary assumption (nUHM).

In making such comparisons,
we emphasize the importance of including radiative corrections
to the relations between these MSSM model parameters $(m_0, m_{1/2}, \mu,
\tan \beta, A)$ and the physical masses of MSSM particles. Radiative
corrections are well-known to be crucial in the MSSM Higgs sector,
but also should not be neglected in the chargino, neutralino, gluino
and squark sectors. As we have emphasized previously~\cite{efgos}, the
differences
between the domains of MSSM parameter space apparently explored
at the tree and one-loop levels are comparable to the differences
between the domains explored in successive years of LEP running at
higher centre-of-mass energies. In view of the intense
experimental effort put into sparticle searches at LEP~II, it is important
that the final results of these efforts be treated with the
theoretical care they deserve. This issue is also relevant if one wishes
to compare the physics reaches for electroweakly-interacting 
sparticles at LEP and for strongly-interacting sparticles at the Tevatron
Collider, in which case one should take into account the important
radiative corrections
to squark and gluino masses~\cite{Pierce:1997zz}, as well as to their
production cross sections.

In addition to direct searches for the production of MSSM
particles, important indirect constraints must also be taken into
account. These include other accelerator constraints, such as the
measured value of the $b \rightarrow s \gamma$ decay
rate~\cite{CLEObsg,ALEPHbsg}, and
non-accelerator constraints related to the possible role of the
lightest supersymmetric particle (LSP) as cold dark matter (CDM). 
The lightest supersymmetric particle (LSP)
would be stable in any variant of the MSSM which conserves
$R$ parity, as we assume here. In gravity-mediated models of
supersymmetry breaking, the framework adopted here, the LSP
is commonly thought to be the lightest neutralino $\chi$, 
and calculations of the cosmological relic density of
LSPs, $\Omega_\chi$, yield values in the range preferred by
cosmology in generic domains of MSSM parameter space~\cite{EHNOS}. The
possibility of supersymmetric CDM
provides one of our principal motivations for seeking a
deeper understanding of the allowed MSSM parameter space, but is
not our only focus in this paper.

The most essential dark-matter constraint is that the relic LSP
density not overclose the Universe. The conditions that the universe
has an age in excess of 12 billion years and that $\Omega_{\rm
  total}\le 1$ imply an upper bound on $\ohsq$ of 0.3.  Further, the
convergent indications from astrophysical structure-formation
arguments and observations of high-redshift supernovae are that
$\Omega_{CDM} < 0.5$\cite{Bahcall:1999xn}, whereas the Hubble expansion
rate $H_0 = 100
h$~km/s/Mpc: $h = 0.7$ with an error of about 10\%\cite{Freedman:1999wh},
so we require
$\Omega_{LSP} h^2 \le \Omega_{CDM} h^2 \le 0.3$. On the other hand,
astrophysical structure formation seems to require $\Omega_{CDM} >
0.2$, so we also require $\Omega_{LSP} h^2 \ge 0.1$, while
acknowledging that a lower value of $\Omega_{LSP}$ could be permitted
if other CDM particles such as axions and/or superheavy relics are
present.

There have recently been some significant developments in the analysis
of supersymmetric CDM. One is that the importance of co-annihilation
effects involving next-to-lightest supersymmetric particles (NLSPs)
such as the $\tilde \tau, \tilde \mu$ and $\tilde e$ for calculations of
the relic density of a gaugino-like LSP has recently been recognized
\cite{EFOSi,glp}.
Another phenomenon whose importance in the CMSSM has recently been underlined is the
possible transition of the electroweak vacuum into a charge-
and colour-breaking (CCB) minimum \cite{bbc}. The absence of such an instability is
not absolutely necessary, since a transition in the future cannot be
excluded. Therefore, we comment on the regions of MSSM
parameter space in which the CCB instability is absent, but do not
focus exclusively on these regions.

The main purpose of this paper is to prepare for the compilation,
comparison and combination of the definitive results from LEP~II
and the Tevatron. We illustrate our analysis with the latest
available
limits from these two experimental programmes~\cite{lepsusy, sugrarept},
supplemented by
educated guesses at their final sensitivities. As we have
explained previously, and discuss in more detail below, a key role
in constraining the MSSM parameter space is provided by the LEP
Higgs search. We express our results as a function of the present
LEP lower limit on $m_H$, currently 107.9~GeV~\cite{Higgs2000}, and the
prospective future sensitivity, which may approach 112~GeV.
We use our analysis to present lower limits on the LSP mass and
on tan$\beta$. We include a discussion of the implications of relaxing
the UHM assumption that the soft supersymmetry-breaking 
contributions to Higgs masses are also
universal. In particular, we investigate whether a light Higgsino LSP
is still a viable dark matter candidate, and 
find that the latest LEP~II data now exclude this
possibility. Finally, we discuss the likely future developments
in the exploration of the MSSM parameter space in the period before the
start-up of the LHC, during which the central role is likely to be
played by Run~II of the Tevatron Collider.

The layout of this paper is as follows. In Section 2 we review in more
detail the theoretical framework we adopt, discussing the issues of
universality and relic coannihilations, and stressing the
importance of Higgs mass constraints. In Section 3 we review our
implementation of the $b \rightarrow s \gamma$ constraint, including,
where applicable, the next-to-leading-order (NLO) QCD corrections.
We discuss the implications of the latest available constraints
from LEP~II in Section 4, combining them in Section 5 with the
cosmological
and astrophysical constraints $0.1 \le \Omega_{CDM} h^2 \le 0.3$
as well as the $b \rightarrow s \gamma$ constraint, and making
the UHM assumption. We find
\beq
m_\chi \ge 51~{\rm GeV}, \; \; \tan \beta \ge 2.2
\label{summarize}
\eeq
and discuss the expanded ranges of $m_\chi$ and $\tan \beta$ that
may be explored by the improved Higgs-mass limits that might be
obtained from the run of LEP~II in the year 2000. 
The limits in (\ref{summarize}) are strengthened when we restrict values
of
$A_0$ to minimize the parameter space with CCB minima, in which case we
find
\beq
m_\chi \ge 54~{\rm GeV}, \; \; \tan \beta \ge 2.8
\label{summarize2}
\eeq

\noindent We further
generalize the discussion to non-universal Higgs masses (nUHM) in Section
6, finding that the limits on $m_\chi$ and
$\tan
\beta$ are relaxed to

\beq
m_\chi \ge 46~{\rm GeV}, \; \; \tan \beta \ge 1.9.
\label{summarize3}
\eeq

\noindent Section 7 is devoted to a discussion of the
possibility of Higgsino dark matter in such a nUHM scenario.
We find that the LEP~II searches for charginos, neutralinos
and Higgs bosons together now  exclude as dark matter an LSP that is more than
about 70\% Higgsino. We turn our
attention to the Tevatron Collider in Section 8.
We compare the LEP~II and Run~I sensitivities to the MSSM
parameters, and discuss and compare the regions of MSSM parameter
space to which Tevatron Run~II data should be sensitive~\cite{sugrarept}.
Finally,
Section 9 summarizes our conclusions and the prospects for future
improvements and extensions of the analysis reported here.

\section{Theoretical Framework}

As already mentioned in the Introduction, we work in the
context of the MSSM with $R$ parity conserved. We assume a
parameterization of soft supersymmetry breaking inspired by
supergravity models with gravity mediation from a hidden sector. 
We assume that the soft supersymmetry-breaking scalar masses $m_0$
are universal at the supersymmetric GUT scale, as are the gaugino
masses $m_{1/2}$ and the trilinear parameters $A$. The
renormalization of the physical values of the soft 
supersymmetry-breaking parameters is then calculated using standard
renormalization-group equations. We use two loop RGEs\cite{mv} to evolve the
dimensionless couplings and the gaugino masses, and one loop
RGEs\cite{ikkt} for the other soft masses, and we include one-loop
SUSY corrections to $\mu$\cite{bbo} and to the top and bottom
masses\cite{Pierce:1997zz} . 

Deviations from scalar-mass universality could easily be
expected, for example in string-motivated models where their
magnitudes could be controlled by flavour-dependent modular
weights \cite{bim}. Upper limits on flavour-changing interactions place restrictions
on the possible generation-dependences of scalar mass parameters,
though these are relatively weak for the third generation. In any case,
these do not constrain non-universalities between 
sparticle fields with
different quantum numbers, namely $\tilde \ell_R$ vs $\tilde \ell_L$ vs
$\tilde q_R$ vs $\tilde q_L$.
Nevertheless, we neglect such possibilities in our analysis.
However, although our default option is that universality
extends also to the soft
supersymmetry-breaking contributions to the Higgs scalar masses
(UHM), we do also allow for the possibility that their soft
supersymmetry-breaking masses may be non-universal (nUHM).

\begin{figure}[htb]
\begin{center}
\mbox{\epsfig{file=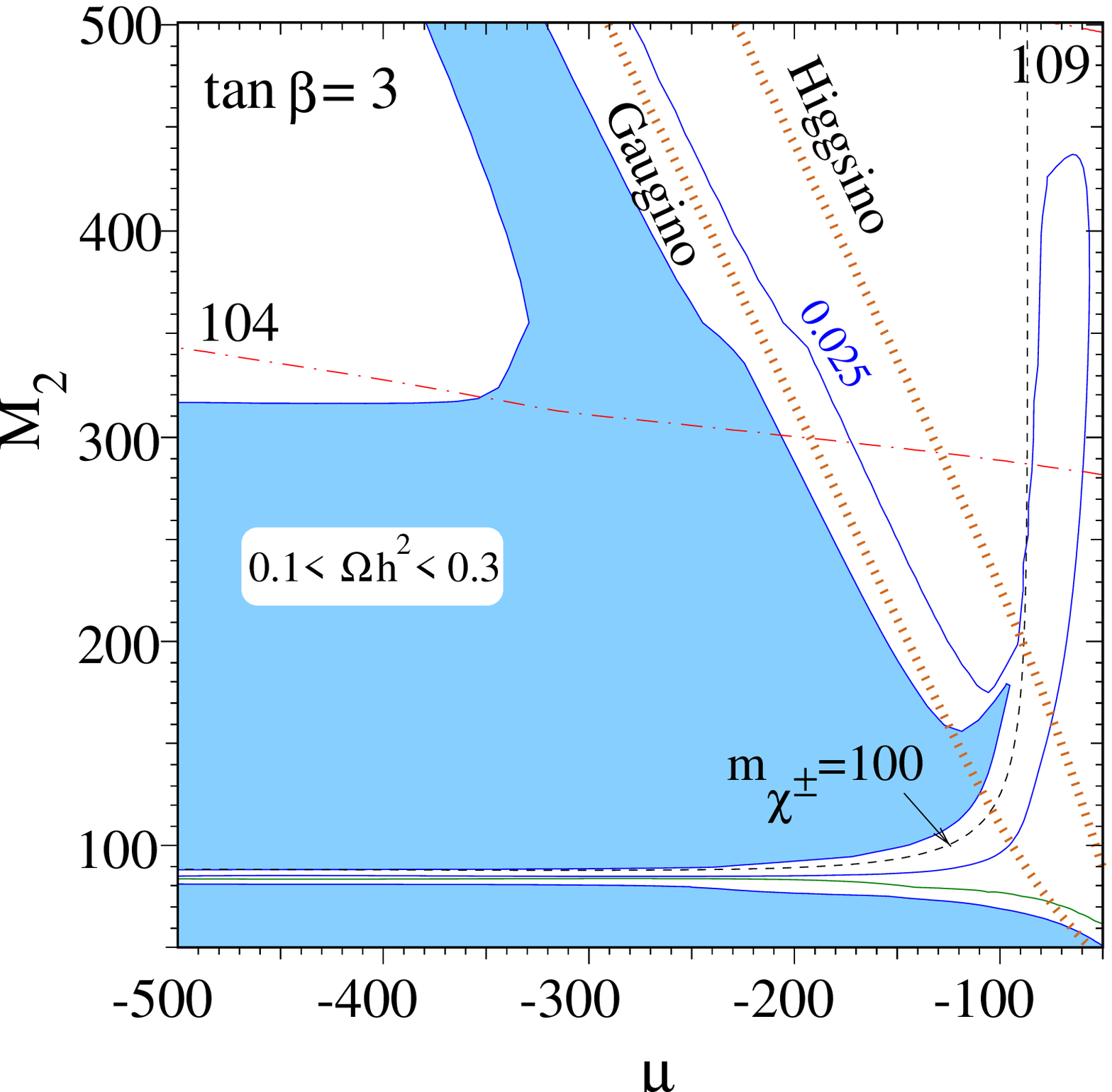,height=7.5cm}}
\mbox{\epsfig{file=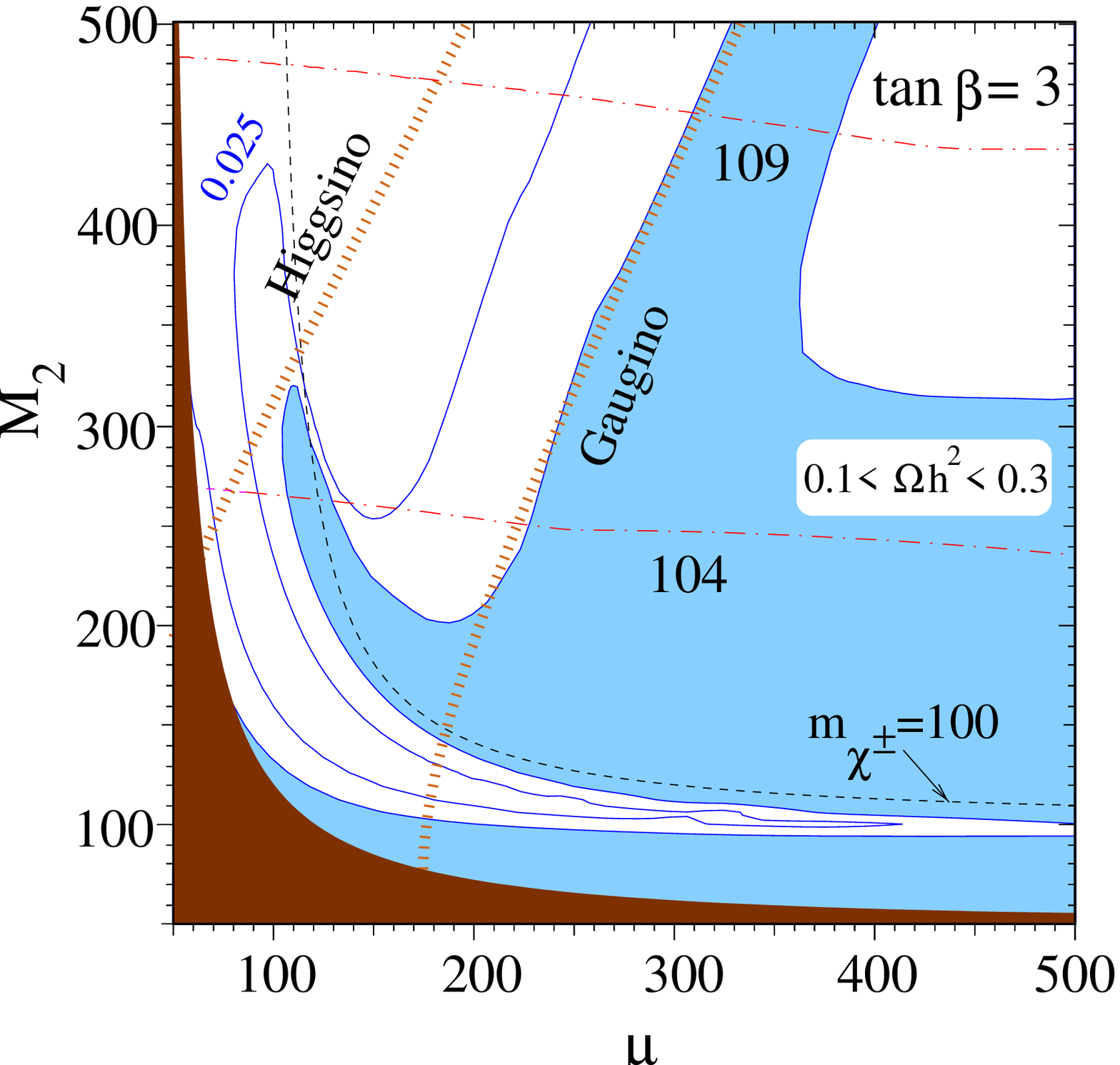,height=7.5cm}}
\end{center}
\caption[.]{\label{fig:big3}\it
  The $\mu, M_2 = (\alpha_2 / \alpha_{GUT}) \times m_{1/2}$ plane for
$\tan\beta = 3$, $m_0=100$ GeV and $m_A
  = 1$ TeV.  Contours of $\Omega_\chi h^2 = 0.025, 0.1$ and $0.3$ are
  shown as solid lines, and the preferred region with
  \mbox{$0.1<\Omega_\chi h^2 < 0.3$} is shown light-shaded.  There are
  also dashed lines corresponding to $m_{\chi^\pm} = 100$~GeV.  The
  near-horizontal dot-dashed lines are Higgs mass contours, and the hashed
  lines are 0.9 Higgsino and gaugino purity contours. The dark shaded
  region has $m_{\chi^\pm}< m_Z/2$.}
\end{figure}

We use the renormalization-group equations and the one-loop
effective potential to implement the constraints of a consistent
electroweak vacuum parameterized by the ratio $\tan\beta$ of
Higgs vev's. We therefore adopt a parameterization of the MSSM in
which $m_0, m_{1/2}, A$, $\tan\beta$ and the sign of $\mu$ are treated as
independent parameters, with the magnitude of the Higgsino mixing
parameter $\mu$ and the pseudoscalar Higgs mass $m_A$ (or, equivalently,
the bilinear soft supersymmetry-breaking parameter $B$) treated as
dependent parameters. 
In the UHM limit,
the correlation between $m_{1/2}$ and $\mu$ is such that the LSP
neutralino $\chi$ typically is mainly a $U(1)$ gaugino $\tilde B$ (Bino). Since
$\mu$ becomes a free parameter (along with $m_A$) in the nUHM case, 
$\chi$ may become Higgsino-like for certain parameter choices (roughly
$M_2 > 2\mu$). Fig.~\ref{fig:big3} gives an overview of the $\mu,
M_2 = (\alpha_2 / \alpha_{GUT}) \times m_{1/2}$ plane for the illustrative
choices $\tan\beta = 3$, $m_0=100$ GeV,
$A_t$ at its quasi-fixed point $\sim2.25 M_2$ \cite{irqfp},  
and $m_A = 1$
TeV, showing various contours of the relic density $\Omega_\chi h^2$, the
contour $m_{\chi^\pm} = 100$~GeV, contours of the mass of the
lightest MSSM Higgs boson, and contours of Higgsino purity.
In this figure we have neglected neutralino-slepton coannihilation (discussed in detail below),
since a small change in $m_0$ can move the masses out of the
coannihilation region.
We see that the LSP is mainly a $\tilde B$ in most of the $\mu, M_2$
plane displayed (where the relic density is of cosmological significance). 
One of the key questions we investigate is whether
a Higgsino LSP is still allowed as a dark matter candidate by LEP~II
data~\cite{efgos}.

The relevance of direct LEP or Tevatron searches for sparticles does
not need emphasis. In fact, it turns out that the indirect constraint
on the MSSM parameter space provided by the Higgs search is also of
great importance, as seen in Fig.~\ref{fig:big3}, particularly in the UHM
case. This constraint 
depends the MSSM mass parameters, because the mass of the lightest MSSM
Higgs $h$ is sensitive, via radiative corrections, to sparticle
masses, in particular the stop masses. This correlation has some
impact even in the nUHM case, as we discuss in more detail later.

It is interesting to confront the range of MSSM parameters
still permitted by the LEP and other direct experimental
searches for MSSM particles with other less direct 
experimental constraints, or with
theoretical prejudices. Among the latter, one might mention
gauge-coupling unification, lepton-quark mass unification and
the absence of fine tuning. Although we consider all these
prejudices appealing, none of them is precise enough to
enable us to draw any firm conclusions. Gauge-coupling unification
cannot be used to constrain $m_0, m_{1/2}$ and $\mu$ in the absence
of a theory of GUT threshold effects. Lepton-quark mass unification
is hostage to uncertainties in neutrino masses and mixing~\cite{CELW}. The
fine-tuning price imposed by LEP data is rising~\cite{tuning},
particularly for
small values of $\tan\beta$, but its interpretation is
subjective and no consensus has been reached on the maximal pain
that can be tolerated. The constraints we apply in this analysis
are rather the indirect experimental ones provided by the
measurement of $b \rightarrow s \gamma$ decay~\cite{CLEObsg,ALEPHbsg},
whose implementation we discuss in the next Section, and the
cosmological relic-density constraint already mentioned in the
Introduction.

In the parameter region of interest,
the relic density $\Omega_\chi h^2$ increases with
increasing $m_0, m_{1/2}$. Therefore the cosmological upper
limit $\Omega_\chi h^2 \le 0.3$ may be used here to set
upper limits on these soft supersymmetry-breaking
parameters.
As discussed in the Introduction, strictly speaking there is
no astrophysical lower limit on $\Omega_\chi h^2$, even if one
accepts that the cold dark matter (CDM) density 
$\Omega_{CDM} h^2 \ge 0.1$, since there might be
other important sources of CDM, such as axions or ultra-heavy
relics. Nevertheless, one may take $\Omega_\chi h^2 \ge 0.1$
as a default assumption.

An important recent development has been the recognition that
coannihilation of the LSP with next-to-lightest sparticles
(NLSPs) may be important~\cite{EFOSi} in the Bino LSP region that is
favoured in the UHM case, in particular. Generically, the NLSP
in this region is the lighter stau $\tilde \tau_1$, with the
$\tilde \mu_R$ and $\tilde e_R$ not much heavier. Since the stable MSSM 
relic cannot possess electric charge, the allowed region of the
MSSM parameter space is bounded by the line $m_\chi = m_{\tilde \tau_1}$.
Close to this line, $\chi - {\tilde \ell}$ and ${\tilde \ell} -
{\tilde \ell}$ coannihilation effects suppress $\Omega_\chi h^2$
below the range that would be calculated on the basis of $\chi - \chi$
annihilation alone. This has the effect, in particular, of
increasing the maximum allowable value of $m_{1/2}$ and hence
allowing $m_\chi \la 600$~GeV for $\Omega_\chi h^2 \le 0.3$.
For larger $m_{1/2} \ga 400$~GeV, the allowed range of $m_0$ has a
typical thickness $\delta m_0 \sim 30$~GeV. On the other hand,
when $m_{1/2} \la 400$~GeV, there is a relatively
broad allowed range for $m_0$ between about 50 and 150~GeV, depending on
$\tan \beta, A$ and the sign of $\mu$.

We have shown previously~\cite{EFOSII,efgos} that the lower limit on
$m_\chi$
imposed by data from LEP and elsewhere may be strengthened by
combining it with additional theoretical constraints such as the
cosmological relic density. The previous analysis included
coannihilation effects only in the Higgsino region. The inclusion of
LSP-NLSP coannihilation in the Bino region is less important for the
inferred lower limit on $m_\chi$, as we discuss later. However,
it is important when one is assessing how much of the preferred range
of MSSM parameter space may escape searches at LEP and elsewhere.

In our analysis below, we consider parameter ranges that highlight 
the current experimental bounds and are consistent with the
relic cosmological density.  We use four default values for $\tan \beta$,
namely
3, 5, 10, and 20.  Lower values of $\tan \beta$ are disfavoured by the
LEP Higgs mass limit, and the study of higher values would require an
improved treatment of the cosmological relic density calculation for
large $\tan \beta$, which lies beyond the scope of this paper. 
Although most of our Figures display more restricted ranges of
$m_{1/2}$, we note that in the UHM case cosmology allows
values of $m_{1/2}$ up to $\sim 1400$ GeV. We 
consider two possible treatments of the trilinear soft
supersymmetry-breaking parameter $A_0$ at the GUT scale
which we assume to be universal. The conservative approach in the UHM
case is to
vary $A_0$ so as to minimize the impact of the accelerator constraints
(UHM$_{\rm min}$),
and the other is to choose \cite{bbc} $A_0 = -m_{1/2}$ 
so as to maximize the area in the $m_0 - m_{1/2}$
parameter plane in the present electroweak vacuum is stable,
and CCB minima are irrelevant.
In the nUHM case, we must also specify values of $\mu$ and $m_A$.
For the purpose of translating Higgs mass limits into limits in the
$m_0 - m_{1/2}$ plane, we fix $m_A = 10$ TeV, so as to maximize the light
Higgs scalar mass and therefore derive the most conservative bound
possible. We allow $\mu$ and $A_0$ to vary as much as possible 
while remaining consistent with the experimental lower bounds on the
sparticle masses. 

\begin{figure}
\begin{center}
\mbox{\epsfig{file=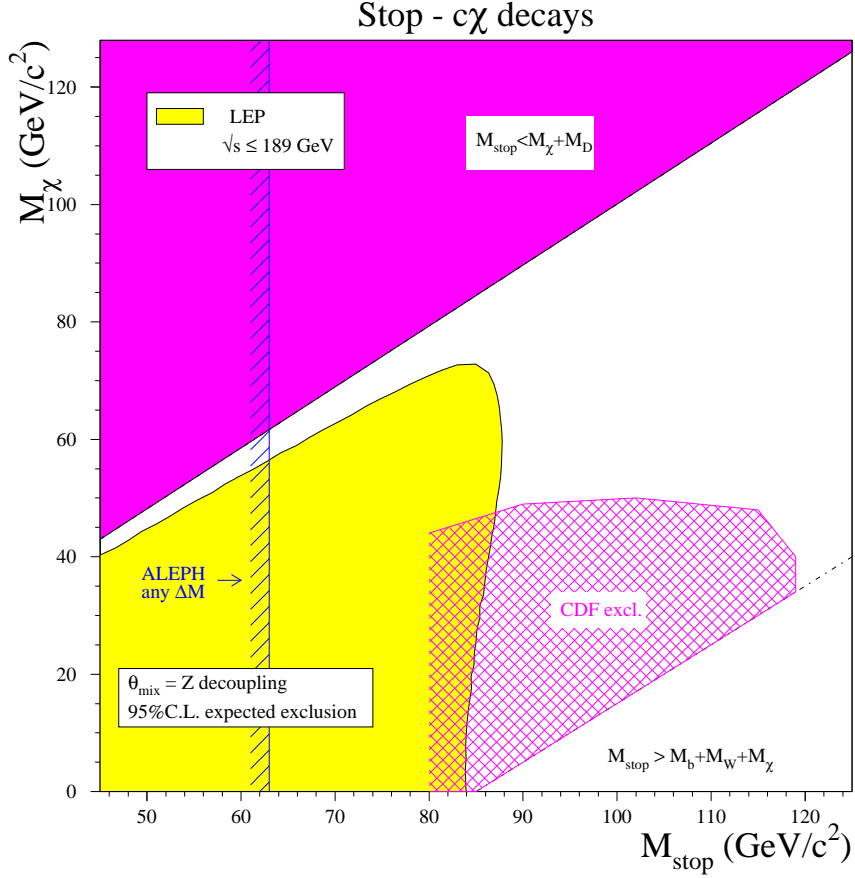,height=12.5cm}}
\end{center}
\caption[.]{\label{fig:stop}\it
Present constraints in the $(m_{\tilde{t}},m_\chi)$ plane 
assuming a 100\% branching ratio for the decay process
$\tilde{t}_1\!\to\!u/c\chi$. 
The vertical hatched band represents the recent ALEPH exclusion 
valid for any $\Delta M$ value presented in ~\cite{ALEPHstop}; 
the light grey region is the LEP-combined excluded region using 
data up to $\sqrt{s}=189$ GeV under the most conservative 
assumption for the coupling $Z\tilde{t}_1\bar{\tilde{t}_1}$ ~\cite{LEPstop};
the cross-hatched 
area is the CDF exclusion using the complete Run~I data sample
~\cite{CDFstop}.}
\end{figure}

\begin{figure}[htbp]
\begin{center}
\mbox{\epsfig{file=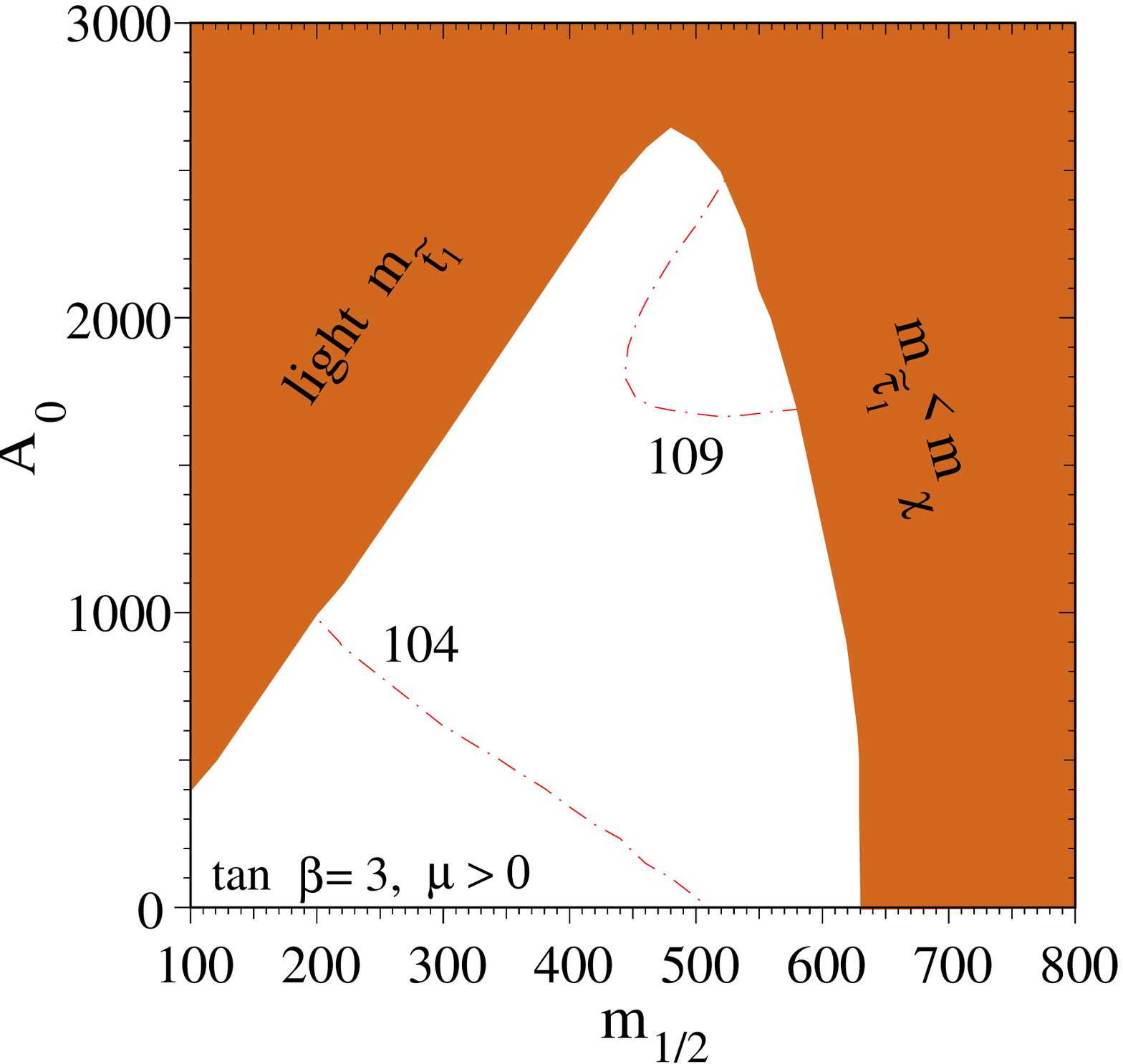,height=6.5cm}}
\mbox{\epsfig{file=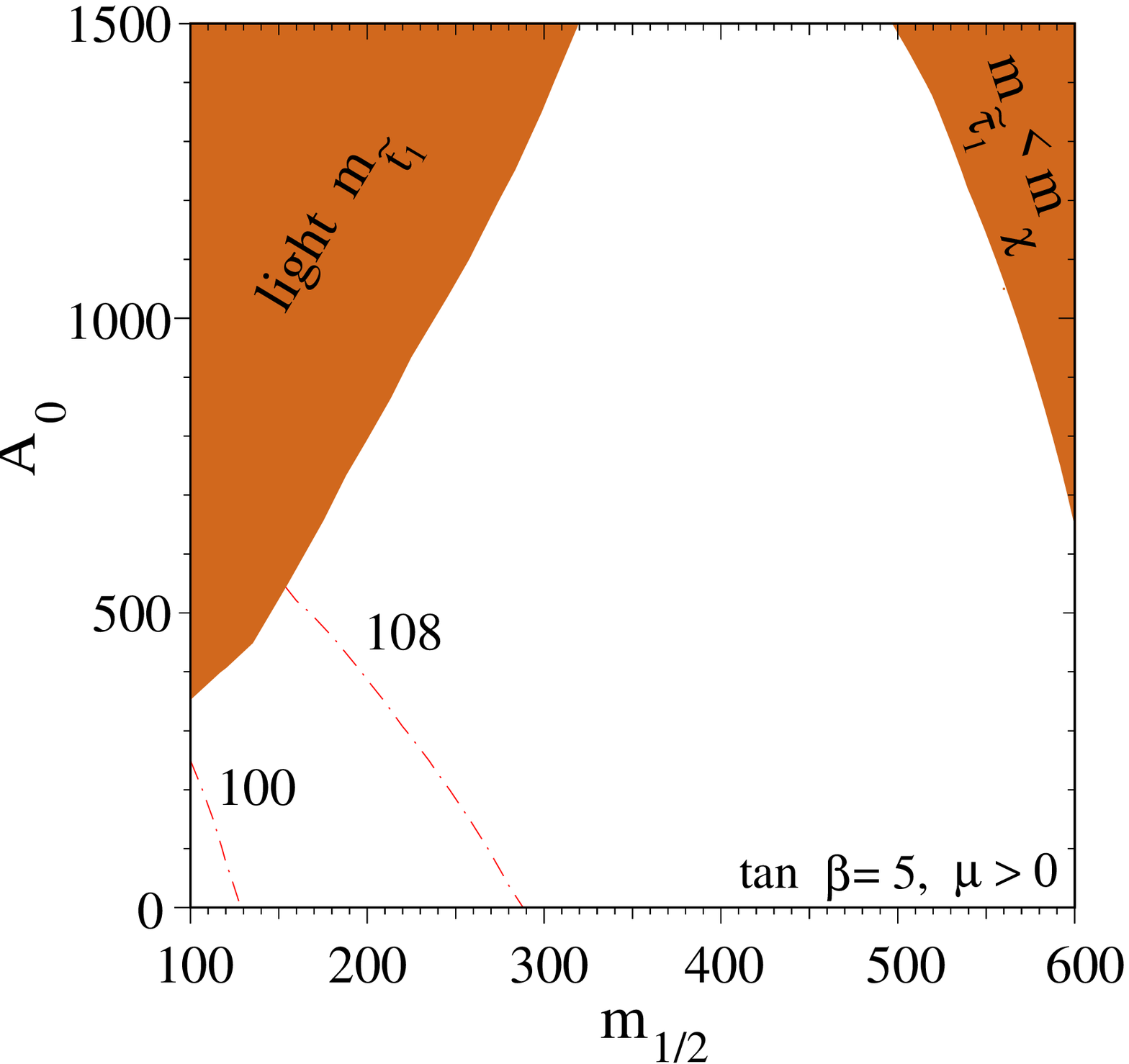,height=6.5cm}}
\end{center}
\begin{center}
\mbox{\epsfig{file=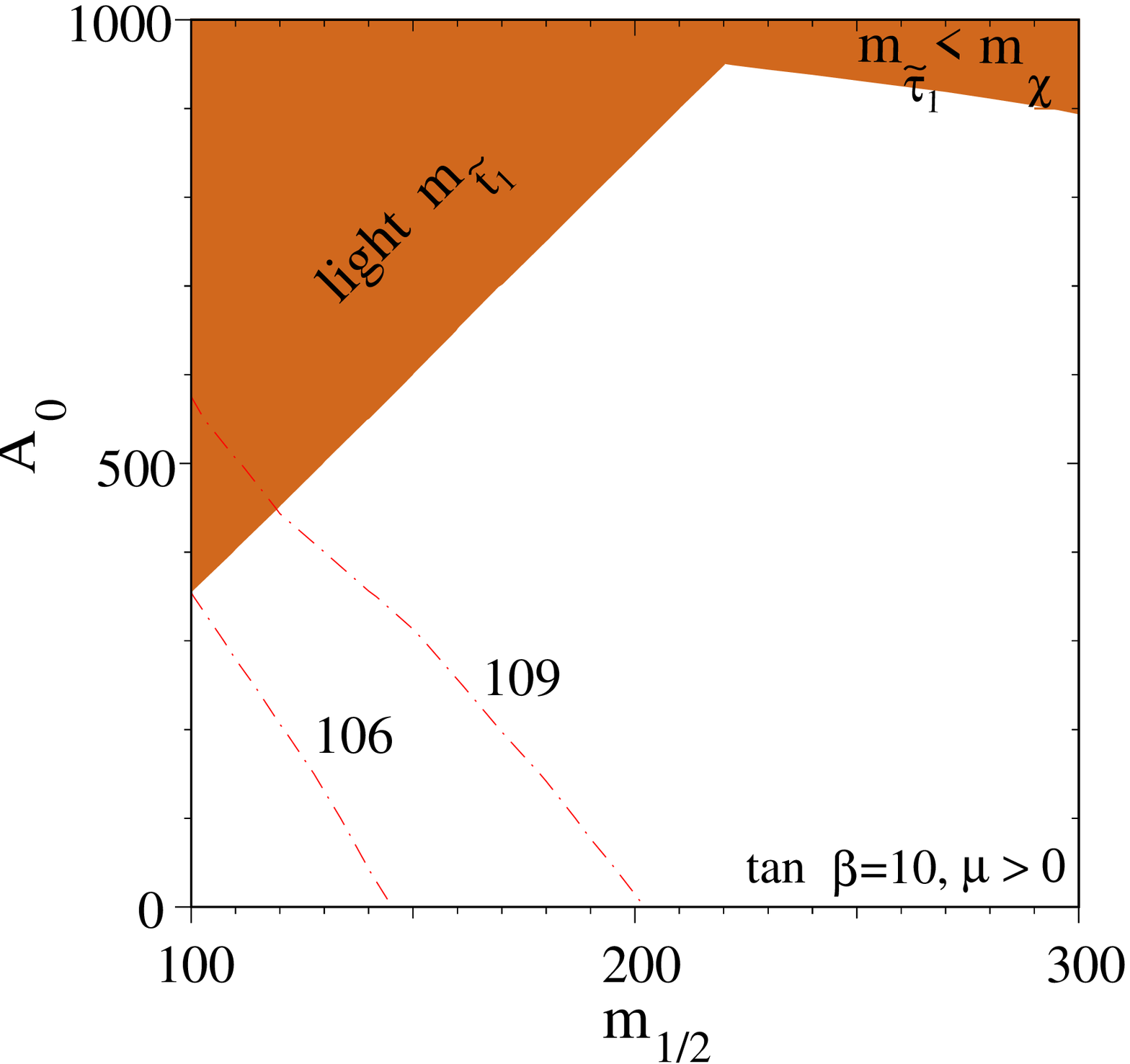,height=6.5cm}}
\mbox{\epsfig{file=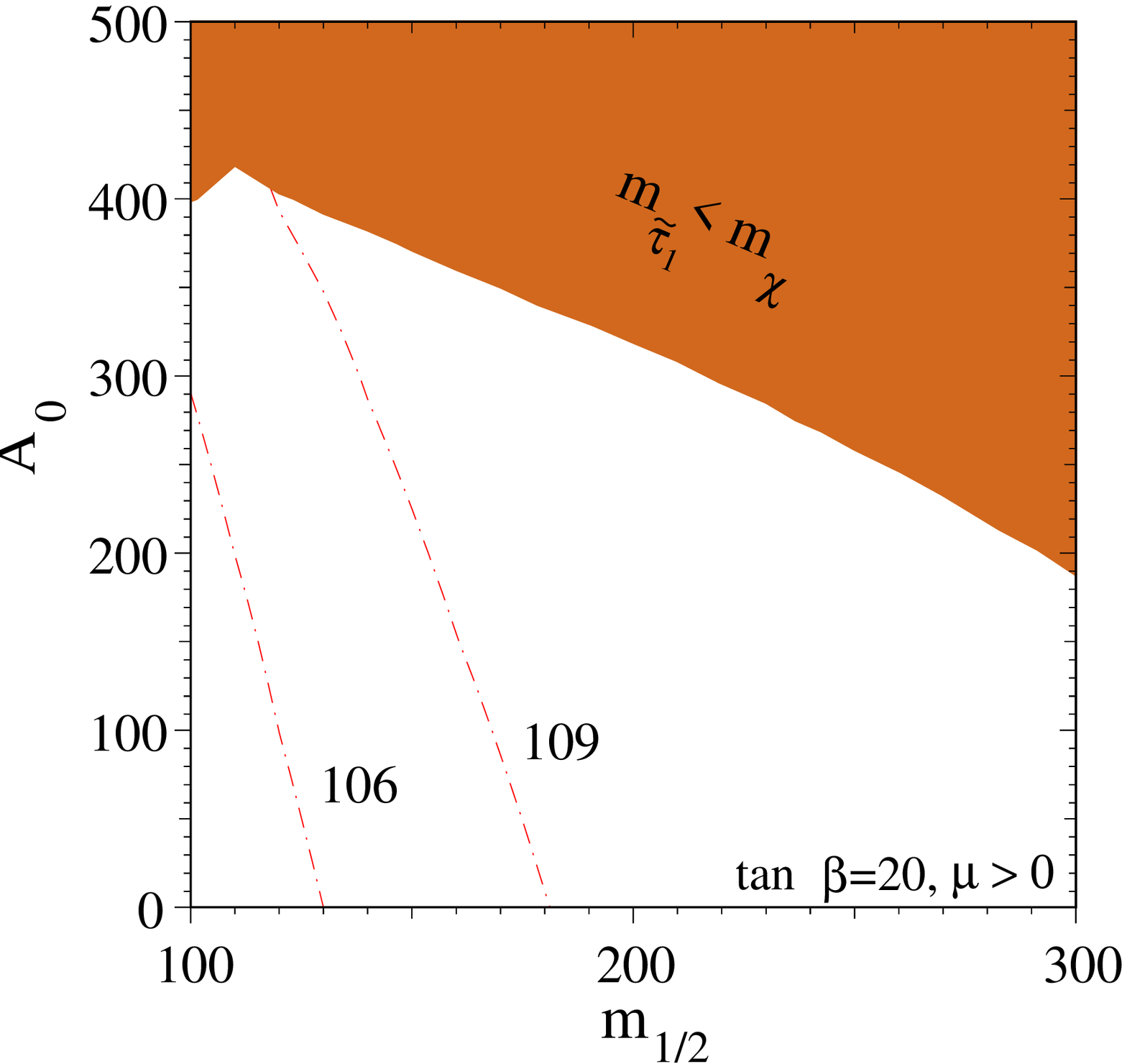,height=6.5cm}}
\end{center}
\caption{\label{fig:avm}\it
Upper bounds on the trilinear supersymmetry-breaking parameter $A_0$, 
for $\tan\beta = 3, 5, 10$ and $20$ and $\mu > 0$ , as a function of
$m_{1/2}$. 
The shaded regions yield either a tachyonic $\tilde t$ or a $\tilde\tau$
LSP. Also shown is the dependence of the lightest Higgs mass on 
$m_{1/2}$ and $A_0$. }
\end{figure}

There are restrictions on large values of
$A_0$, so as to ensure that the sfermion masses
are well-behaved. One of the most stringent bounds
is that imposed by the experimental lower limit on the lighter stop mass,
which depends on $m_\chi$ in the way depicted in Fig.~\ref{fig:stop},
which combines the constraints from~\cite{ALEPHstop,LEPstop,CDFstop}. 
Another important requirement is that the LSP not be a stau:
$m_{\tilde \tau_1} > m_\chi$. The impacts of the stop and stau constraints
are illustrated
in Fig.~\ref{fig:avm}. We show for, $\mu > 0$ and $\tan\beta = 3, 5, 10$
and 20,
the corresponding upper limits on $A_0$ in the UHM case as functions 
of $m_{1/2}$ for $m_0 =
100$ GeV.  Also shown in Fig.~\ref{fig:avm} as broken lines are
representative Higgs mass
contours. It is apparent that the Higgs mass is very 
sensitive to the value of $A_0$.
The corresponding figures for $\mu < 0$ are similar but allow somewhat
higher values for $A_0$. The sensitivity of the Higgs mass to $A_0$
translates into a corresponding sensitivity in the lower limit on
$m_\chi$.

\pagebreak

\section{Constraints from $b \rightarrow s \gamma$ Decay}

The width for the inclusive decay $\bsg$ is determined by
flavour-violating loop diagrams,
and is therefore sensitive to physics beyond the Standard Model. 
In generic models with two Higgs doublets, significant contributions 
come from charged Higgs boson exchange, which always increases 
the SM prediction for $\Bsg \equiv B(\bsg)$, allowing severe lower
limits on the mass of the charged Higgs boson to be set: see~\cite{BSGI}, for
instance.
However, these limits do not apply directly to supersymmetric extensions 
of the Standard Model, because, in addition to the two Higgs
doublets, there are chargino-stop
contributions which can interfere destructively with the charged Higgs 
boson ones, and thereby reduce the predicted rate for
$\Bsg$~\cite{BSGbarbieri}.

Calculations including next-to-leading order (NLO) QCD corrections exist
for both the Standard Model and generic two-Higgs-doublet models 
(see~\cite{BSGI} and 
references therein), whereas in the case of supersymmetry the 
leading order (LO) calculations~\cite{SUSYLObsg} have been complemented with 
NLO QCD corrections that are valid only under certain
assumptions~\cite{BSGII}.

We include in our numerical analysis a $\Bsg$ calculation based 
on the full NLO treatment for the Standard Model and charged Higgs
contributions, and 
the best available\footnote{The code implementing these calculations
has been kindly provided to us by
P.~Gambino, who also helped in designing a recipe 
to determine the applicability of the supersymmetric NLO 
calculations.} treatment of QCD corrections to the supersymmetric
contributions~\cite{BSGII}. 
The latter turned out to
be of limited applicability in our analysis, since
the conditions in which they well approximate the whole 
NLO supersymmetric corrections
are usually not met. Therefore, the results presented below are 
based mostly on the LO supersymmetric contributions only.
The prediction for $\Bsg$ depends on some experimental inputs and 
on three renormalization scales.
The experimental inputs are the top-quark mass, the CKM
mixing-angle factor $|V_{tb}V_{ts}^*/V_{cb}|$, the $c$ and $b$ quark masses, 
the inclusive 
semileptonic branching fraction of $B$ hadrons $B_{lept,X}$, the strong coupling
$\alpha_s(M_Z)$, and the electromagnetic coupling $\alpha_{em}$.
The renormalization scales are 
those relevant to the semileptonic and radiative processes $\sim m_b$, and 
the high-energy matching scale $\sim M_W$. 
We used as nominal values and errors for these quantities those quoted
in Table~1 of~\cite{BSGI}, except for  $B_{lept,X}$, $\alpha_s(M_Z)$ 
and $\alpha_{em}$. For the former 
we used the latest average provided by the LEP electroweak working 
group~\cite{LEPEW2K}, $B_{lept,X}\!=\!0.1058\!\pm\!0.0018$. 
For $\alpha_s(M_Z)$ we took the latest PDG~\cite{PDG98} combination: $0.119\!\pm\!0.002$. 
Finally, for $\alpha_{em}$ we took the value at $q^2=0$ following ~\cite{BSGII}.  
For each given point in the parameter space, 
we determined the theoretical error $\delta\Bsgth$ as the {\tt RMS} of  
1000 $\Bsg$ values obtained by varying 
the experimental inputs with independent and Gaussian errors. 
Moreover, we determined the reference theoretical prediction for
$\Bsgth$ conservatively,
as the value closest to the measured one that we could obtain by varying 
independently the three renormalization scales from half to twice their
nominal value. 

\begin{figure}
\begin{center}
\mbox{\epsfig{file=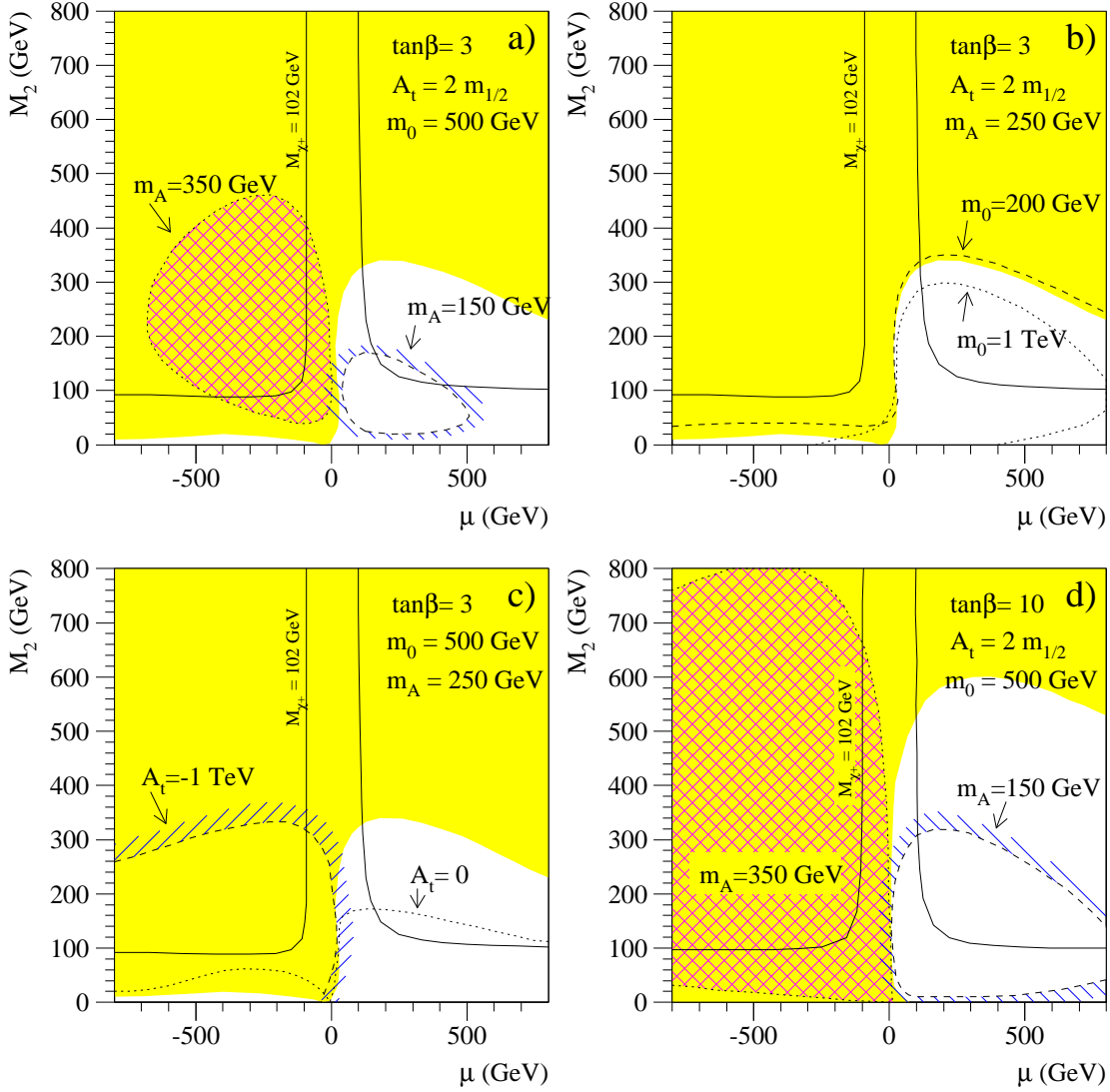,height=16.5cm}}
\end{center}
\caption[.]{\label{fig:bsg1}\it
Constraints on the nUHM parameter space imposed by $\Bsg$:
domains in the $(\mu,M_2)$ plane excluded for $\tan\beta=3$ {\rm (a,b,c)}
and 
$\tan\beta=10$ {\rm (d)}. In all plots the `reference' excluded region 
for $m_A=250$ GeV, $m_0=500$~GeV and the infra-red
quasi-fixed-point value $A_0=2 m_{1/2}$ is shaded, assuming
$m_t=175$~GeV. The effect of varying $m_A$ is shown in panel (a),
the effect of varying $m_0$ is shown in panel (b), the effect of
changing the sign of $A$ is shown in panel (c), and panel (d)
illustrates the effect of increasing $\tan\beta$. Please see the
text for further details.}
\end{figure}

The experimental measurements of the rate for the inclusive process 
$\bsg$~\cite{CLEObsg,ALEPHbsg} are dominated by the latest CLEO result
\begin{equation}
\Bsgme = (3.15\!\pm\!0.35\!\pm\!0.32\!\pm\!0.26)\times 10^{-4}, 
\label{bsgconstraint}
\end{equation}
which is in good agreement with the SM prediction of 
$(3.29\!\pm\!0.33)\times 10^{-4}$~\cite{KNbsg}. There is therefore
no need for any physics beyond the Standard Model, and we
establish only upper limits on the possible supersymmetric contributions.

To determine the 95\% confidence-level exclusion domain in the
supersymmetric parameter space, we treated separately the contributions 
to the CLEO error. 
The three terms in (3) come from limited statistics, 
experimental systematics and model dependence, respectively. 
By adding in quadrature the first two terms we defined $\delta\Bsgme$ 
which we treated as a Gaussian error, whilst 
we considered the third one, $\delta\Bsgmo\!=\!0.26$, as an additional 
scale error. We have then defined a $\chi^2$ function
\begin{equation} 
\chi^2 \equiv \frac{(| \Bsgme\!-\!\Bsgth |\!-\!\delta\Bsgmo)^2}
        {(\delta\Bsgme)^2\!+\!(\delta\Bsgth)^2},
\label{bsgchi2}
\end{equation}
and regard as excluded those points giving 
a $\chi^2$ probability for one degree of freedom smaller than 5\%.

We have investigated the impact of the $\Bsg$ constraint
(\ref{bsgconstraint}, \ref{bsgchi2}) in the
$(\mu,M_2)$ 
plane, without making the UHM assumption. Fig.~\ref{fig:bsg1}a shows the
domains excluded by $\Bsg$ as a
function of $m_A$ for $\tan\beta=3$, 
$m_0=500$~GeV and $A_t=2 m_{1/2}$, the quasi-fixed point
value~\footnote{Note that, in this Figure alone, we display plots for the
renormalized low-energy value $A_t$, rather than the input GUT value
$A_0$.}. The kinematic reach for charginos at $\sqrt{s}=204$~GeV is 
also shown for comparison. As expected, 
the extent of the excluded region depends strongly on 
$m_A\!\simeq\!\sqrt{m^2_{H^\pm}-m^2_{W}}$, 
essentially vanishing when $m_A > 350\div 400$~GeV.  When $m_A=350$ GeV, 
the excluded region collapses to the cross-hatched area shown in Figs.~\ref{fig:bsg1}a and 
\ref{fig:bsg1}d.  Focusing on 
the case of $m_A=250$~GeV, 
at large $M_2$ either the chargino or the stop is heavy enough to
suppress the supersymmetric contributions to $\Bsg$. In this case, the
positive charged-Higgs contribution 
dominates, making the predicted value of $\Bsg$ incompatible with the
measured value. For moderate $M_2$ values, the supersymmetric contribution
becomes sizeable. When $\mu>0$,
it interferes negatively with the charged Higgs contribution, 
reducing the predicted 
value for $\Bsg$, whereas for $\mu<0$ it adds constructively to the
charged Higgs contribution, strengthening the exclusion.
This explains the shape of the excluded domains for $m_A=150,\ 250$~GeV. 
The excluded domains depend only mildly on $m_0$, as shown in
Fig.~\ref{fig:bsg1}b, whilst it can be seen in Fig.~\ref{fig:bsg1}c 
that the dependence on the sign of $A_t$ is significant, and we
also show for comparison the case $A_t = 0$. 
Finally, Fig.~\ref{fig:bsg1}d shows the same exclusion domains as
Fig.~\ref{fig:bsg1}a, but for
for $\tan\beta=10$. While the charged Higgs contributions essentially
saturate for  
$\tan\beta>4\div5$, the supersymmetric contributions contain terms 
of order $1/\cos\beta$, and therefore increase with $\tan\beta$.
When $\mu>0$, this has the effect of further reducing the prediction
and hence the excluded region, 
while, for $\mu<0$, it enhances $\Bsg$, thereby extending the sensitivity
of these constraints to larger $m_A$ values.

\section{Update on constraints from LEP II}

The latest general presentations of results from the four LEP
collaborations were made on Mar.~7th, 2000 \cite{LEPC}. They were based on
the following mean integrated luminosities in each experiment
at the indicated centre-of-mass energies:
\begin{eqnarray}
E_{CM} = 188.6~{\rm GeV} \; &:& \; 171~{\rm pb}^{-1} \nonumber \\
E_{CM} = 191.6~{\rm GeV} \; &:& \; \, 28~{\rm pb}^{-1} \nonumber \\
E_{CM} = 195.6~{\rm GeV} \; &:& \; \, 78~{\rm pb}^{-1}  \\
E_{CM} = 199.6~{\rm GeV} \; &:& \; \, 80~{\rm pb}^{-1} \nonumber \\
E_{CM} = 201.6~{\rm GeV} \; &:& \; \, 38~{\rm pb}^{-1} \nonumber 
\label{intlum}
\end{eqnarray}
No significant signals were announced in any sparticle or
Higgs search channel. Numerical lower limits on the Higgs boson 
masses were presented separately by the four experiments, and a
preliminary 
combination performed by the LEP Higgs Working Group is 
available~\cite{Higgs2000}.
Limits were also presented by the individual experiments on sparticle
production within several frameworks, but the combination of the standard 
channels usually provided by the LEP Supersymmetry Working
Group~\cite{lepsusy} was not made available at this time.
We extrapolate the available combined LEP limits, provided on
the basis of the 
running up to $E_{CM} = 188.6~{\rm GeV}$ for the sparticles and 
up to $E_{CM} = 201.6~{\rm GeV}$ for the Higgses, to 
include the higher-energy/luminosity data.

We consider the following possible scenarios for the future evolution of
the integrated LEP luminosity. The pessimistic one is that no
significant additional high-energy luminosity is accumulated
(remember the beer bottles?). In this case, the sparticle and
Higgs sensitivity will remain essentially as they are at the end of
1999. We believe that a more realistic scenario is for LEP to
accumulate luminosity at the same average rate of 1.3~pb$^{-1}$
as in 1999, but at somewhat higher energies, say 2/3 at
$E_{CM} = 202$~GeV and 1/3 at $E_{CM} = 204$~GeV. This would
result in the following total integrated luminosities per
experiment at energies above $E_{CM} = 200$~GeV:
\begin{eqnarray}
E_{CM} = 202.0~{\rm GeV} \; &:& \; 160~{\rm pb}^{-1} \\
E_{CM} = 204.0~{\rm GeV} \; &:& \; \, 60~{\rm pb}^{-1} \nonumber 
\label{realistic}
\end{eqnarray}
A more optimistic scenario would be that luminosity is
accumulated at a rate sometimes achieved in 1999, but not
consistently, and that 50\% of the running is at $E_{CM} = 204$~GeV
and 206~GeV:
\begin{eqnarray}
E_{CM} = 202.0~{\rm GeV} \; &:& \; 140~{\rm pb}^{-1} \nonumber \\
E_{CM} = 204.0~{\rm GeV} \; &:& \; \, 80~{\rm pb}^{-1} \\
E_{CM} = 206.0~{\rm GeV} \; &:& \; \, 20~{\rm pb}^{-1} \nonumber 
\label{optimistic}
\end{eqnarray}
We do not provide detailed results for this optimistic scenario,
but do make some comments on its potential impact.

The sparticle final states of relevance for this analysis are 
$\chi^+ \chi^-, \chi \chi', \chi' \chi'$ and
${\tilde \ell}^+ {\tilde \ell}^-$. In addition, the experimental limits 
on the squark-production processes ${\tilde t} {\bar {\tilde t}}$ and 
${\tilde b} {\bar {\tilde b}}$ can be used to infer constraints on the 
$A$ parameters, as discussed in Section 3. We contrast two
approaches in the following, either we take the CCB constraint
into account and fix $A_0$ so as to minimize its impact, 
or we take a conservative approach, allowing
any value of $A$ consistent with the experimental limits on
$m_{\tilde{t}_1}$ and other constraints (UHM$_{\rm min}$). 

The experimental efficiency for sparticle detection and hence the
cross-section upper limit depends on other parameters besides the
target sparticle mass, for example the mass difference $\Delta M$ in the
sparticle decay, e.g., $\chi^+ \rightarrow \chi + X$. We have modeled
these varying detection efficiencies using a multistep function, 
with a lower $\Delta M$ cutoff, 
low and high $\Delta M$ regions. We have used the available publications
by the
LEP collaborations and the documentation provided by the LEP Supersymmetry
Working Group~\cite{lepsusy} to derive reasonable values for the
transitional
values of $\Delta M$
and the average efficiency values within the two regions. We did
the same for the background contaminations, except in the case of 
slepton production, in which case we modeled the dominant $W^+W^-$
background in different regions of
the $(M_{\tilde{l}},M_\chi)$ plane using its detailed kinematics.
We have checked that our parameterization reproduces the available published 
results. 
In each of the `realistic' and `optimistic' scenarios
(\ref{realistic}, \ref{optimistic}), we make the assumption that 
the experimental efficiencies and contaminations remain similar 
to those at $E_{CM} \le 189$~GeV, based on the fact that the properties 
of the standard process do not change dramatically in the spanned  
energy range. Examples of the estimated upper limits 
on sparticle production cross sections that we obtain from our
extrapolation to the three
LEP running scenarios discussed above are given in Table~1. 

\begin{table}
\begin{center}
\begin{tabular}{|l||c|c|c|}
\hline
 LEP scenario        &  1999   & `realistic' 2K & `optimistic' 2K \\
 Max $\sqrt{s}$ (GeV)&  201.6  &  204         &   206         \\
\hline\hline
$\chi^+\chi^-$       &  0.19   &  0.11        &  0.11  \\
$\tilde{e}^+\tilde{e}^-$ &  0.06   &  0.05        &  0.06  \\
$\tilde{\tau}^+\tilde{\tau}^-$ &  0.09   &  0.08       &  0.08  \\
$\chi^0_i\chi^0_j$   &  $\sim$0.08   &  $\sim$0.08        &  $\sim$0.07  \\
\hline
\end{tabular}
\caption{\label{tab:xslim}
\it Examples of estimated upper limits on sparticle 
cross sections (in pb): for charginos and sleptons, assuming
$M_\chi=50$ GeV,
$M_{\chi^\pm}=100$, $M_{\tilde{e}^\pm}=95$ and  $M_{\tilde{\tau}^\pm}=90$~GeV, 
respectively. In the case of neutralino production, typical values  are
given.}
\end{center}
\end{table}

\begin{figure}[htb]
\begin{center}
\mbox{\epsfig{file=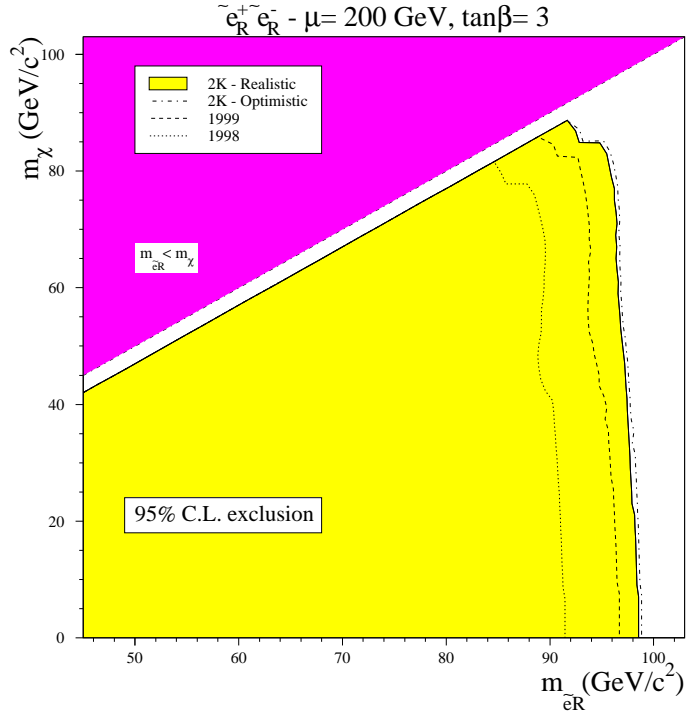,height=10cm}}
\end{center}
\caption[.]{\label{fig:selectron}\it
Constraints in the $m_{{\tilde e}_R}, m_\chi$ plane imposed by the
combined LEP data at $E_{CM} \le 189$~GeV (dotted line), our
estimates for the limits obtainable by combining the LEP data taken in
1999 (\ref{intlum}) (dashed line), and our estimates for the possible
2000 exclusions in the `realistic' (\ref{realistic}) (solid line) and
`optimistic' (\ref{optimistic}) (dot-dashed line) scenarios described
in the text. }
\end{figure}

For charginos, we conservatively assume no detection efficiency for 
$\Delta M<5$ GeV. For larger $\Delta M$ values,
the estimated upper limits allow one to exclude 
chargino production up to a few hundred MeV below the kinematic 
limit, unless sneutrino masses, and hence $m_0$, are very small. 
In the case of associated neutralino production,
we combine all the kinematically accessible channels, weighted 
by their {\it visible} cross sections, i.e., we take into account
the branching fractions into $\chi\nu$ final states, and the estimated
efficiencies. 
We found that the estimated upper limits depend only weakly on the point 
in the MSSM parameter space, and typical values are given in Table~1. 
In the case of slepton production, we set the $\Delta M$ cutoff at 3 GeV.
As an example, Fig.~\ref{fig:selectron} shows the exclusions we obtain in 
the plane ($m_{{\tilde e}_R}, m_\chi$) for $\tan\beta = 3$, $\mu=200$~GeV
and BR(${\tilde e}^\pm_R\to\chi e^\pm$=100\%), under 
the different hypotheses for LEP running in 2000.

We now turn to the estimation of upper limits on Higgs production.
In order to estimate the prospects for Standard Model Higgs limits from
the reaction $e^+ e^- \rightarrow h Z^0$, we take the simple
parameterization of the LEP limits obtained from data up to 
$E_{CM} \sim 189$ discussed in~\cite{Janot1999}. 
Extrapolating this parameterization to include the 1999 data set (\ref{intlum})
leads to the estimated limit $m_H \ge 109$~GeV. 
This result is in good agreement with the expected limit reported 
in~\cite{Higgs2000}. However, the observed limit quoted in the same
reference is 107.9 GeV: the difference is explained as a statistical 
fluctuation in the data at the level of one standard deviation.

\begin{figure}[hbt]
\begin{center}
\mbox{\epsfig{file=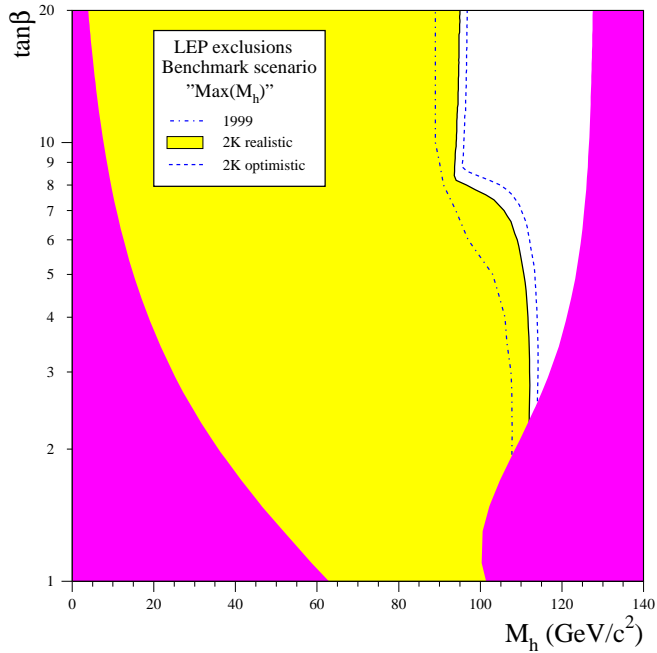,height=9.5cm}}
\end{center}
\caption[.]{\label{fig:Higgs}\it
Constraints in the $m_h, \tan\beta$ plane imposed 
in the `Max($M_h$)' benchmark scenario by
combining the LEP data taken in
1999 (\ref{intlum}) (dot-dashed line), and our estimates for the
possible
2000 exclusions in the `realistic' (\ref{realistic}) (solid line)
and `optimistic' (\ref{optimistic}) (dashed line) scenarios described
in the text. Also shown (dark-shaded) are the regions of the $m_h,
\tan\beta$
plane excluded by theoretical calculations~\cite{Carena9912223}.}
\end{figure}

The corresponding limits for the `realistic' and `optimistic'
running scenarios for the year 2000 are 112 and 114~GeV, respectively,
following roughly the empirical rule $M_H \ge E_{CM} - 93$~GeV. Similar
estimates apply to the MSSM for small $\tan\beta \la 5$, as shown in
Fig.~\ref{fig:Higgs}~\footnote{We have verified that these limits are not
weakened by the appearance of invisible decay modes
$h, A \rightarrow \chi \chi$.}.
For larger values of $\tan\beta$, we use the same
limiting cross section for $e^+ e^- \rightarrow h Z^0$, which gives a
weaker
lower limit on $m_h$, because of the smaller $Z^0 Z^0 h$ coupling.
When $\tan\beta \ga 8$, the production mechanism $e^+ e^- \rightarrow
h A$ becomes important, which we include in our analysis following again 
the prescription given in~\cite{Janot1999}. Our estimated limiting curves in
the $m_h, \tan\beta$ plane shown in Fig.~\ref{fig:Higgs}
have been calculated in the `Max($M_h$)' benchmark scenario
suggested in~\cite{Carena9912223}. We have not attempted to
combine the $h Z^0$ and $h A$ analyses, but have only overlapped them,
so our results could be considered conservative in the
intermediate-$\tan\beta$ region. We indicate in Fig.~\ref{fig:Higgs}
the LEP limit for the full 1999 data set (dot-dashed line), our estimate
for the `realistic' 2000 running scenario (\ref{realistic}) (shaded) and
the `optimistic' scenario (\ref{optimistic}) (dashed line). Also shown
(shaded) are the regions of the $m_h, \tan\beta$ plane excluded by
theoretical calculations~\cite{Carena9912223}.

Some caution is required when we compare our results
directly with the lower limits given by the LEP experiments because of
theoretical uncertainties in the MSSM Higgs mass calculations.
Conservatively, we allow for an error of $\sim$3~GeV in these, so that 
we translate the experimental limits into the supersymmetric parameter
space using the theoretical contours for
104~GeV (for $\tan \beta =3$) and 100~GeV (for $\tan \beta =5$) in 
the case of the complete 1999 data LEP scenario, and 
109~GeV  and 108~GeV, respectively for the two values of $\tan \beta$,  
in the `realistic' 2000 LEP scenario. 
At the higher values of $\tan \beta$ considered, the Higgs mass limits
do not provide strong constraints and are not used.

\begin{figure}[htbp]
\begin{center}
\mbox{\epsfig{file=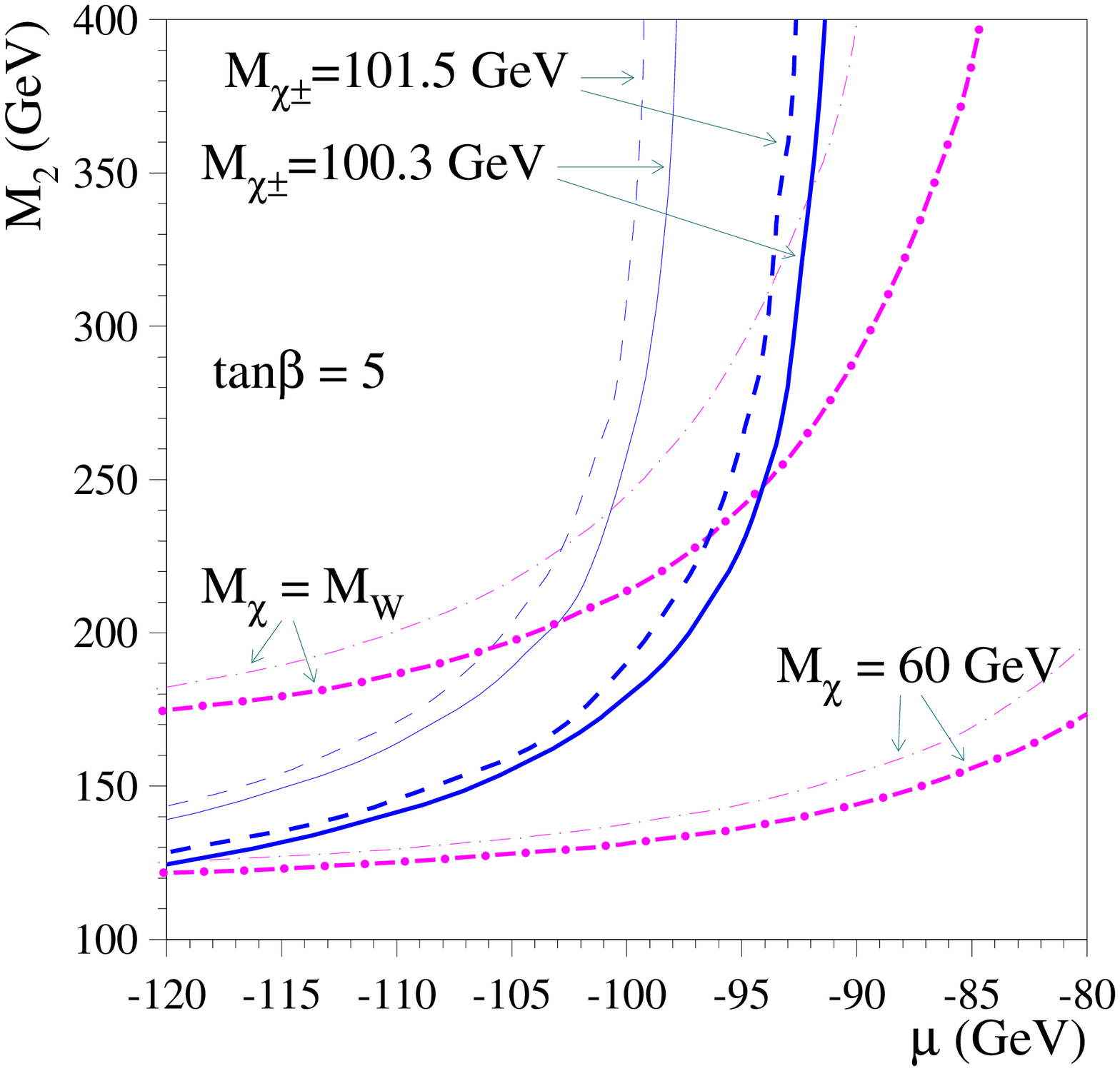,height=6.5cm}}
\mbox{\epsfig{file=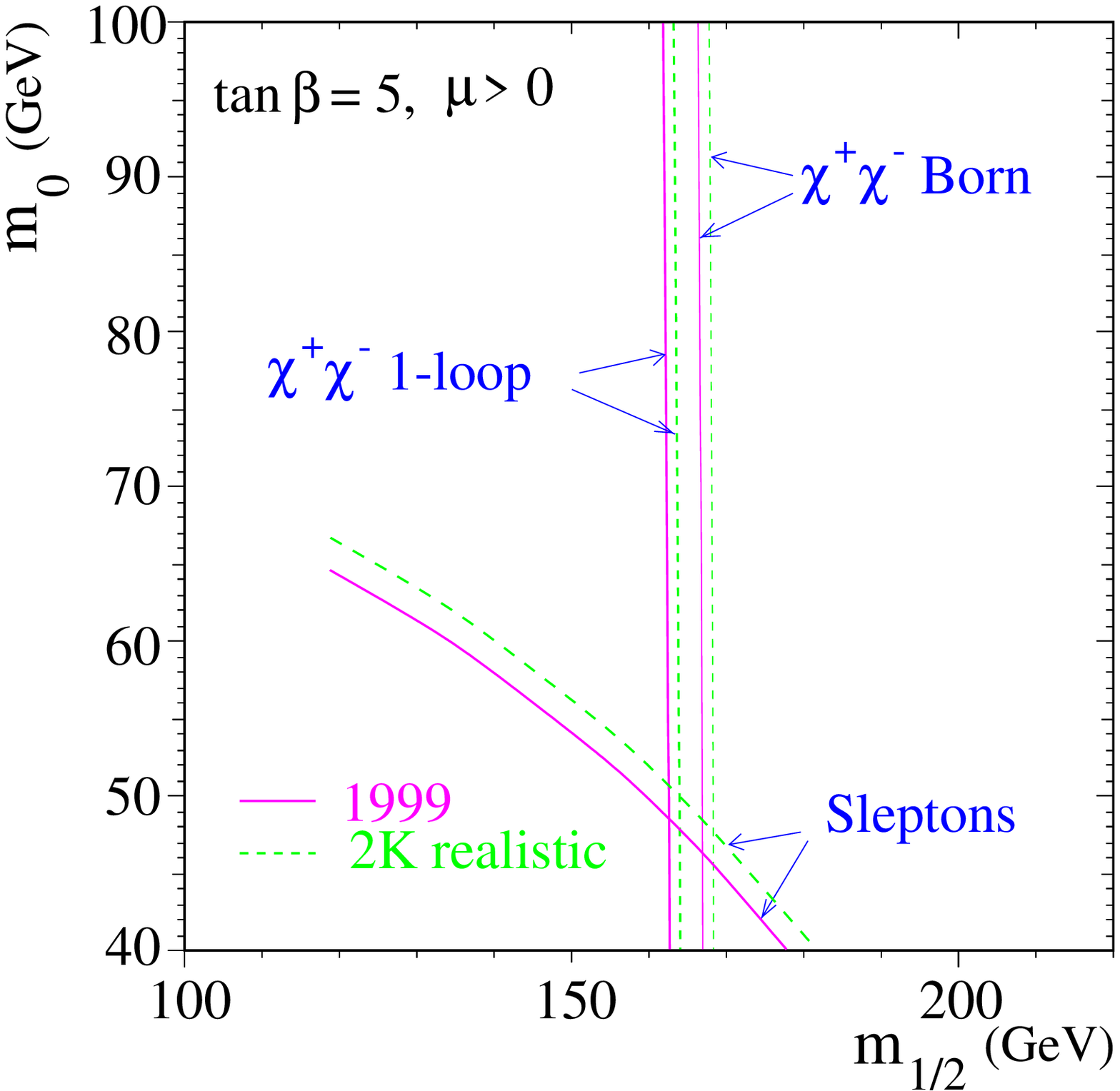,height=6.3cm}}
\end{center}
\caption[.]{\label{fig:radcorr} \it
The effects of radiative corrections to chargino masses in (a) the $(\mu,
M_2)$ plane for $\tan\beta = 5$ and (b) the $(m_{1/2}, m_0)$ plane for $\tan\beta =5$
and $\mu > 0$. The contours obtainable from the 1999 and `realistic' 2K
data are indicated, both with (thicker lines) and without 
radiative corrections (thinner lines). We notice that the differences
between the lines with and without radiative corrections are larger
than those between the 1999 and 2K lines.}
\end{figure}

We stress that radiative corrections to chargino and neutralino
masses, though less dramatic than to Higgs masses, are also
relevant to the interpretation of experimental limits on
physical particle masses in terms of constraints on MSSM
parameters such as $\mu$, $m_{1/2}$ and $m_0$~\cite{efgos}. Two such
effects
are seen in Fig.~\ref{fig:radcorr}. 
We see that the differences between the 1999 and 2K limits are
considerably smaller than the shifts induced by the radiative corrections.
In particular, as shown in the $\mu, m_{1/2}$ plane in panel (a),
radiative corrections are very significant in the delicate Higgsino region
discussed in Section 7. Their inclusion is indispensable for the accurate 
interpretation of the LEP data, as we do throughout this paper.

\section{The Case of Universal Higgs Masses}

We next apply the above accelerator constraints under the
assumption that the soft supersymmetry breaking masses are
universal, including the Higgs multiplets (UHM), exploring
their impact in the
the $m_{1/2}, m_0$ parameter plane and comparing them with the
constraints from cosmology on the relic abundance of the LSP.
We remind the reader that, in this UHM context,
for fixed $\tan \beta$ and sign of $\mu$, the
only parameter choice remaining is the value of $A_0$. We discuss
below two cases, one in which we require the absence of charge and
color breaking (CCB) minima~\cite{bbc}, but fix
$A_0 = -m_{1/2}$ so as to minimize their impact,
and the other in which we disregard CCB minima, and allow
$A_0$ to vary freely (UHM$_{\rm min}$).

\begin{figure}[htbp]
\begin{center}
\mbox{\epsfig{file=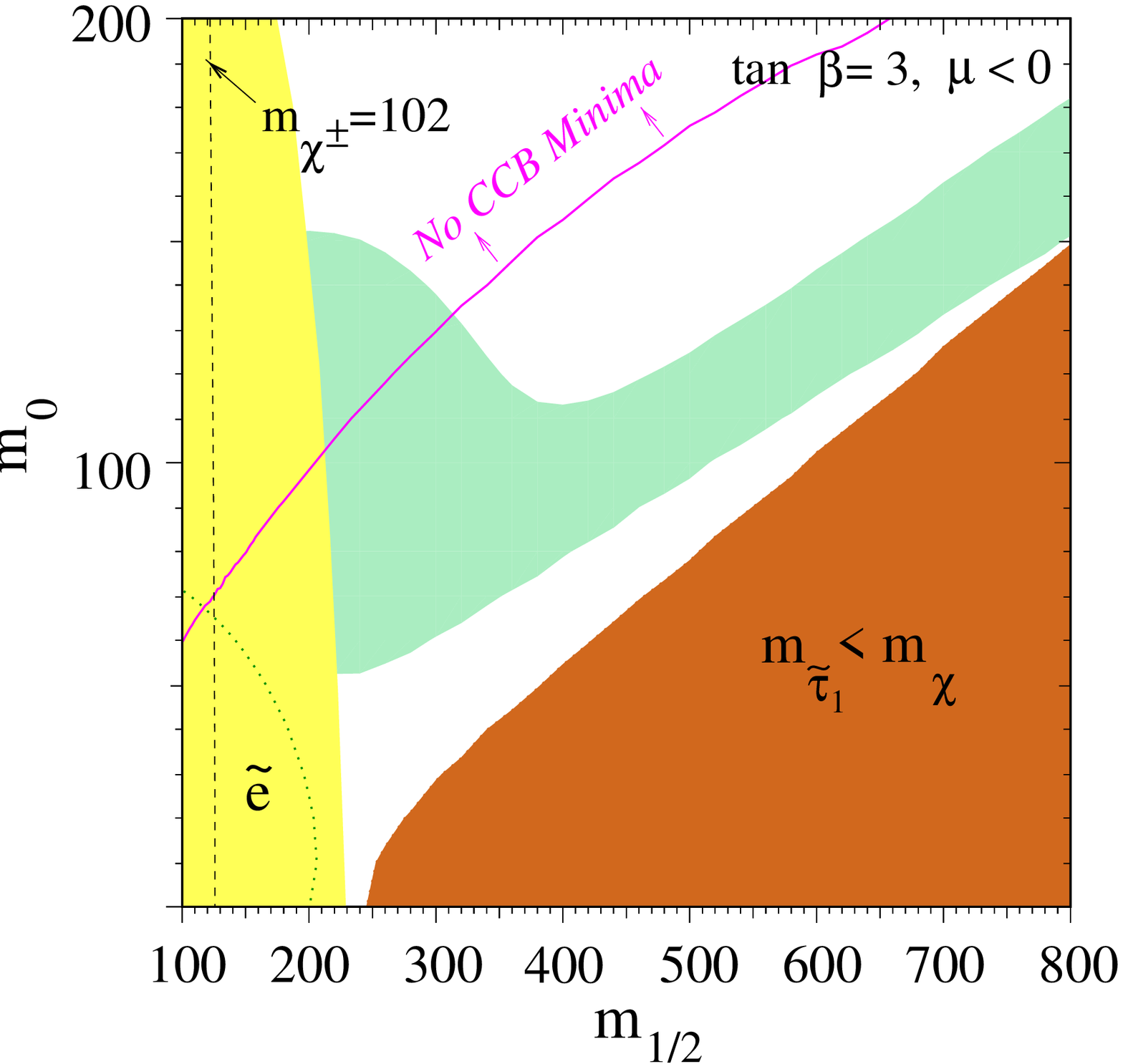,height=7.5cm}}
\mbox{\epsfig{file=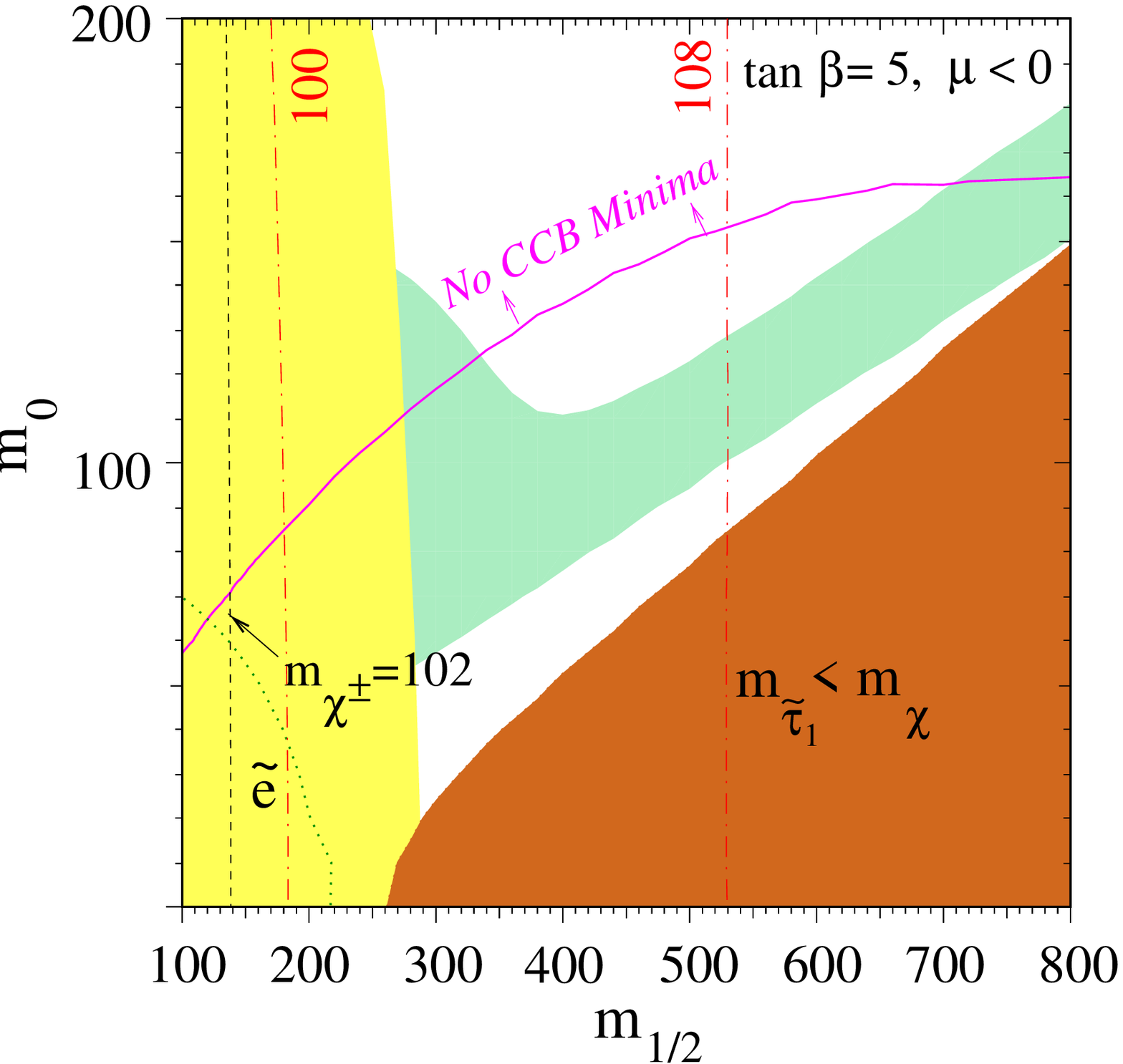,height=7.5cm}}
\mbox{\epsfig{file=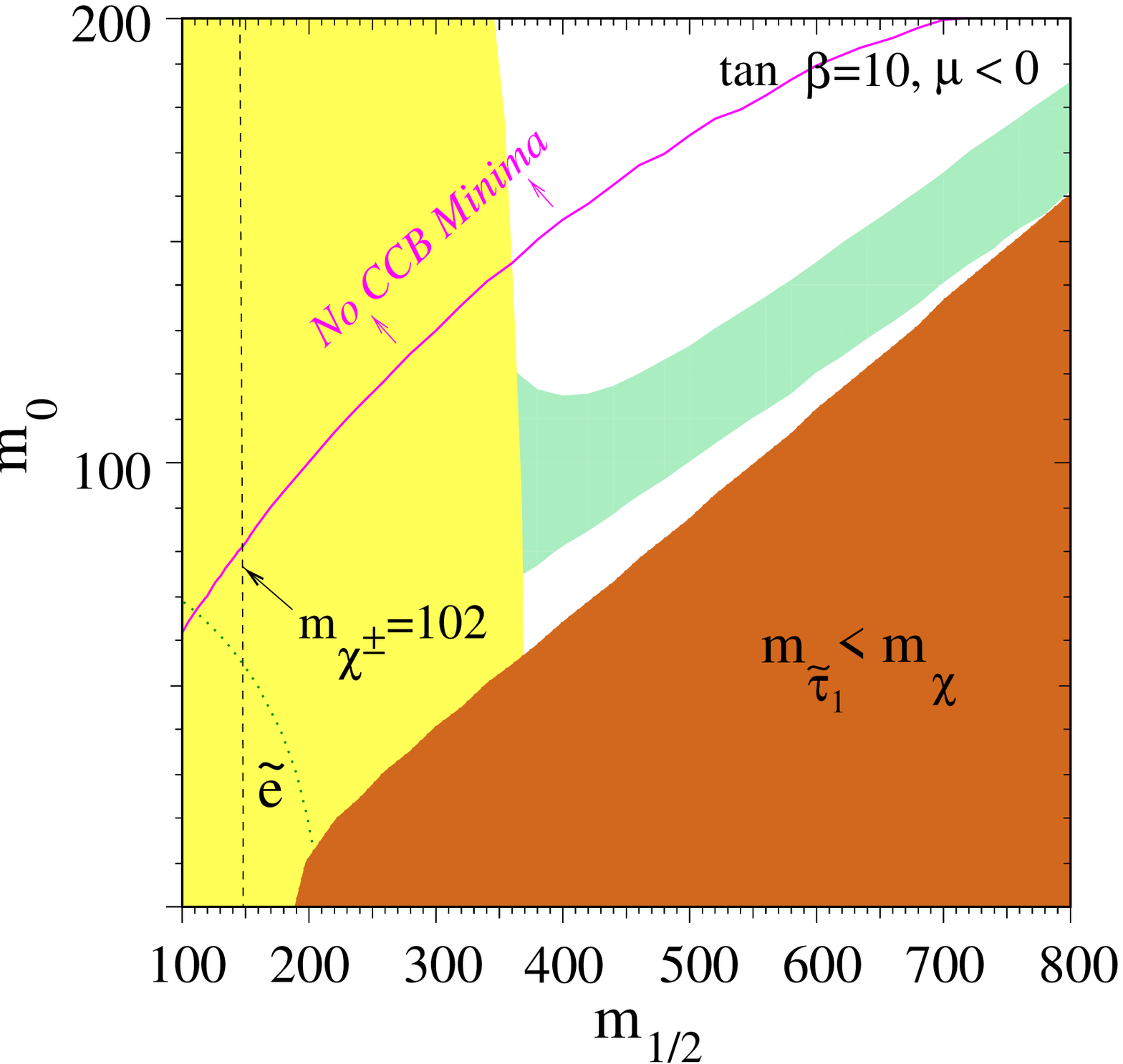,height=7.5cm}}
\mbox{\epsfig{file=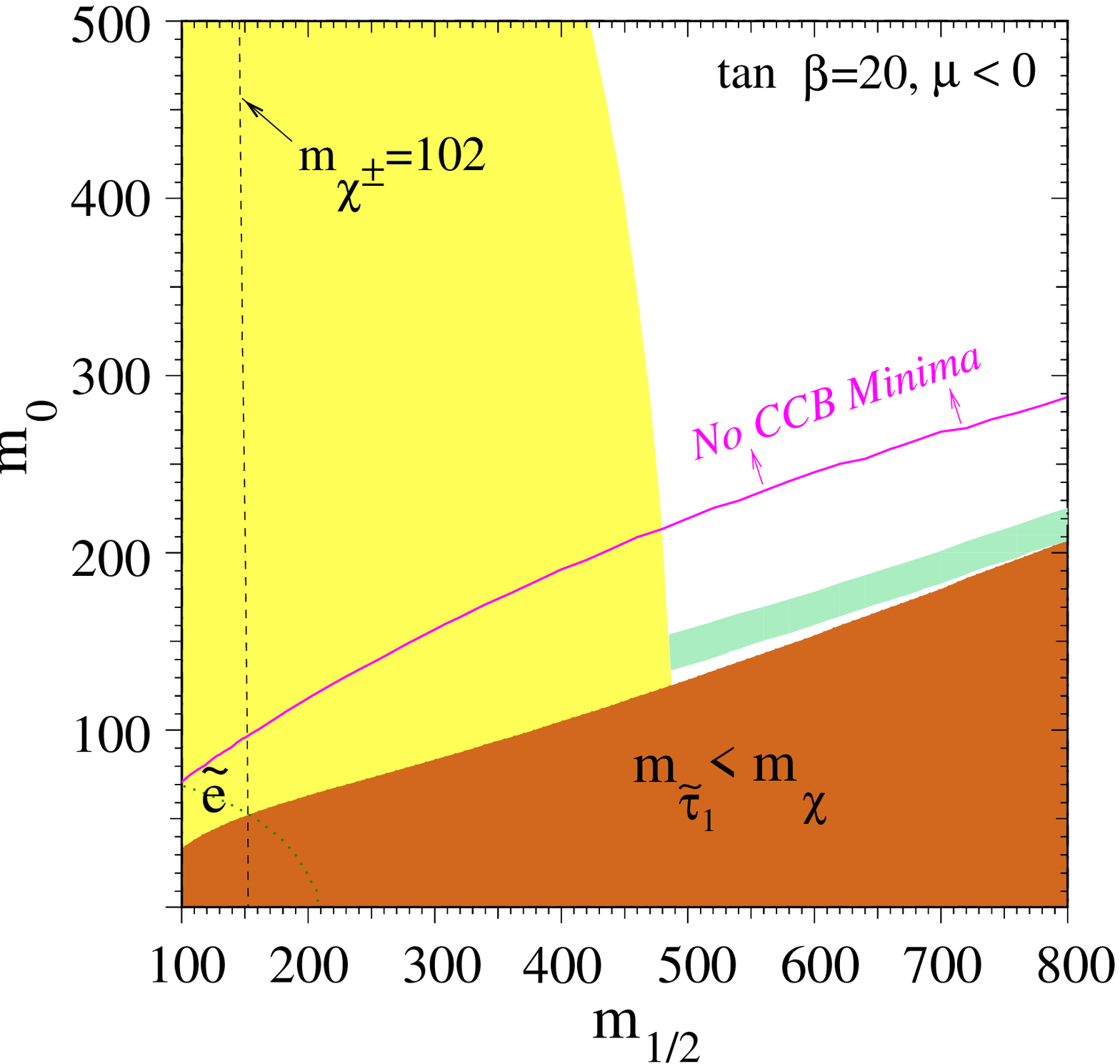,height=7.5cm}}
\end{center}
\caption[.]{\label{fig:m0mhalfnegative}\it
The $m_{1/2}, m_0$ plane for $\mu < 0$, $A = - m_{1/2}$ so as to
minimize the impact of the CCB constraint (indicated by a solid line) and
$\tan\beta =$ (a) 3, (b) 5, (c) 10 and (d) 20. The region
excluded by our $b \rightarrow s \gamma$ analysis has light shading. The
region allowed
by the cosmological constraint $0.1 \le \Omega_\chi h^2 \le 0.3$,
after including coannihilations, has medium shading. Dotted lines
delineate the announced LEP constraint on the $\tilde e$ mass and 
the disallowed region where $m_{\tilde \tau_1} < m_\chi$ has dark shading.
The contour $m_{\chi^\pm} = 102$~GeV is shown
as a near-vertical dashed line in each panel. Also shown as
dot-dashed lines are relevant Higgs mass contours.}
\end{figure}

\begin{figure}[htbp]
\begin{center}
\mbox{\epsfig{file=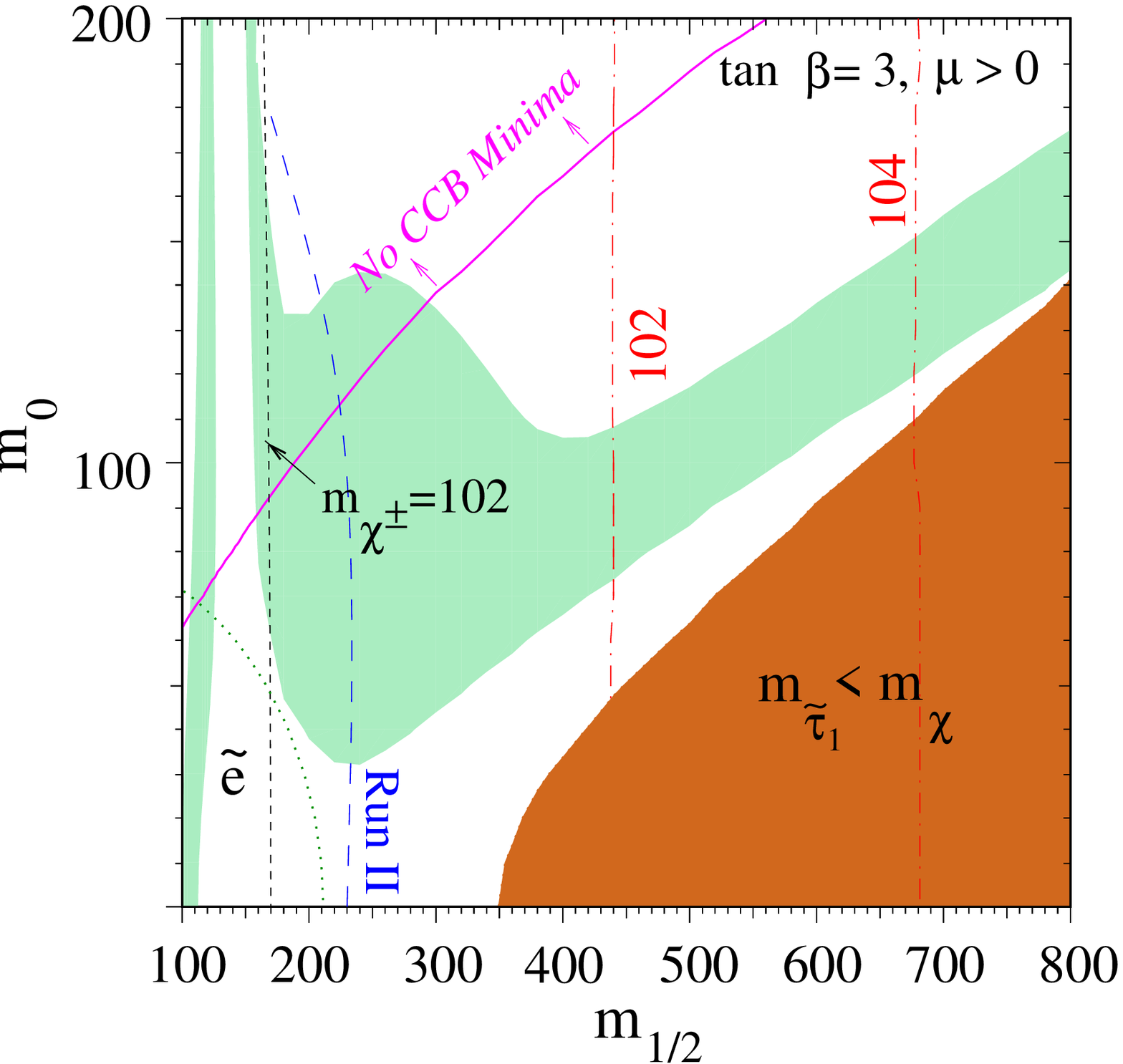,height=7.5cm}}
\mbox{\epsfig{file=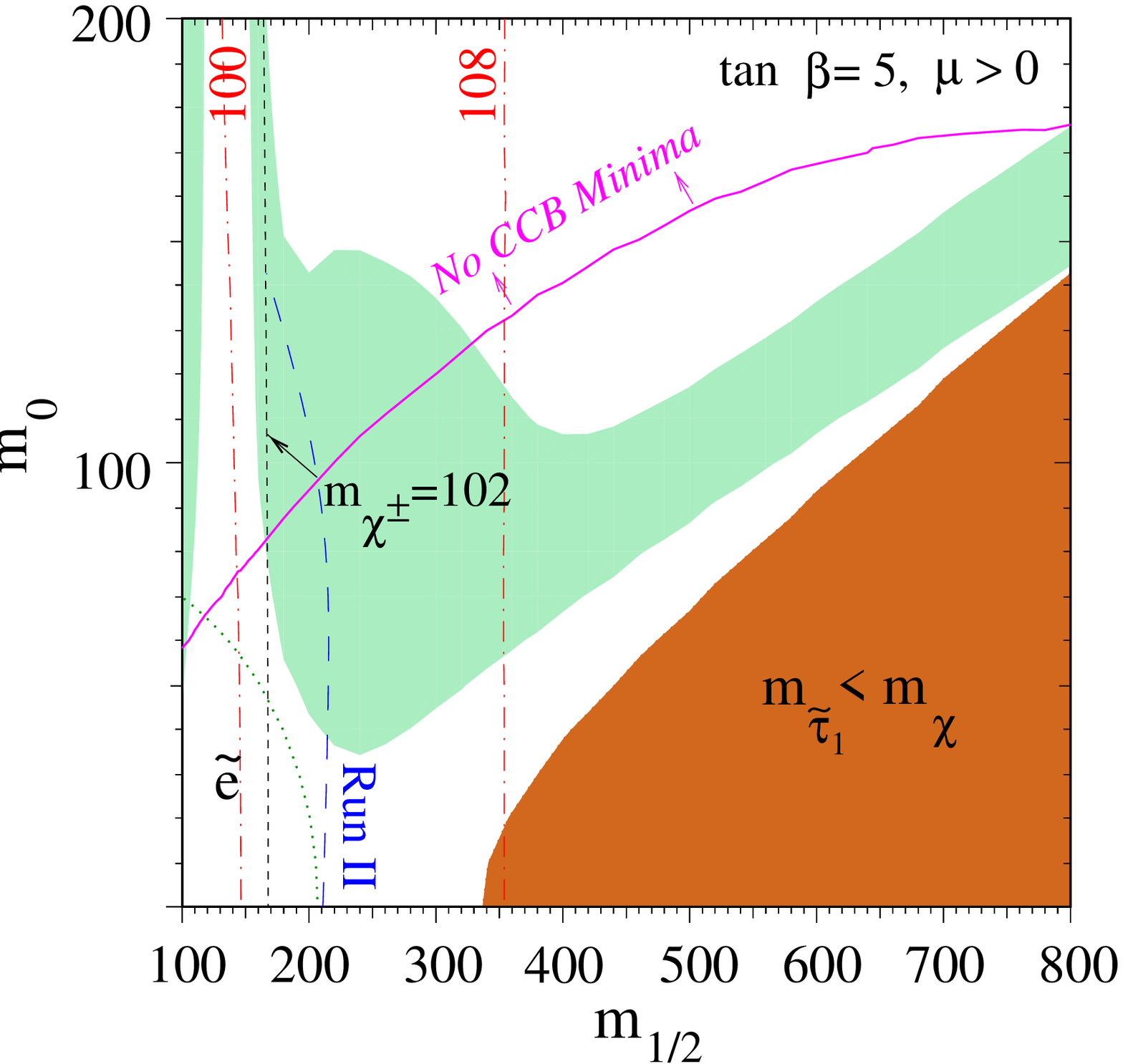,height=7.5cm}}
\mbox{\epsfig{file=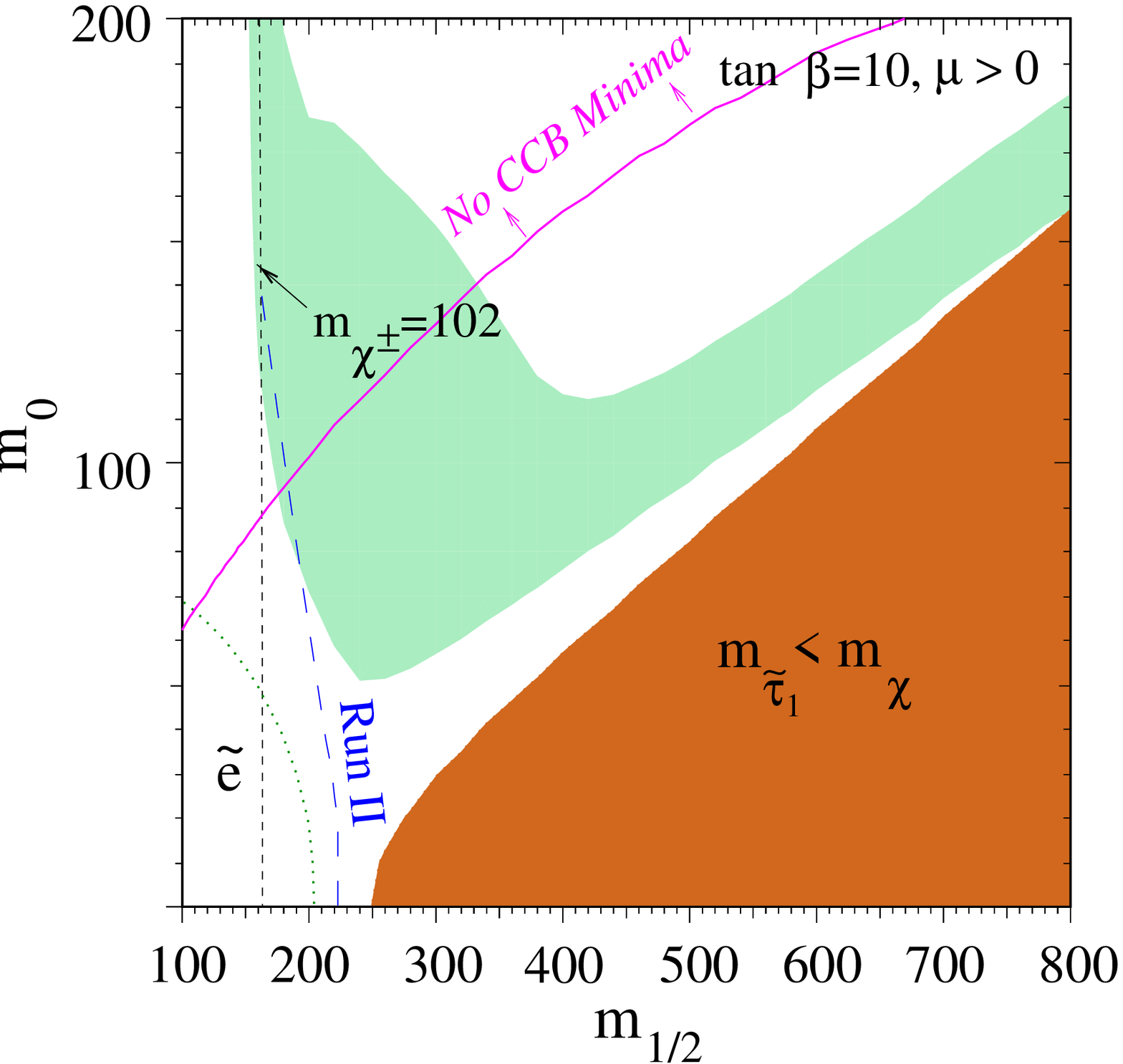,height=7.5cm}}
\mbox{\epsfig{file=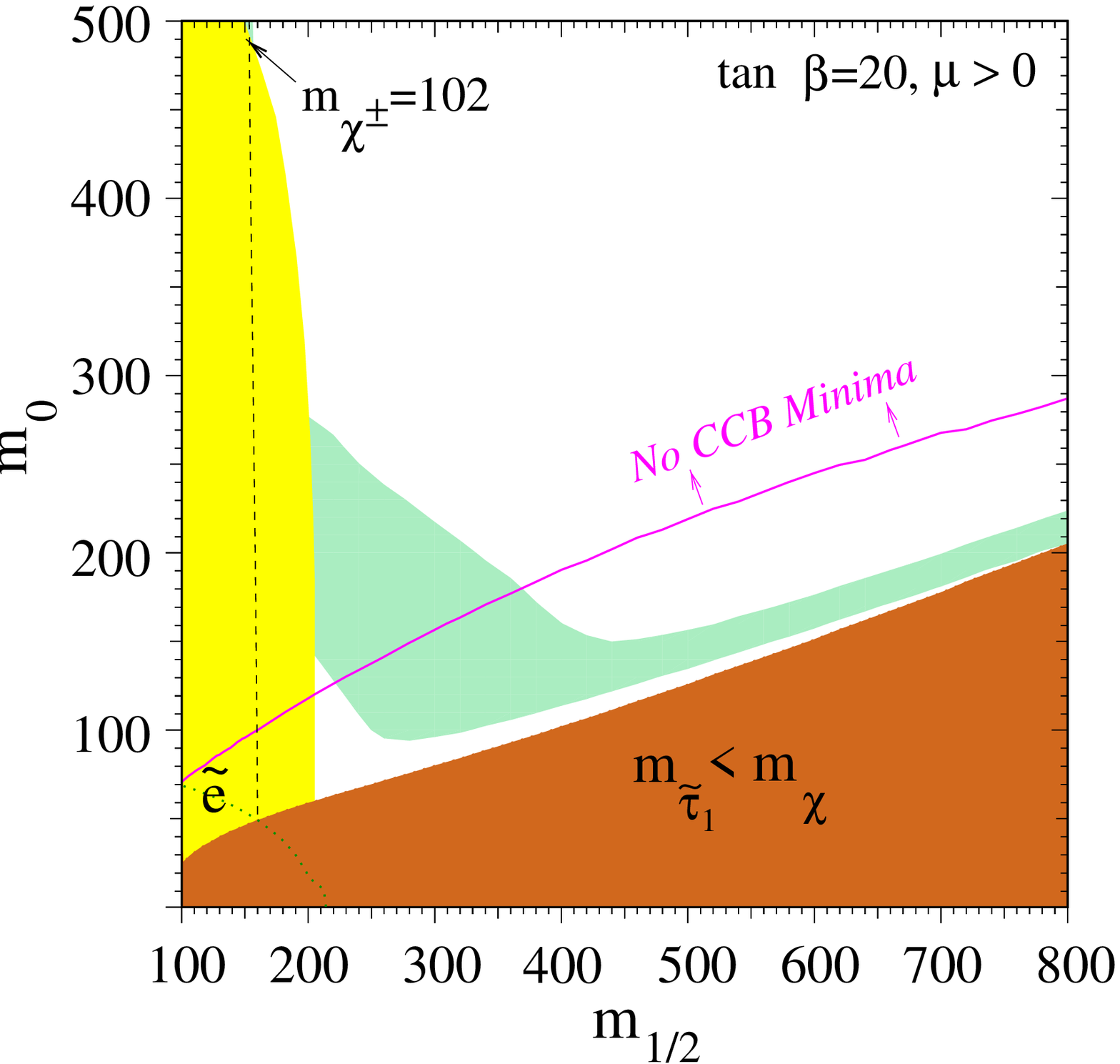,height=7.5cm}}
\end{center}
\caption[.]{\label{fig:m0mhalfpositive}\it
The $m_{1/2}, m_0$ plane for $\mu > 0$, $A = - m_{1/2}$ and
$\tan\beta =$ (a) 3, (b) 5, (c) 10 and (d) 20. The 
significances of the curves and
shadings are the same as in Fig.~\ref{fig:m0mhalfnegative}.
The light-shaded region in panel (d) is excluded by the
$b \rightarrow s \gamma$ constraint.
The long dashed curves in panels (a), (b) and (c) represent the
anticipated limits from
trilepton searches at Run II of the Tevatron \cite{sugrarept}.
}
\end{figure}

We start with the CCB UHM case shown in
Figs.~\ref{fig:m0mhalfnegative} and ~\ref{fig:m0mhalfpositive}.
One is safe from CCB minima above the curved solid lines in these plots,
which are calculated with $A_0 = - m_{1/2}$, so as to minimize the
impact of this prospective constraint. 
Plotted as near-vertical dashed lines are
the chargino mass contours: $m_\chi^\pm = 102$. 
At the higher values of
$\tan \beta$, the bound on $m_{1/2}$ from the chargino mass limit is
nearly independent of
$\tan \beta$, and the dependence is always very slight for $\mu > 0$.
For completeness we also show the limit from the selectron
mass bound.

The light shaded regions in Fig~\ref{fig:m0mhalfnegative} are
those excluded by the $b \rightarrow s \gamma$ constraint
discussed in Section 3.
We see that, for $\mu < 0$, the impact of $\Bsg$   
constraints increases sizeably with
$\tan\beta$, and covers in these cases a significant fraction of
the region otherwise preferred for dark matter reasons. The cut-off at
large $m_0$, $m_{1/2}$ is due to
the corresponding increase of $m_A$, and hence $m_{H^\pm}$, in the
UHM, which reduces the charged
Higgs contribution. The exclusion is not  
very sensitive to the $A_0$ value chosen.
As we can see in Fig~\ref{fig:m0mhalfpositive}, for $\mu>0$ the interference 
between the supersymmetric and charged Higgs contributions cancel
the effect of new physics in the low-medium $\tan\beta$ range; 
however, there is still some sensitivity at large $\tan\beta$ 
where the large negative supersymmetric contribution 
make $\Bsg$ significantly smaller than the measured value, as seen in
panel (d) of Figs.~\ref{fig:m0mhalfpositive}.

\begin{figure}[htbp]
\begin{center}
\mbox{\epsfig{file=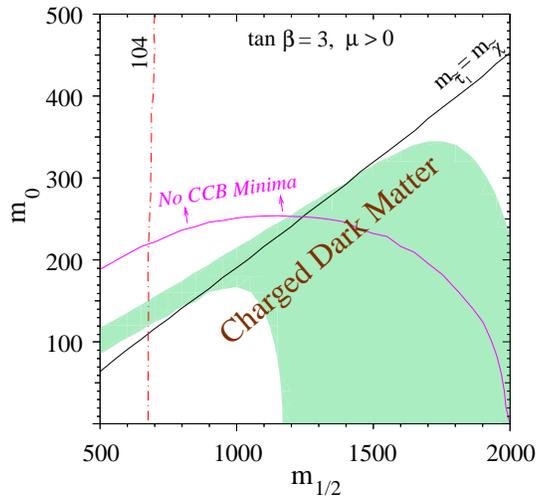,height=6.5cm}}
\end{center}
\caption[.]{\label{fig:bigone}\it
An extension of the $m_{1/2}, m_0$ plane for $\mu > 0$, $A = - m_{1/2}$
and
$\tan\beta =$ 3. Below the solid diagonal line, the LSP is charged and
hence excluded. The absolute upper bound to $m_{1/2}$ is found when the
shaded region drops entirely below this contour.}
\end{figure}

Also shown in these figures by near-vertical dot-dashed lines are
the limits coming from the Higgs mass bounds. Note that for $\mu <0$,
these contours only appear for $\tan \beta = 5$, where we display the 100
and 108 GeV contours corresponding to the 1999 and prospective
`realistic' 2K experimental limit, allowing a safety margin of 3 GeV as
discussed earlier. At $\tan
\beta = 3$, the position of these contours is far off to the right,
excluding the entire region displayed. At $\tan \beta = 10$ and 20, the
contours would appear to the left of the chargino bound and are not
shown. For $\mu > 0$, the limits are weaker, i.e., the contours move to
the left. In the case of $\tan \beta = 3$, the 
102 GeV and 104 GeV contours
have now moved into the displayed range of $m_{1/2}$, and the
contours for 100 GeV and 108 GeV are shown for $\tan\beta = 5$. In
Fig.~\ref{fig:bigone},
we show an extended range in $m_{1/2}$ and the position of the 104 GeV
contour for $\tan \beta = 3, \mu > 0$. 

The Higgs mass contours depend on $m_0$, because the radiative
corrections to the Higgs mass cause these curves to bend left at very
large $m_0$. Thus ultimately, the only pure accelerator constraint on
$m_{1/2}$ comes from the chargino mass limit.  However, as we have
demonstrated previously, cosmology excludes such high values of
$m_0$, thus
maintaining the importance of the Higgs mass bound in limiting $m_{1/2}$,
and ultimately $m_\chi$ and $\tan \beta$. The medium-shaded regions
in
Figs.~\ref{fig:m0mhalfnegative} and ~\ref{fig:m0mhalfpositive} show the
areas in the $m_0, m_{1/2}$ plane for which the relic cosmological
density falls between 0.1 $< \Omega h^2 <$ 0.3 when co-annihilation
effects are included. 
We note that the chargino mass constraint now essentially excludes the
re-entrant parts of the dark matter density contours caused by
resonant direct-channel annihilations when $m_{1/2} \le 160$~GeV,
which were visible in Fig.~\ref{fig:big3} as well as
Fig.~\ref{fig:m0mhalfpositive}.
The dark shaded regions in Figs.~\ref{fig:m0mhalfnegative} and
\ref{fig:m0mhalfpositive} correspond to a charged LSP, as indicated.

We show in Fig.~\ref{fig:bigone} the `tail' of the cosmological region
where $m_\chi \sim m_{\tilde \tau_1}$ for $\tan\beta = 3$. As can be seen
in Fig.~7 of
\cite{EFOSi}, the tip of this region is allowed by the CCB constraint for
$\tan\beta = 3, 10$, and we can see in Fig.~\ref{fig:m0mhalfnegative}
that the CCB constraint is weaker for $\tan\beta = 5$. We show in
Fig.~\ref{fig:bigone} the $m_h = 104$~GeV contour, corresponding to 
the 1999 bound on
the Higgs mass after allowing for a 3 GeV theoretical
uncertainty in the prediction.
We recall that the Higgs mass limit is even stronger for
negative $\mu$.  We can safely
set the limit $\tan \beta > 2.8$ for $\mu > 0$ in the UHM. Overall, the 
limits we obtain on $\tan \beta$ in different LEP scenarios 
for both signs of $\mu$ are shown in Table 2.

\begin{table}[htbp]
  \begin{center}
    \begin{tabular}{|l|c|c|}
\hline
&1999&`realistic' 2K\\
\hline
$\mu<0\;\;\;\;\;$&3.2&4.0\\[.2em]
$\mu>0$&2.8&3.6\\
\hline
    \end{tabular}
    \caption{\label{tab:tbvmh}
    {\it Limits on $\tb$ imposed in the UHM by the 1999 and
`realistic' expected 2K Higgs mass limits.}}
  \end{center}
\end{table}

\begin{figure}[htbp]
\begin{center}
\mbox{\epsfig{file=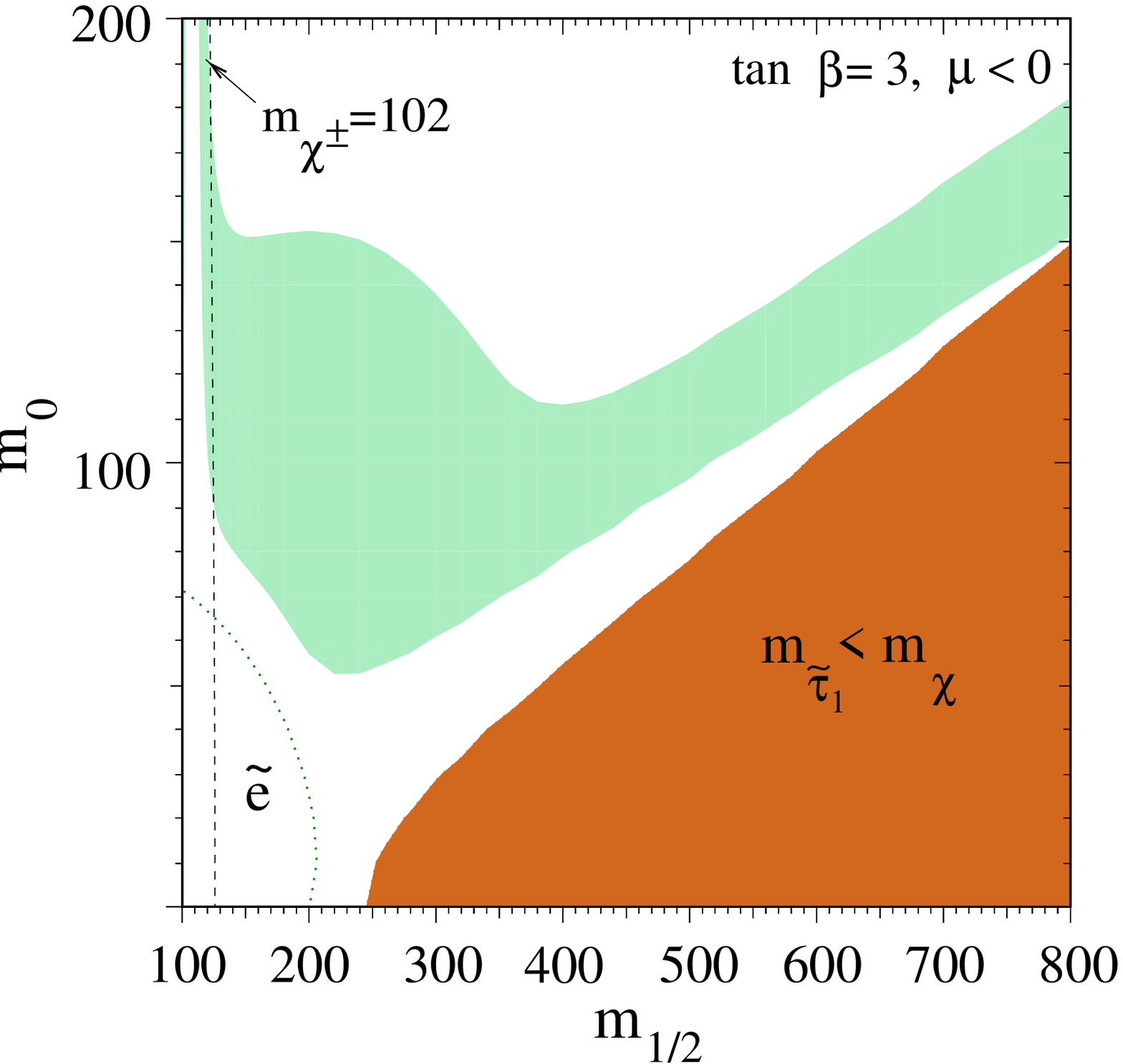,height=7.5cm}}
\mbox{\epsfig{file=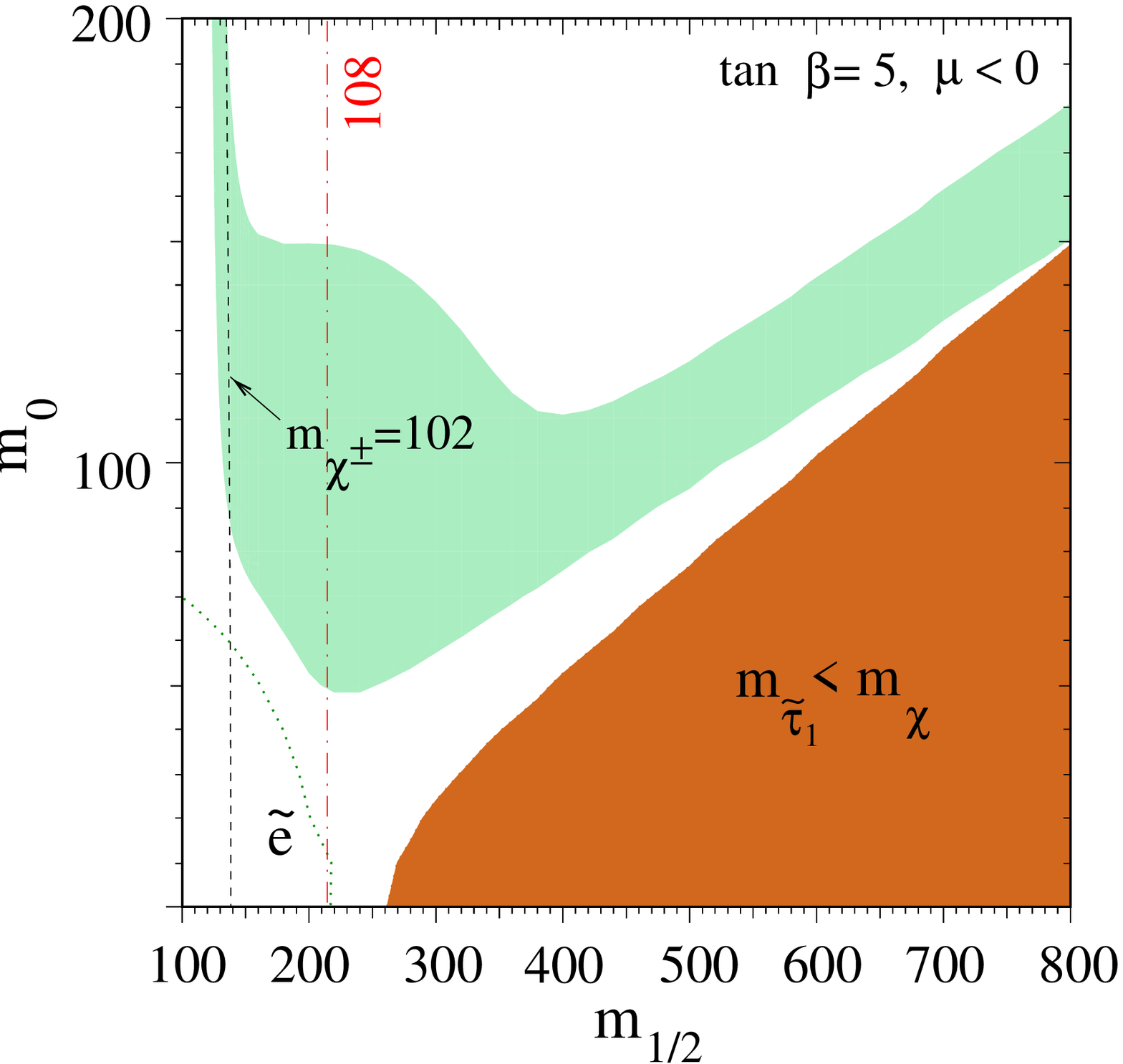,height=7.5cm}}
\mbox{\epsfig{file=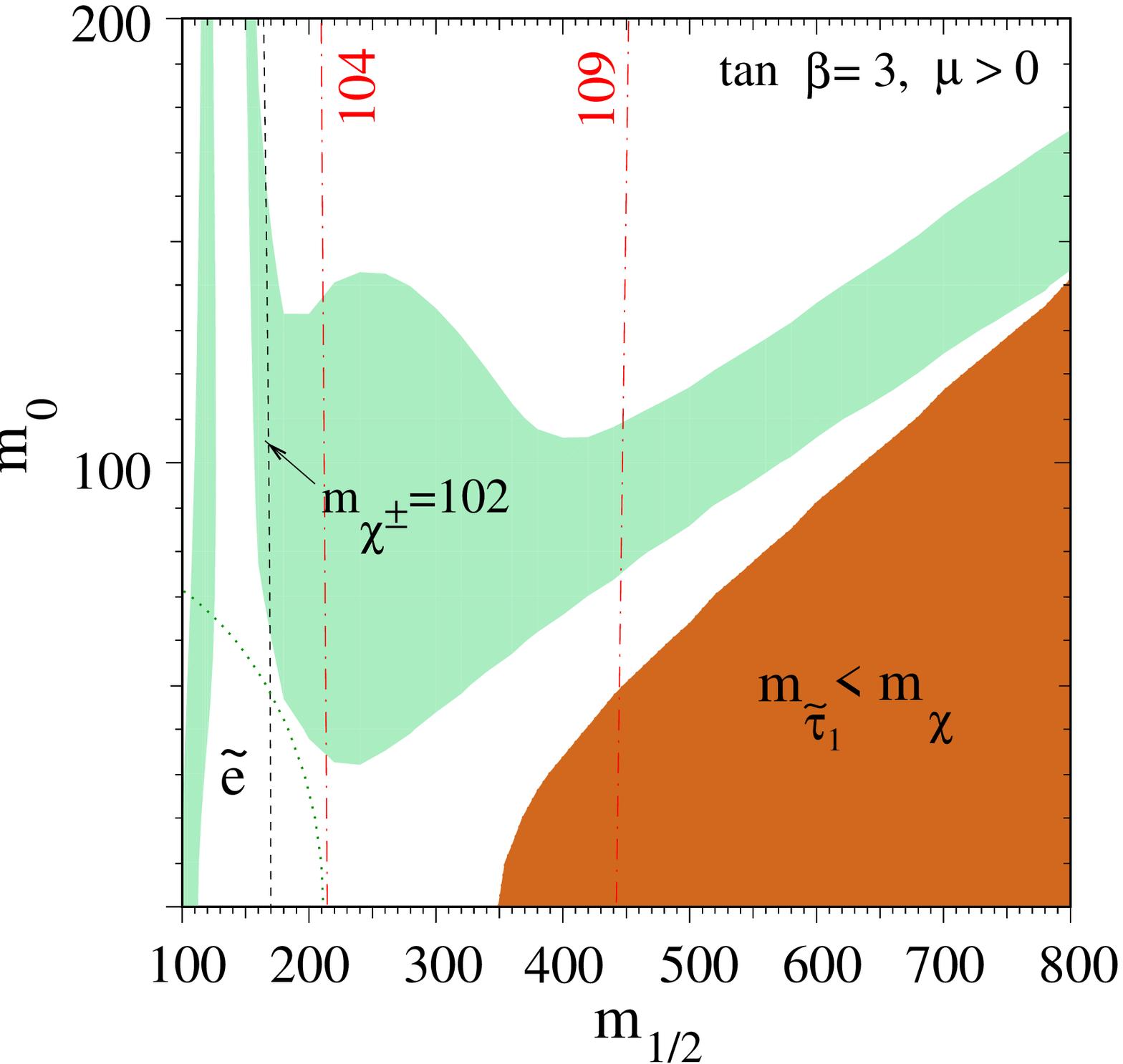,height=7.5cm}}
\mbox{\epsfig{file=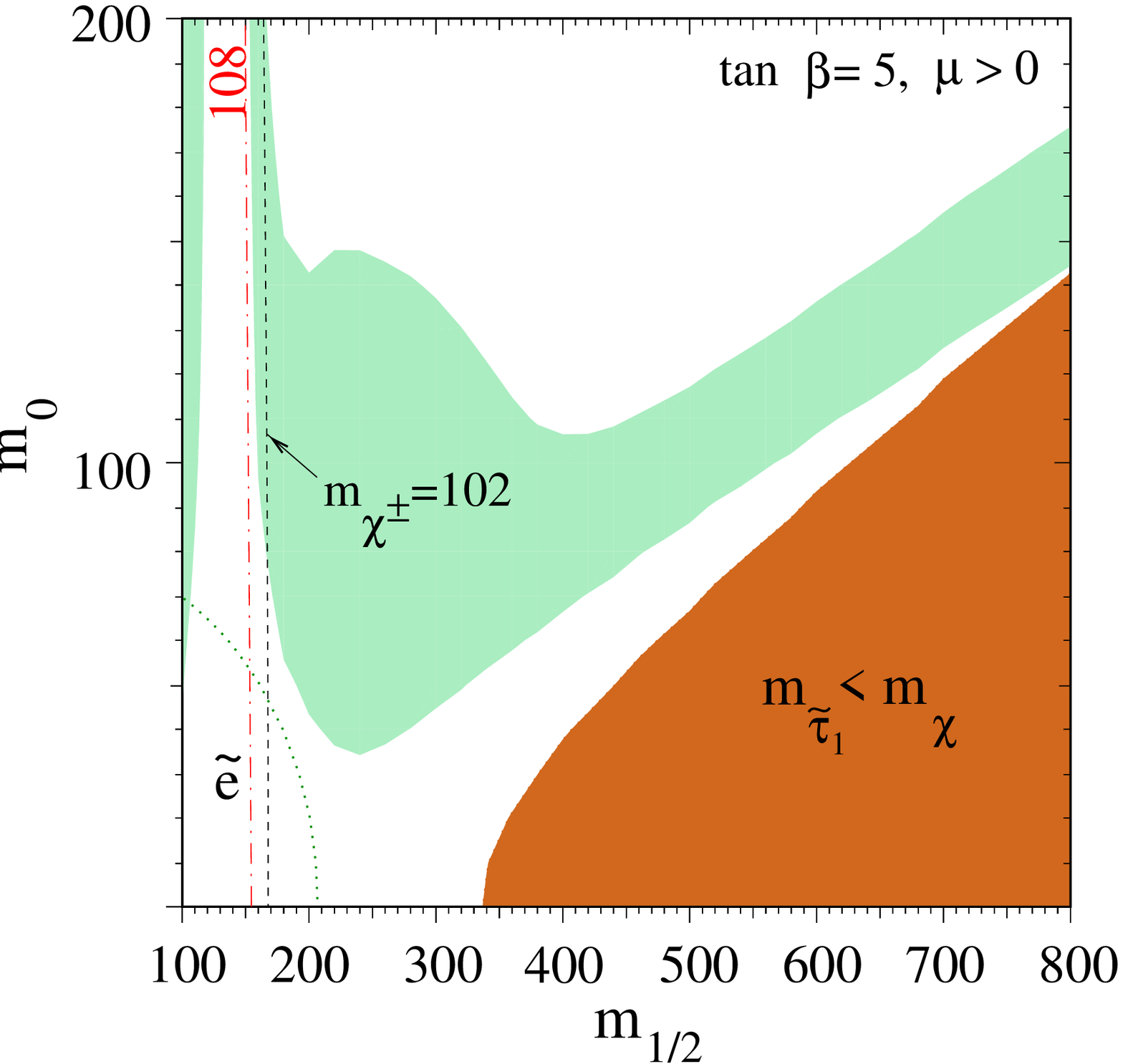,height=7.5cm}}
\end{center}
\caption[.]{\label{fig:UHMnonCCB}\it
The $m_{1/2}, m_0$ plane for $\mu < 0$ and
(a) $\tan\beta =$ 3, (b) $\tan\beta =$ 5, and $\mu > 0$ and  (c)
$\tan\beta =$ 3, (d) $\tan\beta =$ 5,
with $A_0$
allowed to vary, and no CCB constraint applied. For
clarity, the region excluded by the $b \rightarrow s \gamma$
constraint has not been shaded: it is essentially identical
to that in Fig.~\ref{fig:m0mhalfnegative}. The region
allowed after including coannihilations,
by the cosmological constraint $0.1 \le \Omega_\chi h^2 \le 0.3$
has medium shading. Dotted lines
delineate the announced LEP constraint on the $\tilde e$ mass and
the disallowed region where $m_{\tilde \tau_1} < m_\chi$ has dark shading.
The contour $m_{\chi^\pm} = 102$~GeV is shown
as a near-vertical dashed line in each panel. Also shown as
dot-dashed lines are relevant Higgs mass contours.}
\end{figure}

We now repeat the above UHM analysis for the more conservative case
(UHM$_{\rm min}$)
in which we do not require stability against collapse into
a CCB vacuum, and we allow $A_0$ to vary as far as possible, consistent
with the experimental constraints on $m_{\tilde t_1}$ in particular.
Fig.~\ref{fig:UHMnonCCB} shows
options for $\tan\beta$ and the sign of
$\mu$ which exhibit interesting differences from the previous CCB case.
For simplicity, we have chosen not to include the regions that are
excluded by the $b \rightarrow s \gamma$ constraint: these turn out to be
essentially independent of $A_0$, and hence may be taken from the
corresponding panels in Fig~\ref{fig:m0mhalfnegative}. The interesting
and significant differences are in the contours of the Higgs mass.
The Higgs mass is sensitive to $A_0$, and may be significantly lower than
in the previous CCB case, with corresponding implications for the lower
limits on $m_\chi$ and $\tan\beta$ that we quote below.

Clearly, by allowing $A_0$ to vary (rather than restrict its value to 
$-m_{1/2}$), we expect weaker bounds from the Higgs mass than those 
found when the CCB constraints were incorporated. 
Since the Higgs mass constraint was only important for lower values of
$\tan \beta$, we show results only for $\tan \beta = 3, 5$ in
Fig.~\ref{fig:UHMnonCCB}.  Whilst, for $\tan \beta  = 3$ and $ \mu < 0$,
the 104 and 108 GeV Higgs mass contours are still to the right of the
displayed region in the figure, we see that in the other cases shown,
all of the contours are moved substantially to the left. In fact for
$\tan \beta=5$ and $\mu > 0$, the Higgs mass bound is no longer
competitive with the chargino bound. 
As in the previous UHM case, we also find lower bounds on $\tan \beta$
in this case where the CCB constraint is relaxed, as shown in Table 3
for different LEP running scenarios and the two signs of $\mu$.

\begin{table}[htbp]
  \begin{center}
    \begin{tabular}{|l|c|c|}
\hline
&1999&`realistic' 2K\\
\hline
$\mu<0\;\;\;\;\;$&2.7&3.1\\[.2em]
$\mu>0$&2.2&2.7\\
\hline
    \end{tabular}
    \caption{\label{tab:tbvmhc}
    {\it Limits on $\tb$ imposed in the UHM by the 1999 and
`realistic' expected 2K Higgs mass limits, relaxing the requirement
that there be no CCB vacuum.}}
  \end{center}
\end{table}

\section{Bounds for Non-Universal Scalar Masses}

In the previous Section, we have derived stringent limits on the
$m_{1/2}, m_0$ plane from the absence of sparticles and Higgs
bosons at LEP, assuming universality for scalar masses including the soft
Higgs masses (UHM). These limits are particularly strong at low values of
$\tan
\beta$, and in fact exclude $\tan \beta \la 2.8$. One should expect
these limits to weaken when the assumption of UHM is relaxed (nUHM). 
In this Section, we rederive the appropriate limits in the  $m_{1/2},
m_0$ plane for the more general nUHM case. 
We now treat both $\mu$ and $m_A$ as independent
parameters, in addition to the free parameters $m_{1/2}, m_0, A$, and
$\tan \beta$ of the previous Section.
As indicated in Section 2, we take $m_A = 10$ TeV, so that the Higgs
mass limits give the most conservative bounds on $m_{1/2}$.
As before, we restrict the values of $A$ by requiring that
$m_{{\tilde t}_1}$ be consistent with the experimental lower limit,
and $\mchi>m_{\tilde\tau_1}$, as shown in Fig.~\ref{fig:avm}. 

\begin{figure}[htbp]
\begin{center}
\mbox{\epsfig{file=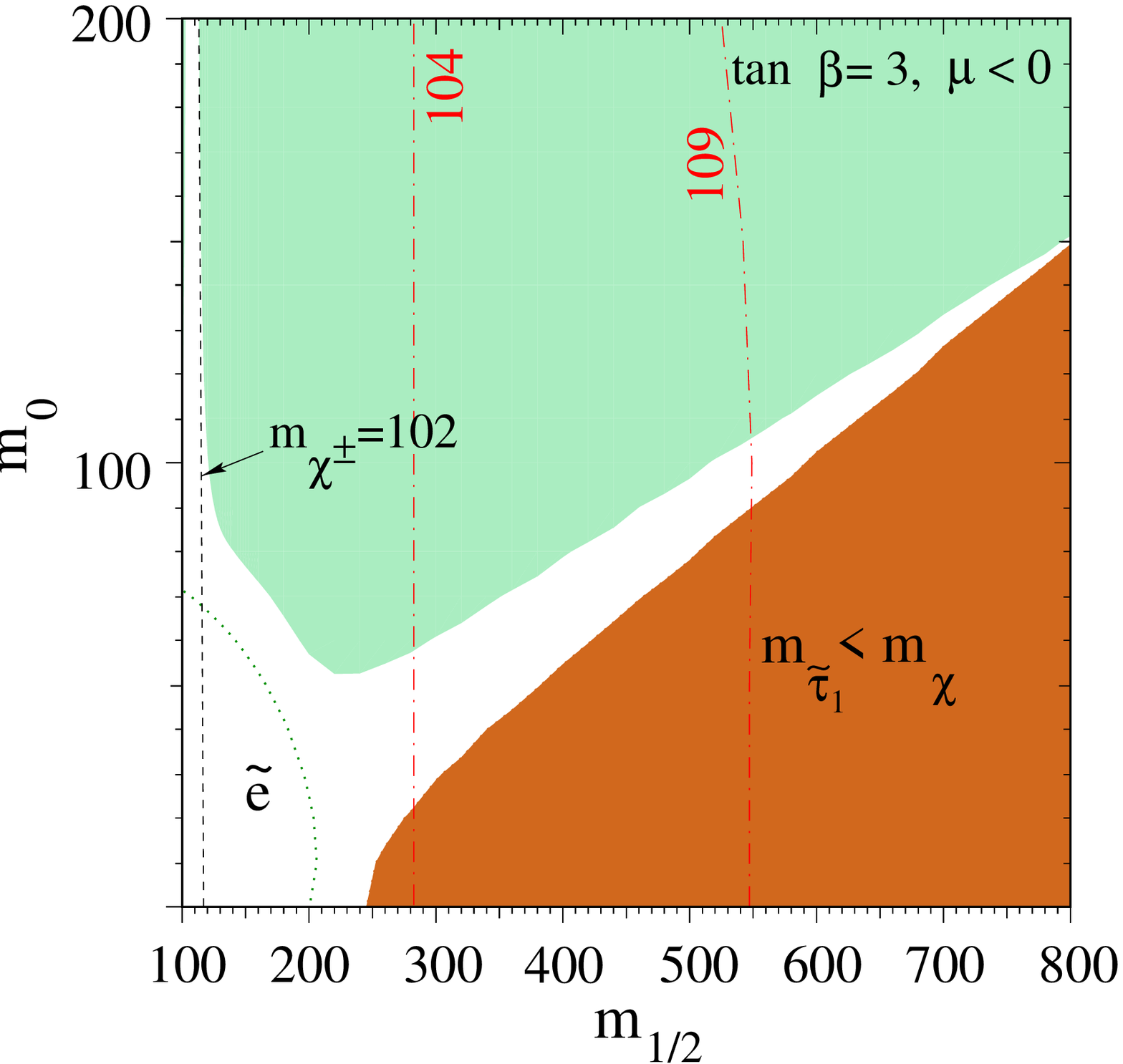,height=7.5cm}}
\mbox{\epsfig{file=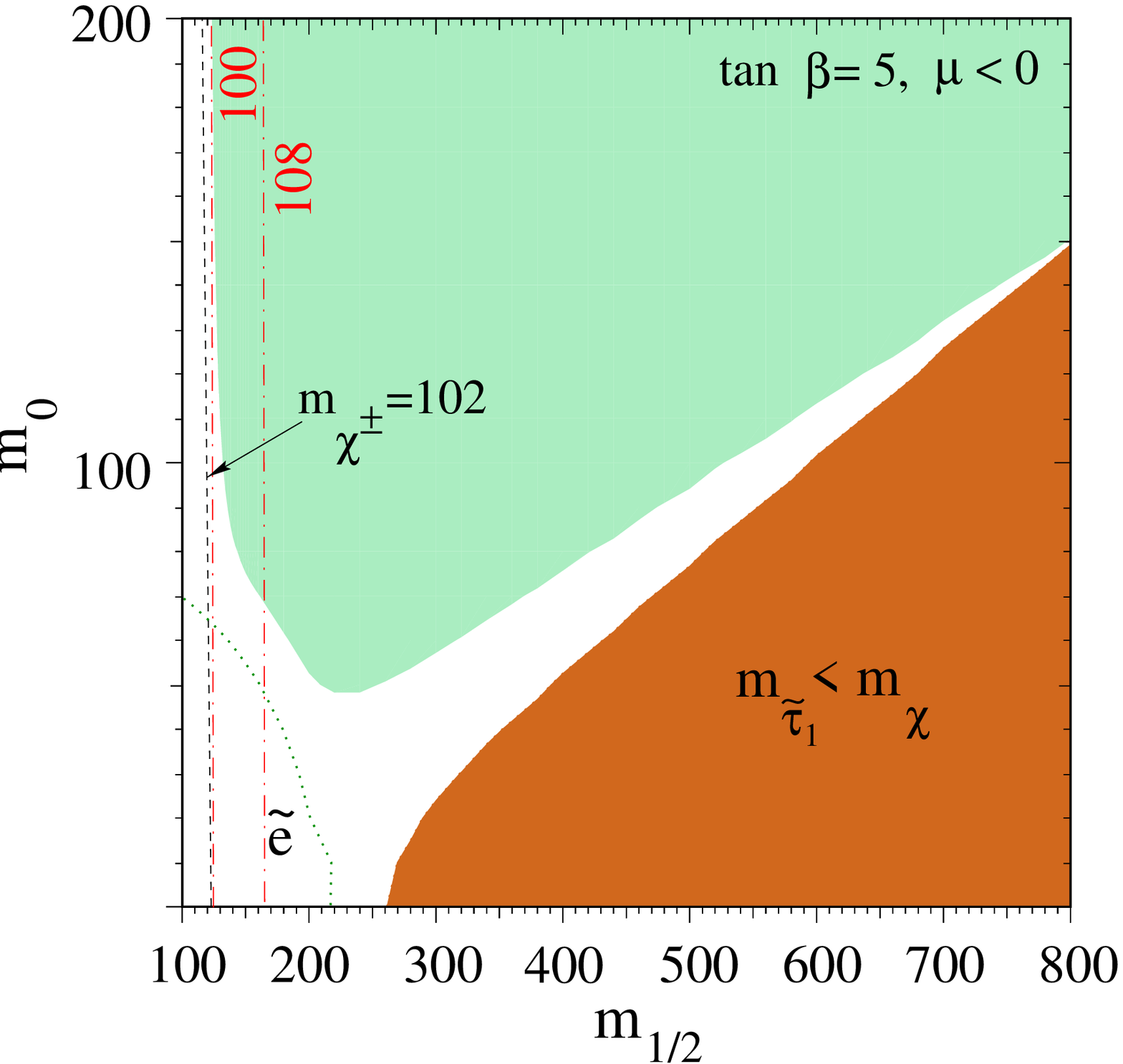,height=7.5cm}}
\mbox{\epsfig{file=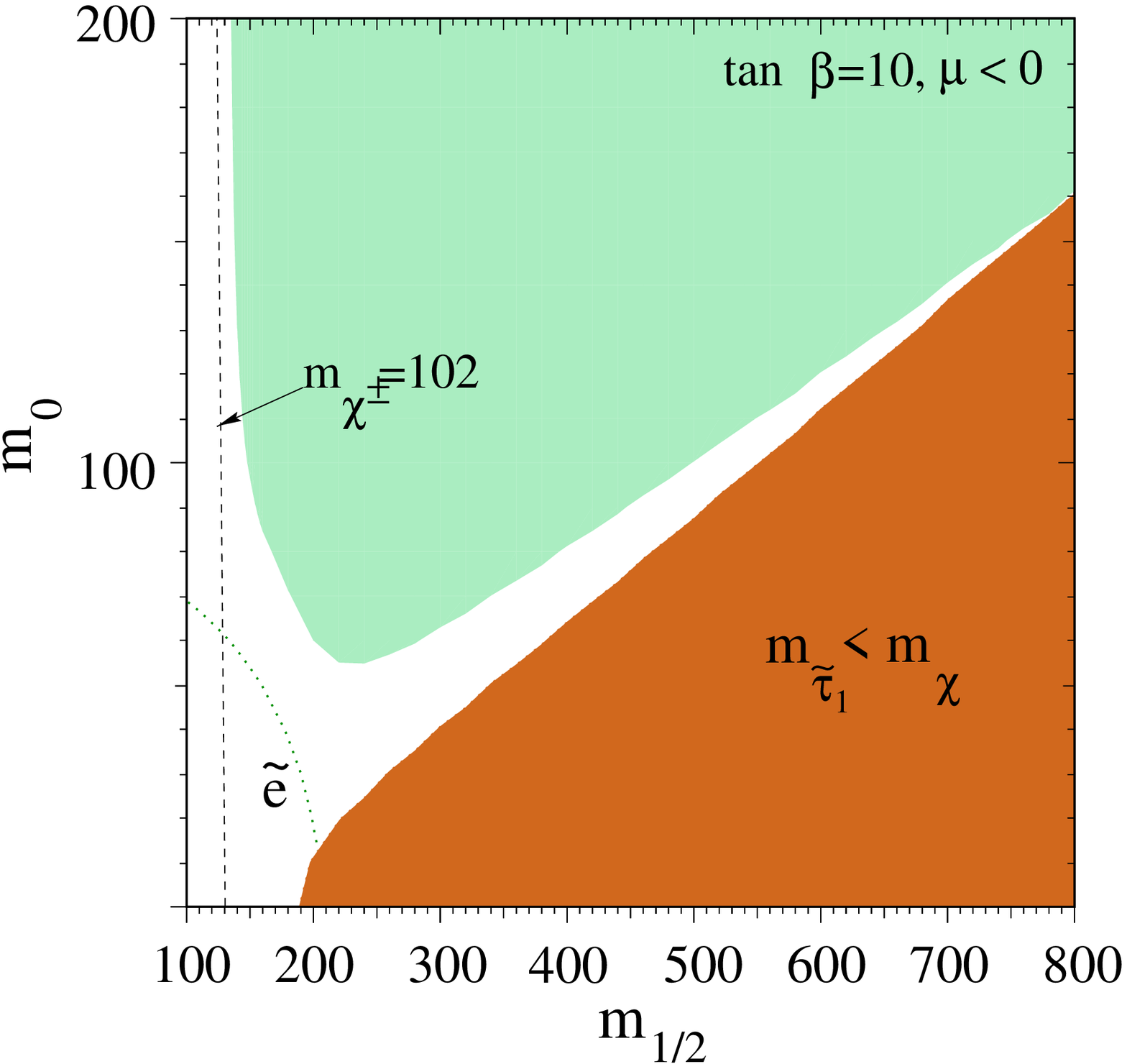,height=7.5cm}}
\mbox{\epsfig{file=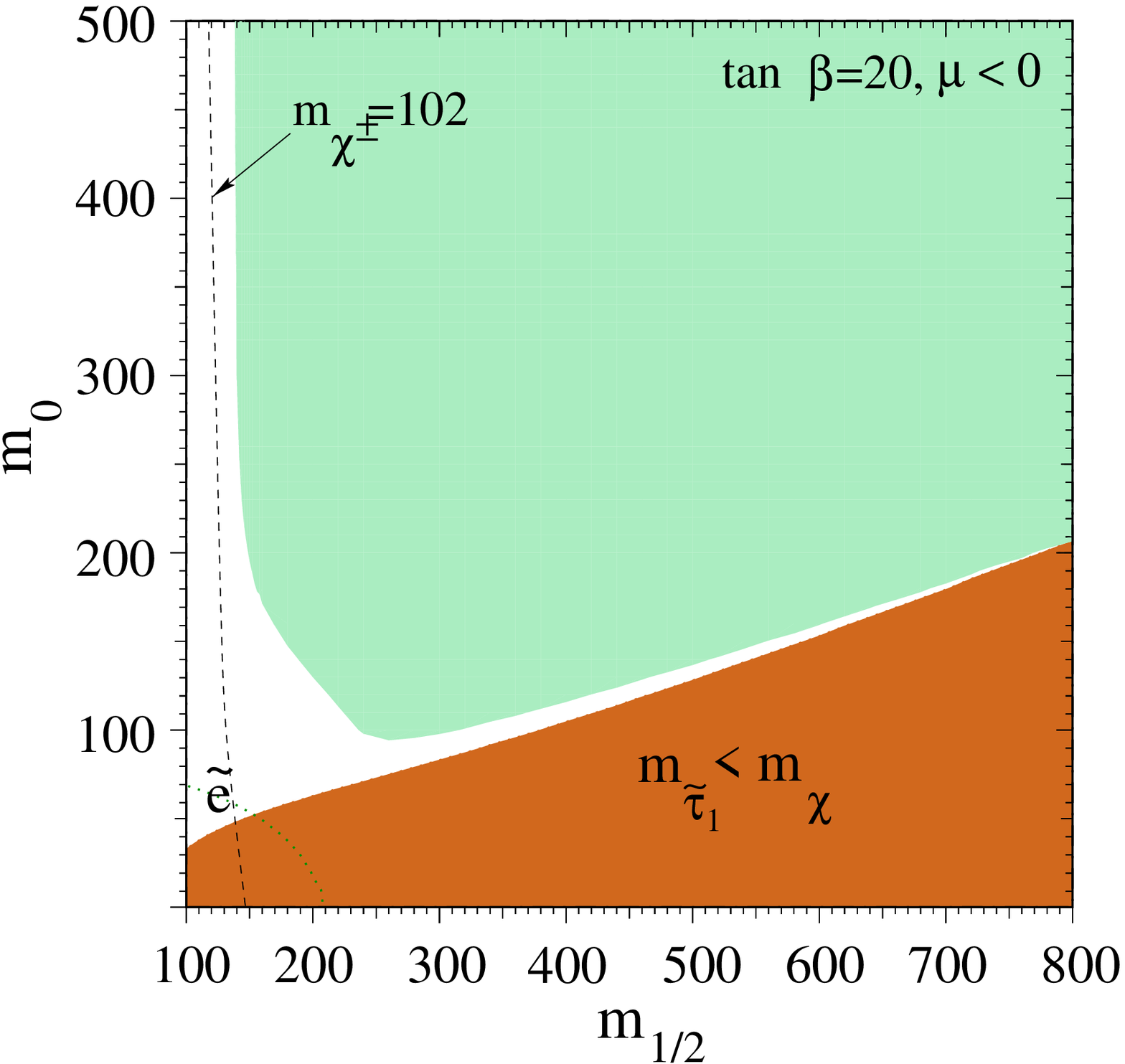,height=7.5cm}}
\end{center}
\caption[.]{\label{fig:nUHMnegative}\it
The $m_{1/2}, m_0$ plane for $\mu < 0$ in the nUHM case, for
$\tan\beta =$ (a) 3, (b) 5, (c) 10 and (d) 20. The region allowed
after including coannihilations,
by the cosmological constraint $0.1 \le \Omega_\chi h^2 \le 0.3$
is shown shaded. Dotted lines
delineate the announced LEP constraint on the $\tilde e$ mass and
the disallowed region where $m_{\tilde \tau_1} < m_\chi$ has dark shading.
The contours $m_{\chi^\pm} = 102$~GeV are shown
as near-vertical dashed lines. Also shown as
near-vertical dot-dashed lines are Higgs mass contours.}
\end{figure}

\begin{figure}[htbp]
\begin{center}
\mbox{\epsfig{file=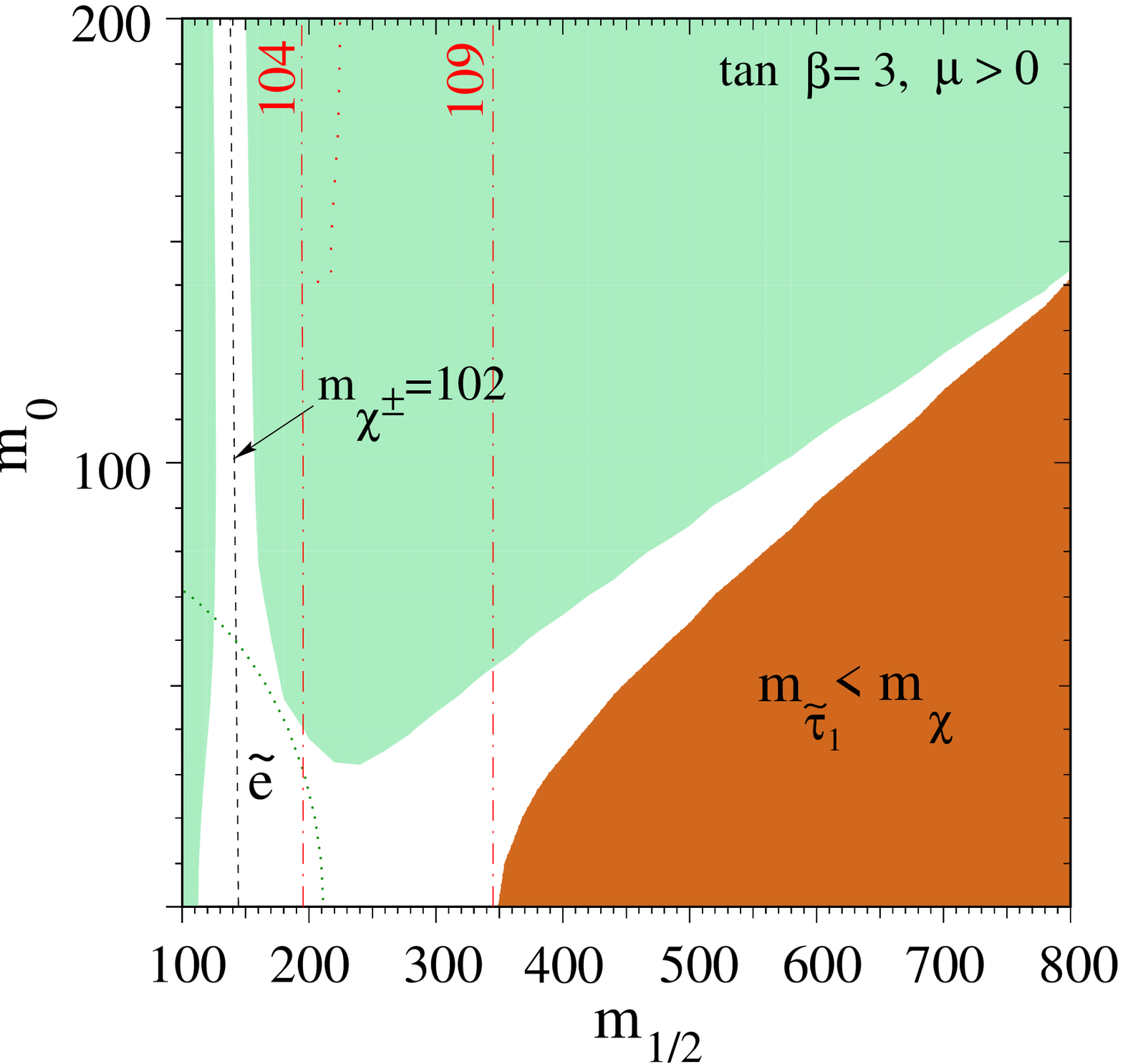,height=7.5cm}}
\mbox{\epsfig{file=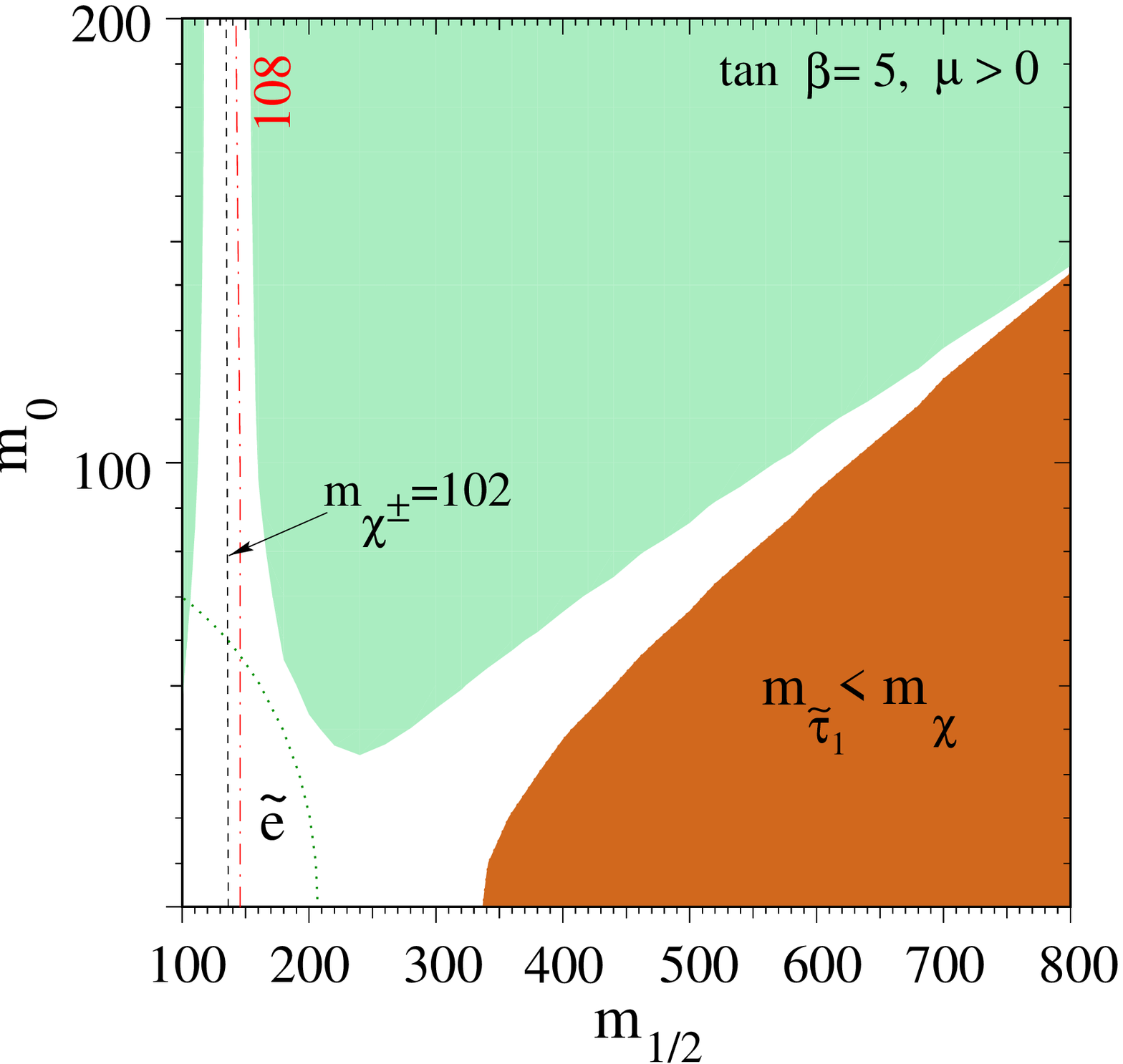,height=7.5cm}}
\mbox{\epsfig{file=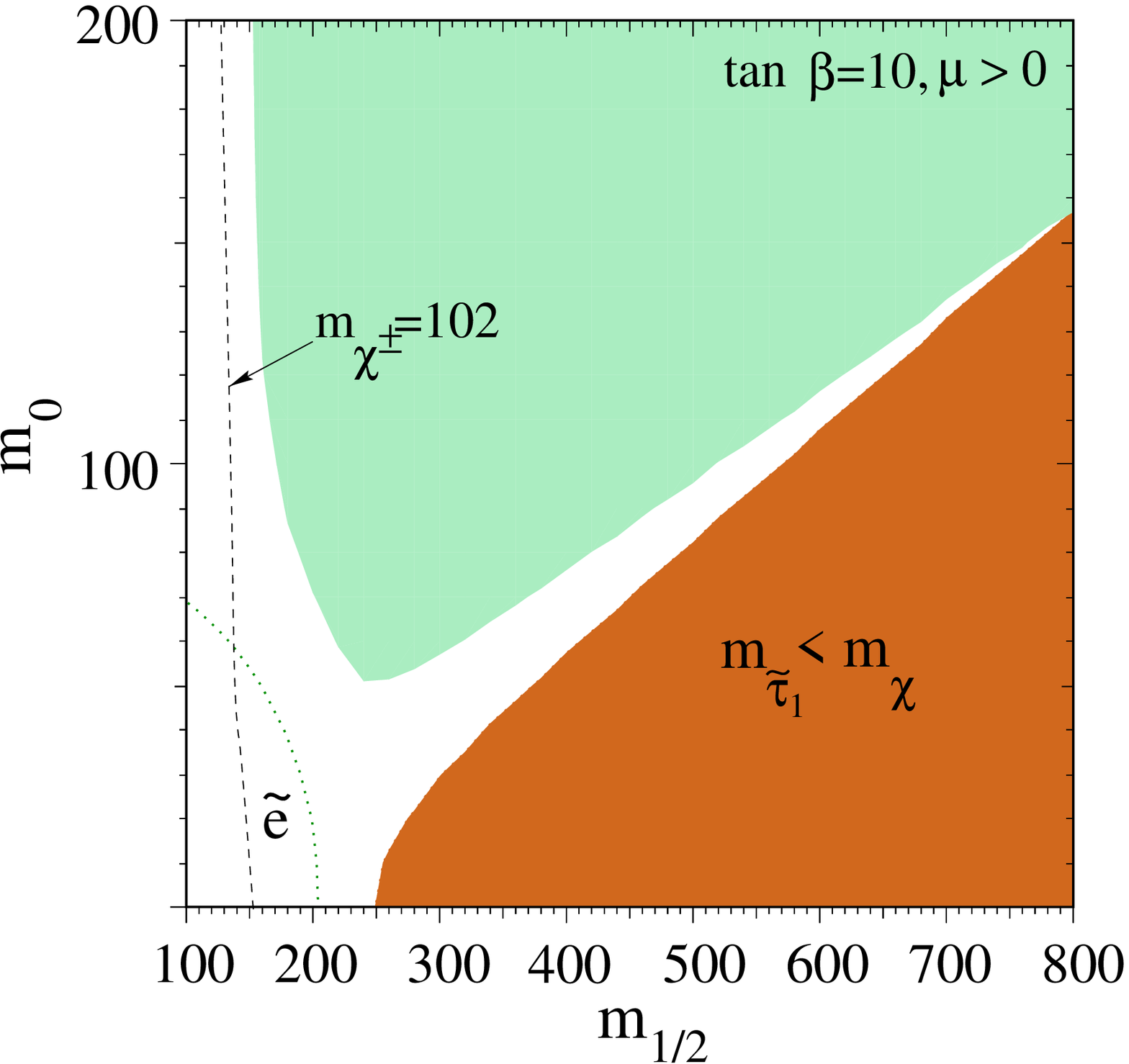,height=7.5cm}}
\mbox{\epsfig{file=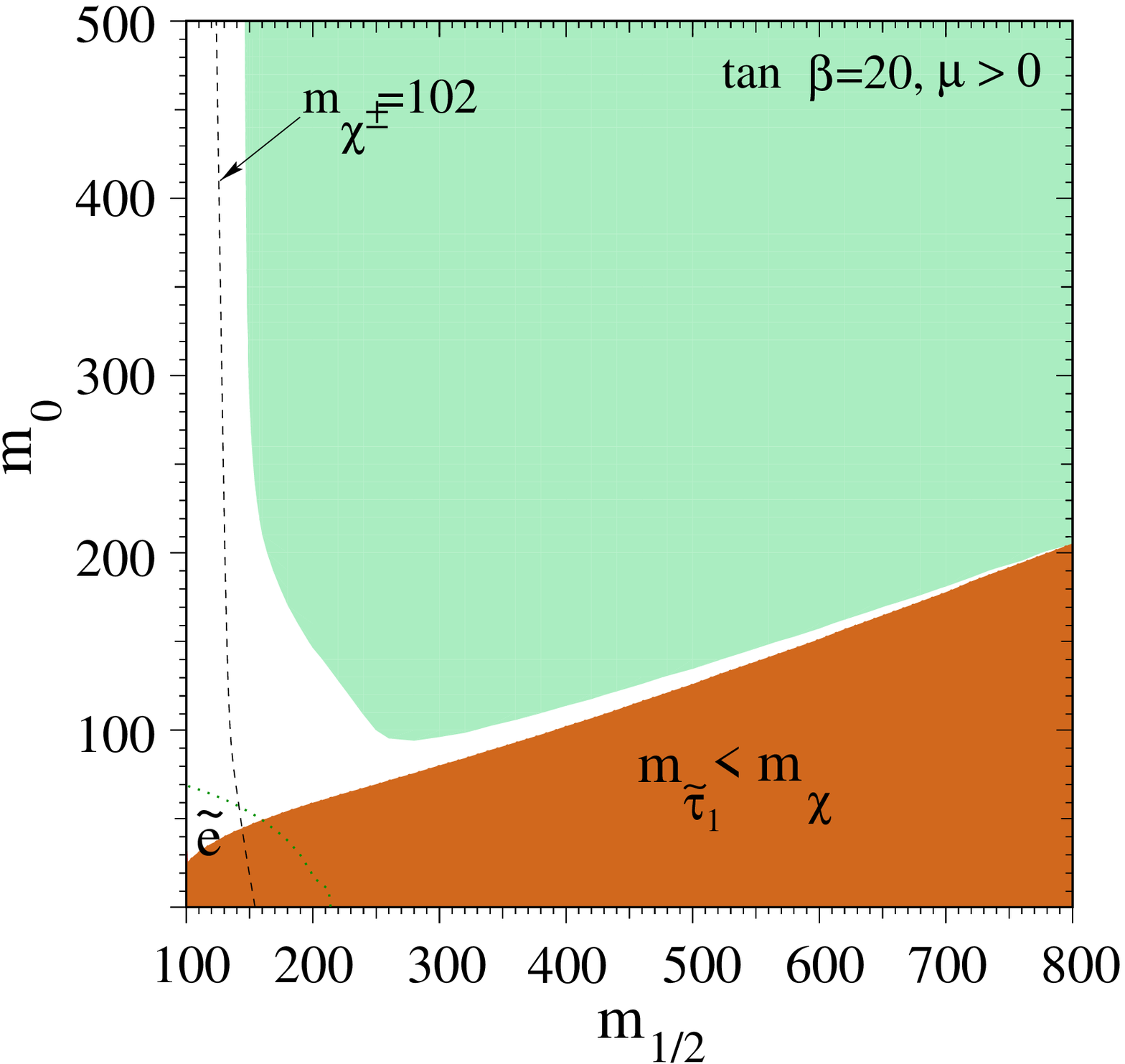,height=7.5cm}}
\end{center}
\caption[.]{\label{fig:nUHMpositive}\it
The $m_{1/2}, m_0$ plane for $\mu > 0$ in the nUHM case, for $\tan\beta =$
(a) 3, (b) 5, (c) 10 and (d) 20. The
significances of the curves and
shadings are the same as in
Fig.~\ref{fig:nUHMnegative}.  The dotted extension of the $m_h=104$
GeV contour corresponds to the shift in the Higgs contour when
the cosmological limit on the relic density is imposed (visible only
in (a)).
}
\end{figure}

Our results for the nUHM case are shown in
Figs.~\ref{fig:nUHMnegative} and ~\ref{fig:nUHMpositive}.
We again show the kinematical limit on the chargino mass: $m_\chi^\pm =
102$ by the near-vertical dashed line. 
Again, for $\mu > 0$, these lines are essentially vertical,
because the chargino mass is independent of
$m_0$, apart from the effects of radiative corrections. 
As seen in the different panels
of Fig.~\ref{fig:nUHMnegative}, the lower bound on $m_{1/2}$ from the
chargino  bound increases  from $m_{1/2} = 112$~GeV to 145 GeV as $\tan
\beta$ is increased from 3 to 20 for $\mu<0$.  In contrast,  as seen in
Fig.~\ref{fig:nUHMpositive}, the bound  on $m_{1/2}$ for
$\mu > 0$ lies near 140 GeV over the same range in $\tan
\beta$. As expected, the nUHM curves for the chargino mass limits
always lie to the left of the UHM ones.

As in Figs.~\ref{fig:m0mhalfnegative}, \ref{fig:m0mhalfpositive}, and 
\ref{fig:UHMnonCCB}, the shaded region corresponds to the parameter values
in which it is possible to achieve $0.1 \le \Omega h^2 \le 0.3$.
This region is unbounded from above, since it is possible to adjust $\mu$
to insure an acceptable relic density even if the sfermion masses are
large.  This is because by lowering $\mu$, the LSP can become a mixed
state (rather than an almost pure bino) and annihilation channels via
$Z^0$-exchange open up. 

We also show in Figs.~\ref{fig:nUHMnegative}a (b) and
\ref{fig:nUHMpositive}a (b) the nUHM contours for Higgs masses of
104 (100) and 109 (108) GeV,
the 1999 and 2K `realistic' bounds for $\tan \beta = 3$
($\tan \beta = 5$). For $\tan \beta = 5$ and $\mu > 0$, the 100 GeV
contour is to the left of the chargino mass contour and is not shown. We
recall that, in the nUHM case, there are no unique Higgs mass contours in
the $m_0, m_{1/2}$ plane, due to the freedom in choosing
$\mu$, $m_A$ and $A_0$. The contours shown correspond to parameter choices
giving the weakest bound. 

Because the Higgs mass bound is
weakest for relatively large values of $\mu$ in the nUHM case (implying
that the neutralino is a bino), the relic
LSP density increases with $m_0$ as one moves upward along the Higgs mass
contours and the sfermion masses increase. 
In some cases, e.g., that in Fig.~\ref{fig:nUHMpositive}a, the relic
density along the 104 GeV contour exceeds 0.3 at about $m_0 = 140$ GeV.
At higher values of
$m_0$, the value of $m_{1/2}$ must be increased to remain consistent with both
the Higgs mass limits and 
cosmology. This adjustment is shown by the dotted curve to the right of
this contour. Similar behavior was seen in
\cite{EFOSII}. For the other Higgs mass contours, either the shift (in
$m_{1/2}$) is insignificant at $m_0 \le 200$ GeV, or the relic density
does not exceed $\Omega h^2 = 0.3$ for $m_0\le 200$ GeV.   For $\mu < 0$, 
the relic density is never saturated for the values of $m_0$ shown. 
As before, for $\tan \beta = 10$ and 20, the chargino bound is always
stronger than the Higgs mass bound. 

\section{Higgsino Dark Matter}

We now turn to the question whether there is any room left for
Higgsino dark matter.  We update the analysis of~\cite{efgos},
including the improved experimental limits discussed above, and we
explore the sensitivity of our conclusions to the 
possible range of LSP relic density. We begin by
reviewing briefly the analysis of \cite{efgos}.

We recall that a general neutralino is a linear combination of the
Higgsinos and neutral gauginos, $\chi = \beta {\tilde B} + \alpha
{\tilde W_3} + \gamma {\tilde H}_1 + \delta {\tilde H}_2$.  In this
notation, the Higgsino {\it purity} is defined to be
$p=\sqrt{\gamma^2+\delta^2}$, a state that is half gaugino and half
Higgsino has Higgsino purity $1/\sqrt{2}$, and, as in~\cite{efgos}, we
take as our working definition that a neutralino is a `pure'
Higgsino if $p>0.9$, even though such a state already has a
sizeable gaugino
fraction: $\sqrt{\alpha^2+\beta^2} \sim 0.44$.  The lightest neutralino
tends to be
Higgsino-like if $\mu < M_2/2$ and gaugino-like when $\mu > M_2/2$. This
was already shown in Fig.~\ref{fig:big3}, where we plot as hashed lines
contours of Higgsino purity in the $\mu, m_{1/2}$ plane for
$\tan\beta = 3$. We have also plotted thin solid contours for
$\Omega_\chi
h^2 = 0.025, 0.1$ and 0.3, thick solid lines corresponding to
$m_{\chi^\pm} = 100$~GeV, and dot-dashed Higgs mass contours. In
this illustration we have
taken $m_A=1$ TeV and $m_0=100$ GeV.  It is apparent that the
bulk of the cosmological region with $0.1 \le \Omega_\chi h^2 \le 0.3$
has larger $|\mu|$ (for given $M_2$) than do the Higgsino purity
contours, indicating that LSP dark matter is generically a gaugino: in these
regions, it is mainly a Bino.  There are, however, small regions at
smaller $|\mu|$ (for given $M_2$), where the LSP is mainly a Higgsino.
However, as can be seen in Fig.~\ref{fig:big3}, this Higgsino
possibility is under severe pressure from several LEP constraints,
including the chargino, $\chi \chi'$ and Higgs searches, and it is the
possible 
exclusion of the Higgsino dark matter regions we explore in this
Section.

\begin{figure}[htbp]
\begin{center}
\mbox{\epsfig{file=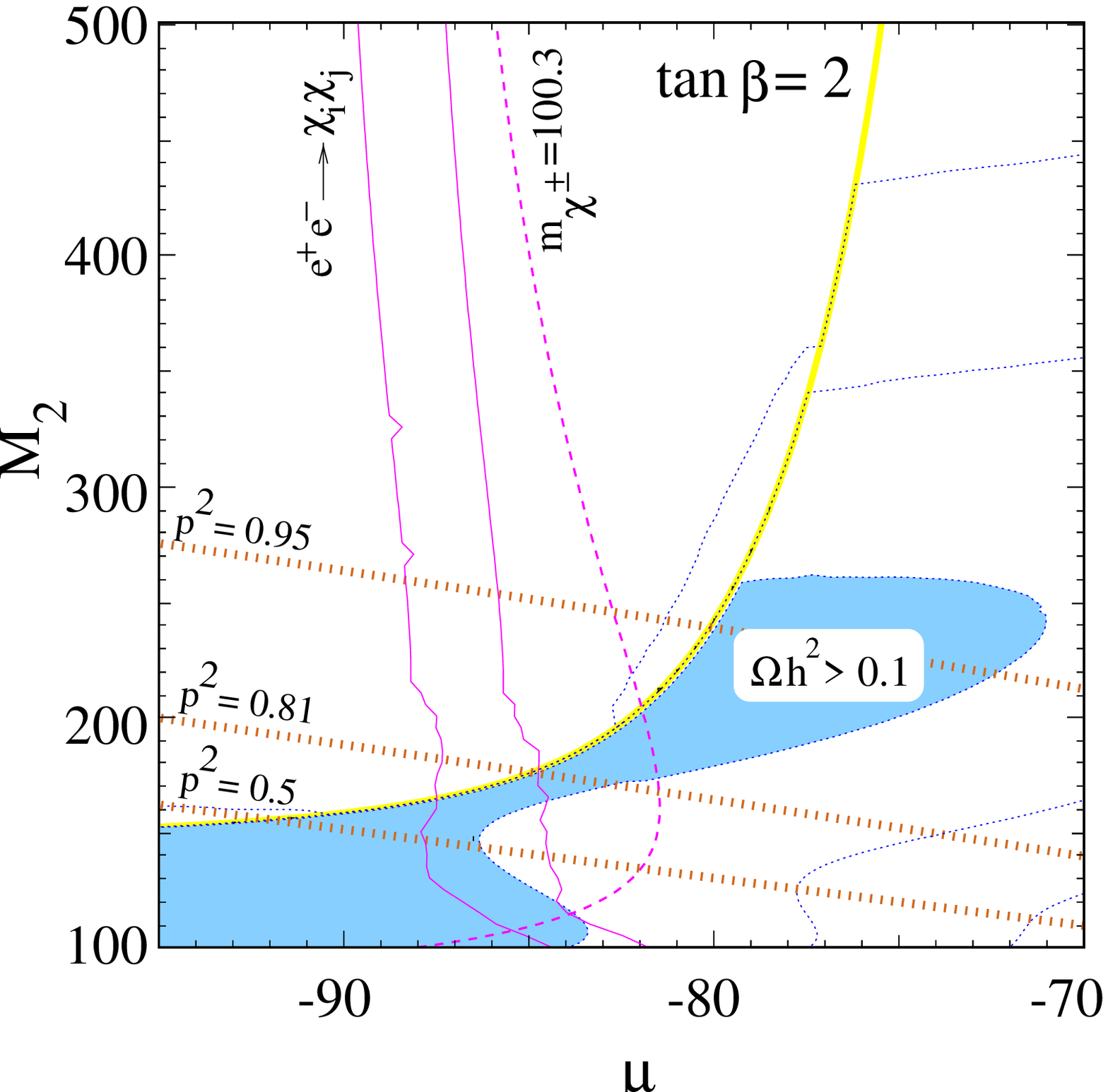,height=7.5cm}}
\mbox{\epsfig{file=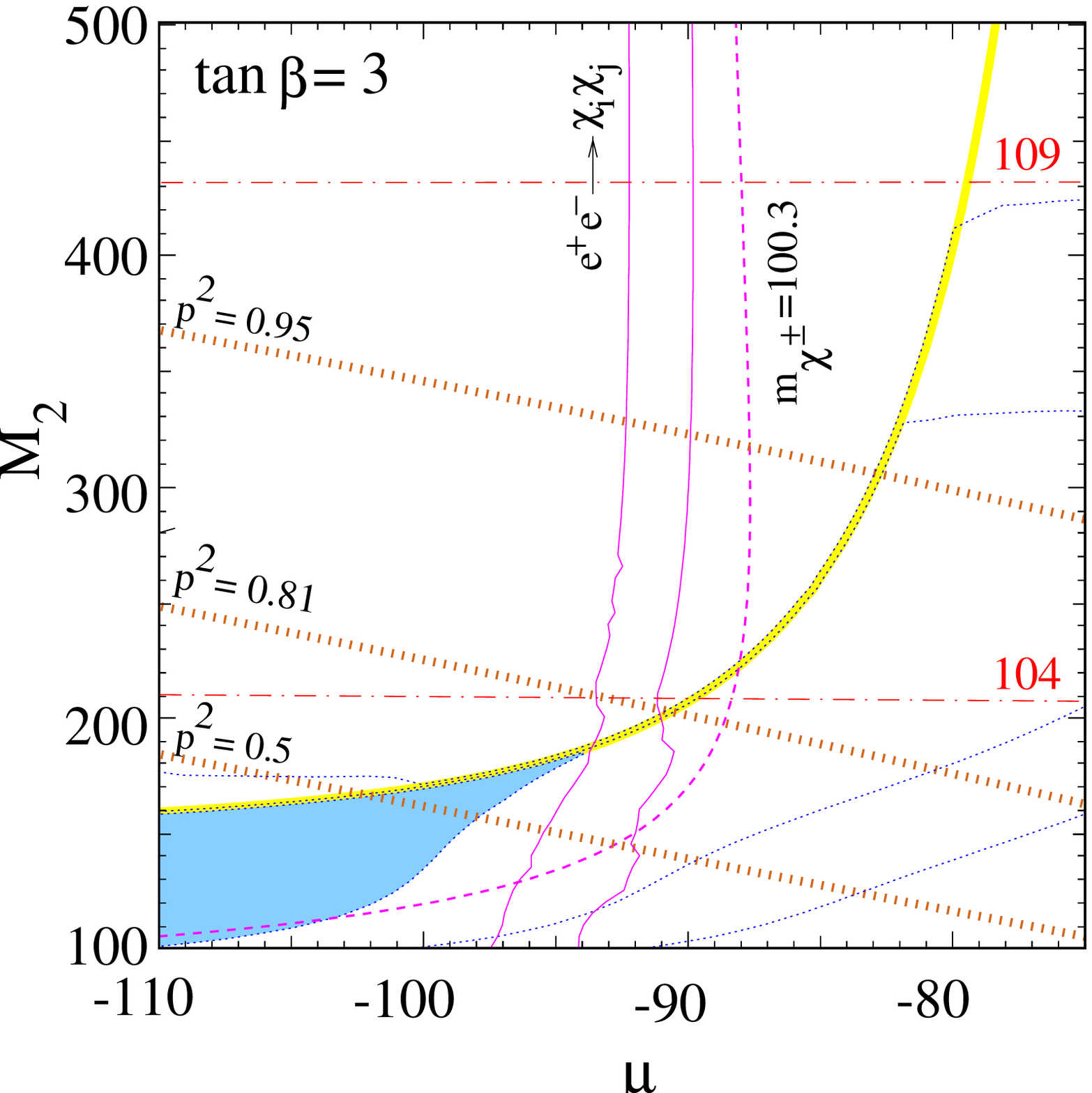,height=7.5cm}}
\mbox{\epsfig{file=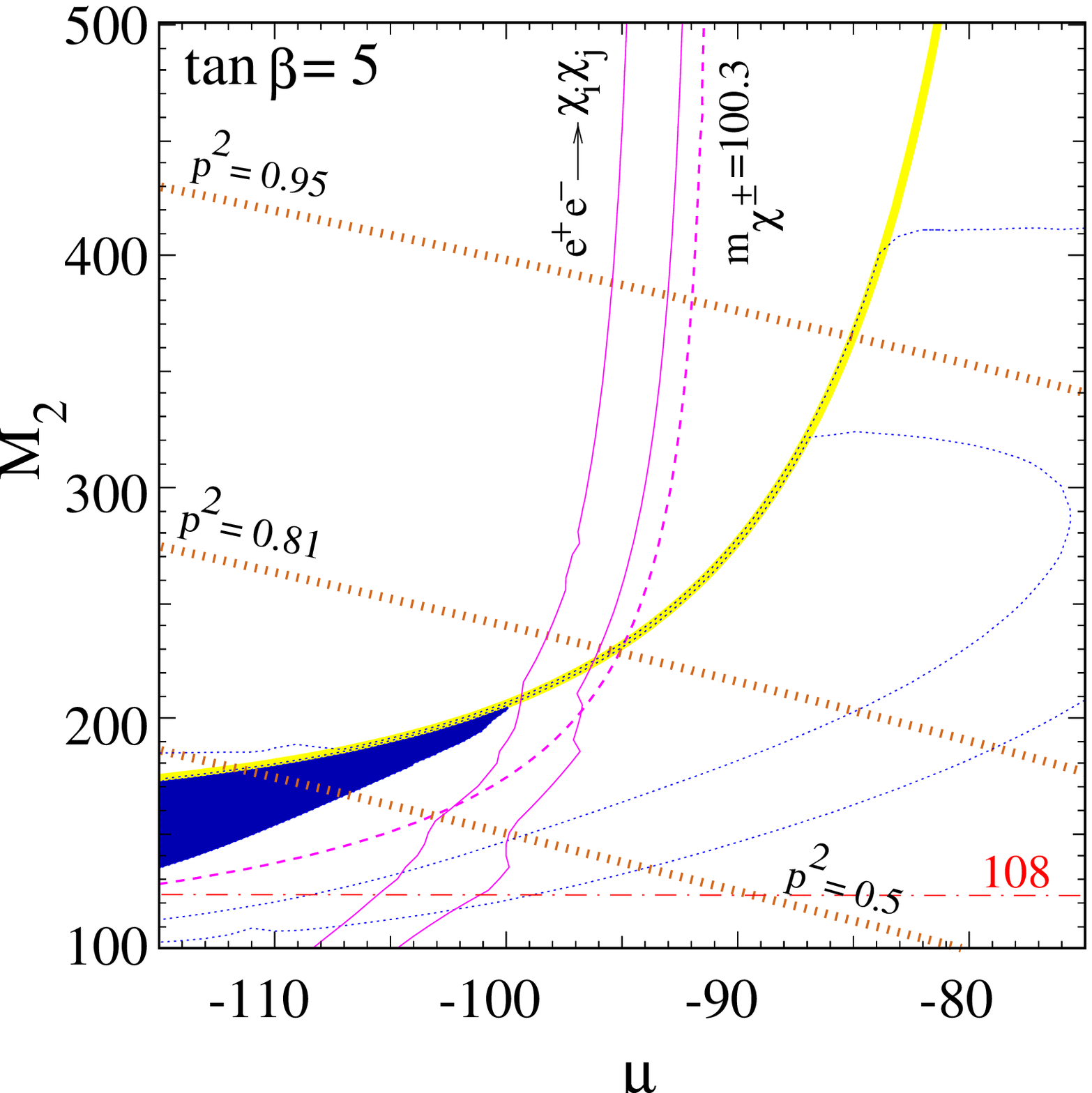,height=7.5cm}}
\mbox{\epsfig{file=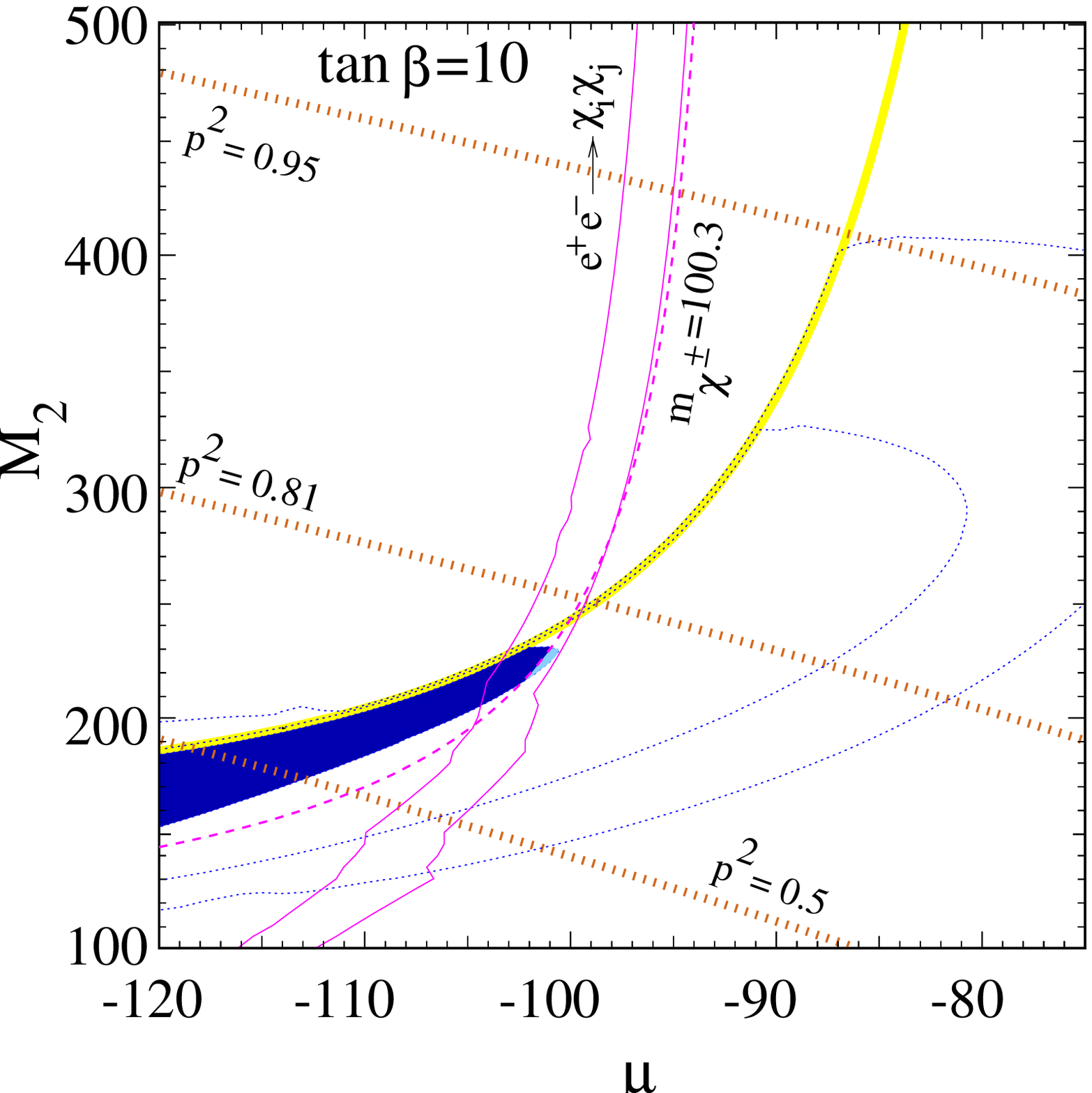,height=7.5cm}}
\end{center}
\caption[.]{\label{fig:teardropn}\it
  Small regions of the $\mu, M_2$ plane for $\tan\beta = 2, 3, 5, 10$
  and $m_A = 10$ TeV.  The thin solid lines are contours for
  $\Omega_\chi h^2 \ge 0.1, 0.5, 0.025$, and the light-shaded region is
cosmologically preferred. The dashed (light solid)
  lines correspond to $m_{\chi^\pm}=100.3$ {\rm GeV} ($\mchi=m_W$),
  dot-dashed horizontal lines correspond to the indicated Higgs masses, and
  the two darker near-vertical solid contours indicate the current 
  neutralino associated production bound $\sigma (e^+ e^- \rightarrow
\chi \chi_2, ...)_{\rm vis}$ and our estimate of the `realistic'
final bound (see Table \ref{tab:xslim}). Hashed contours represent
  Higgsino purity. In panel (a) [(d)], the Higgs mass is everywhere less
  than 106 GeV [greater than 109 GeV]. The dark-shaded regions in panels
(c) and (d) are those surviving all the constraints.}
\end{figure}

\begin{figure}[htbp]
\begin{center}
\mbox{\epsfig{file=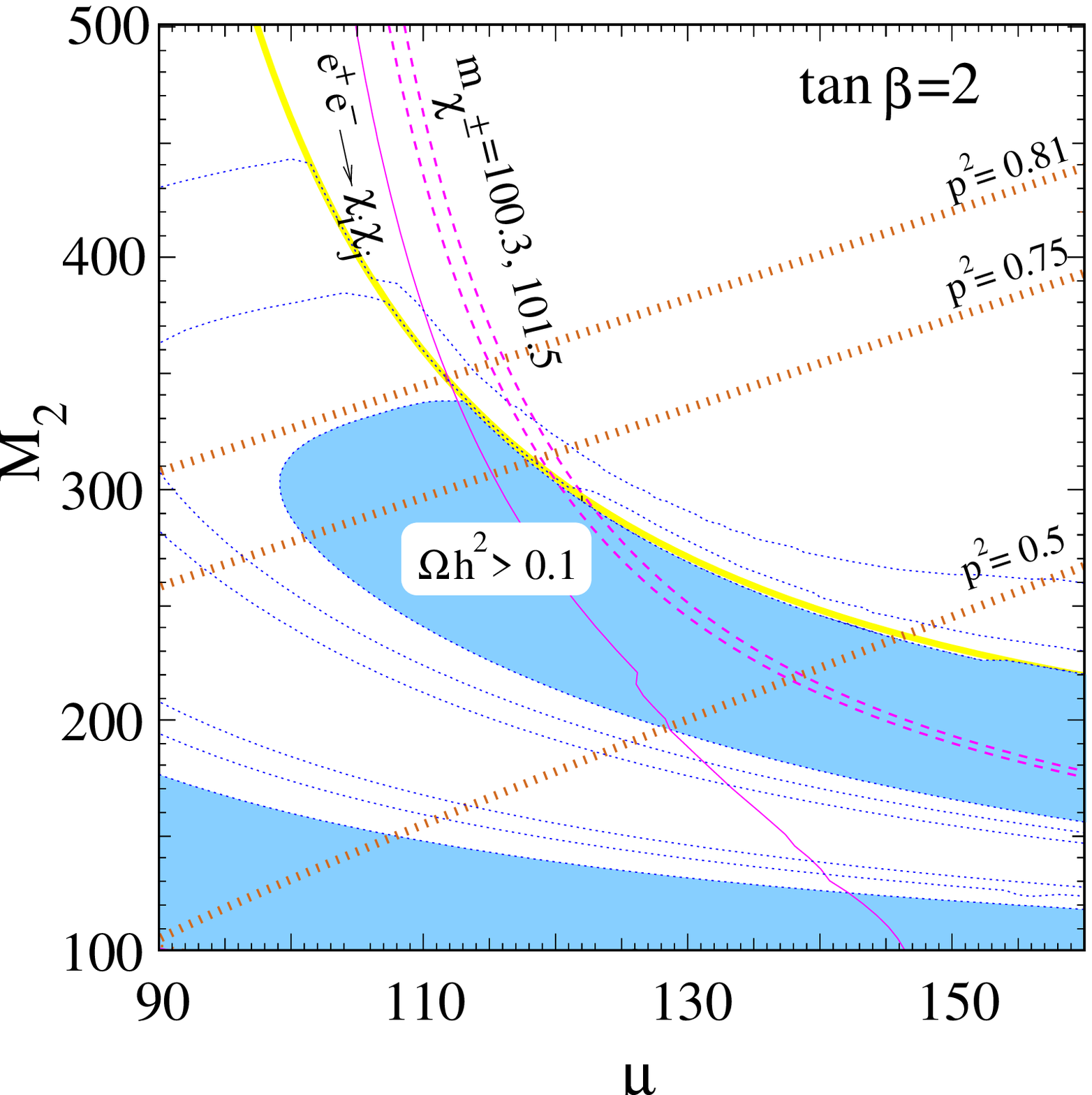,height=7.5cm}}
\mbox{\epsfig{file=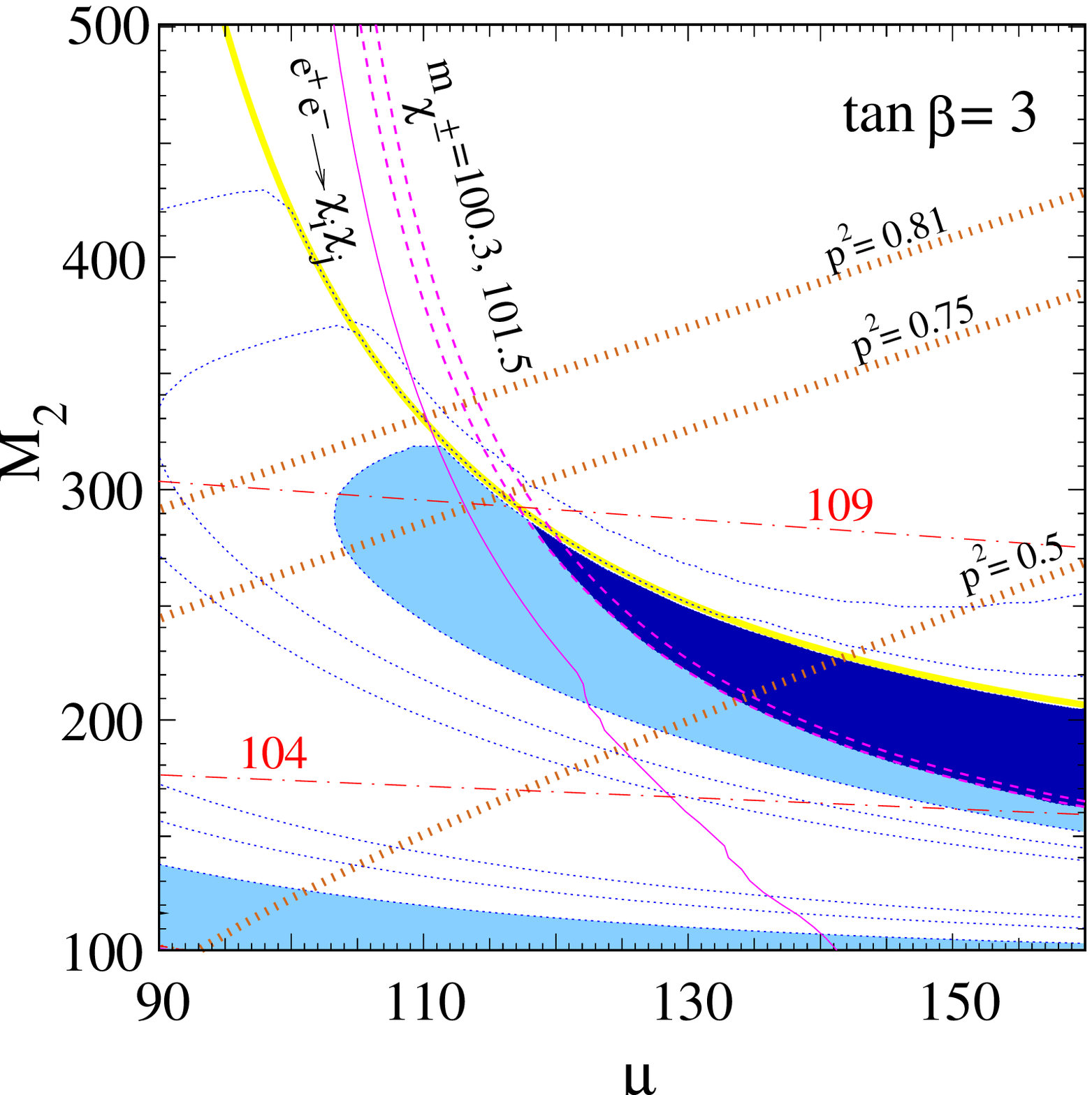,height=7.5cm}}
\mbox{\epsfig{file=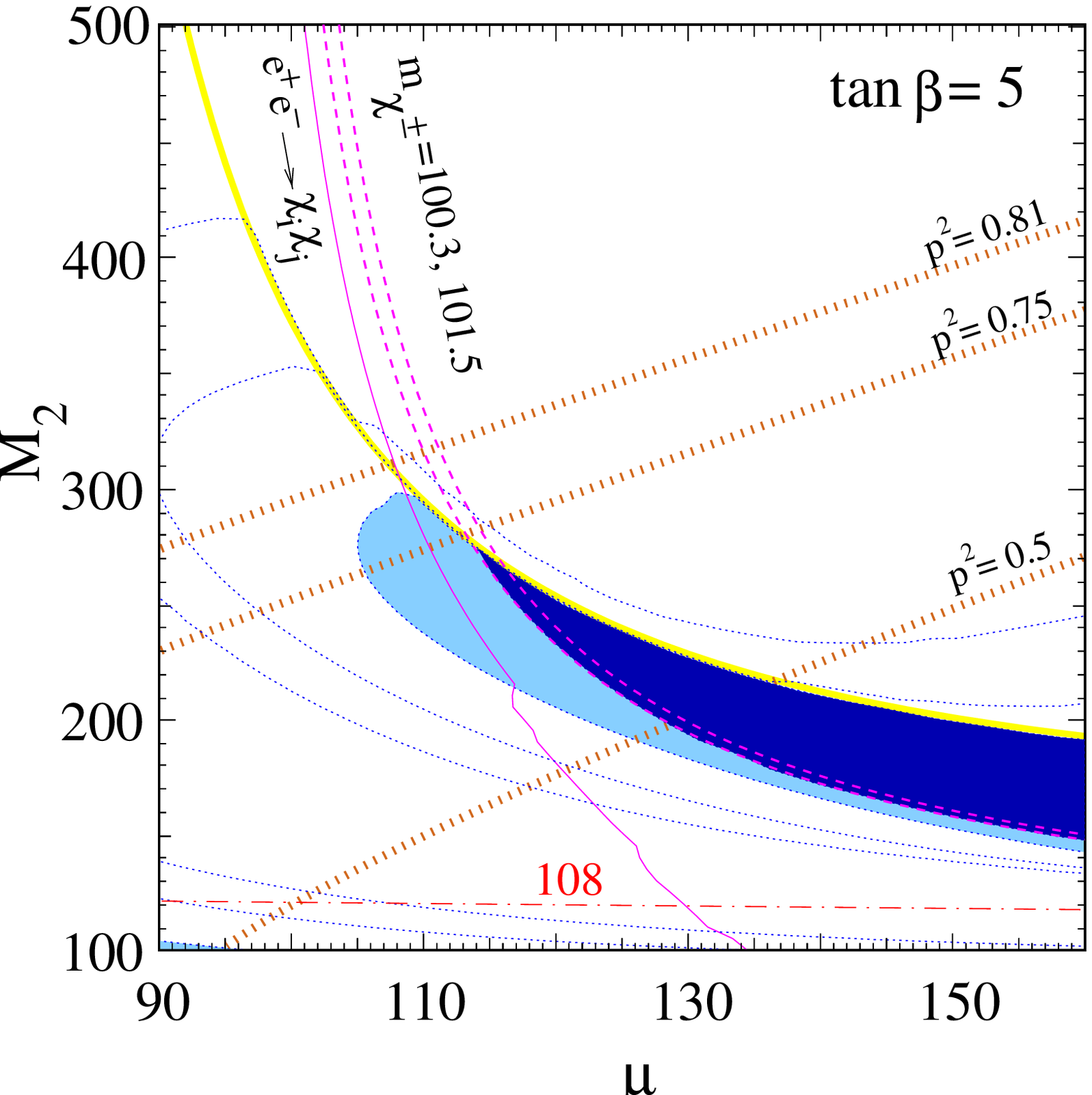,height=7.5cm}}
\mbox{\epsfig{file=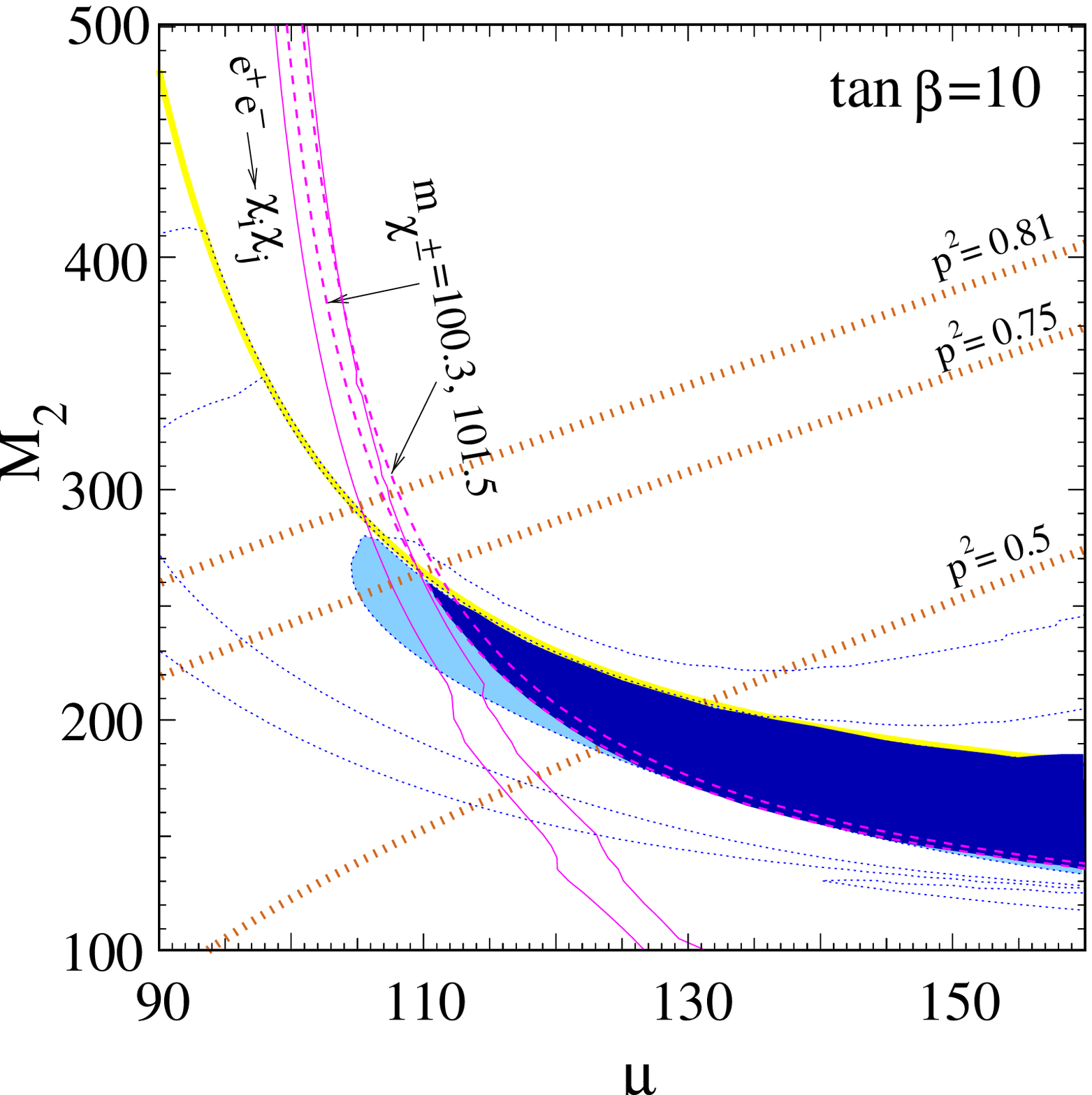,height=7.5cm}}
\end{center}
\caption[.]{\label{fig:teardropp}\it
  As in Fig.~\protect\ref{fig:teardropn}, for $mu>0$.}
\end{figure}

Accordingly, we now focus in more detail on the Higgsino regions, as
illustrated in Figs.~\ref{fig:teardropn} and \ref{fig:teardropp},
where detailed views of the Higgsino parts of the $\mu, M_2$ plane are
shown for $\tan\beta = 2, 3, 5$ and 10.  Consider in particular
Fig.~\ref{fig:teardropn}a, for $\tan\beta=2$ and $\mu<0$.  Here we
have taken
$m_A=10$ TeV to minimize the effect of the Higgs mass limit,
and
$m_0=1$ TeV to maximize the neutralino relic density.  In contrast to Fig.~\ref
{fig:big3}, we have also adjusted $A_t$ to maximize the Higgs mass and produce
the weakest Higgs constraint.
The hashed,
dot-dashed, thin solid and dark thick solid contours are as in
Fig.~\ref{fig:big3}. We plot as a dashed line the current chargino
mass limit. We also show as two solid contours the most
recent 1999 LEP~2 bounds on the 
summed visible cross section for associated neutralino
production $\sigma (e^+ e^- \rightarrow \chi_i \chi_j)_{\rm
vis}$ and our `realistic'
estimate for the final 2K bounds (see Table \ref{tab:xslim}). We recall that
the associated production bounds are more constraining for
smaller values of the scalar masses~\footnote{For the sake of 
exposition, we forget
  for the moment that, at this low value of $\tan\beta$, the entire
  displayed region has a Higgs mass less than 106 GeV and can be
  excluded on this basis alone.}.
It is evident that the entire
region with $\ohsq > 0.1$ and Higgsino purity $p>0.9$ is excluded by
the current experimental limits, for this value of $\tan\beta$.

The general shape of the $\ohsq$ contours can be understood as
follows.  As one moves to large values of $M_2$, the neutralino
becomes more pure Higgsino, which leads to an approximate three-way
mass degeneracy between the lightest and next-to-lightest neutralinos
$\chi,\chi_2$ and the lightest chargino $\chi^{\pm}$. Since the $\chi_2$
and $\chi^{\pm}$ are therefore abundant at the time when the $\chi$
freezes out of chemical equilibrium in the early Universe \cite{co1}, their
coannihilations with the $\chi$ act to bind the
$\chi$ more tightly in chemical equilibrium with the thermal bath, and
delay the freeze-out of the $\chi$ relic density. Since the $\chi$,
$\chi_2$ and $\chi^{\pm}$ annihilate very efficiently, this greatly
reduces the relic density of neutralinos for larger values of $M_2$ \cite{co2}.
We have included the one-loop radiative corrections to the chargino
and neutralino masses, which can significantly affect both the
experimental chargino limits and the neutralino relic density, when the
chargino and neutralino are closely degenerate. Whenever the
mass degeneracy is sufficiently tight to affect the chargino bounds,
coannihilation suppresses the relic density to very small values,
below those of cosmological interest~\cite{efgos}. 

Similarly, as one moves to larger values of $|\mu|$, the mass of the
$\chi$ increases, until $\mchi> m_W$.  At this point, the $\chi$ can
annihilate efficiently into $W$ pairs, and the relic density drops
dramatically as one crosses this threshold \footnote{Sub-threshold
  annihilation into $W$ pairs smoothes out this sudden drop~\cite{co1},
  and shifts the left edge of the $\ohsq=0.1$ contour a few GeV to the
  right.}.  The light solid contour corresponds to $\mchi=m_W$, and
the falloff of the relic density above this threshold is evident. It
is the dramatic decrease in $\ohsq$ for $\mchi>m_W$ that not only
excludes Higgsino dark matter, but also implies that we are not very
dependent on our default choice of lower relic
density cutoff: $\ohsq > 0.1$.  In fact, one cannot even supply enough
neutralinos to provide the galactic dark matter: $\ohsq\sim0.025$
above the $W$ threshold, while still satisfying the experimental
constraints.  We see the same effect in all four panels of
Figs~\ref{fig:teardropn}, and, although we only display four values of
$\tan\beta$, we have verified that this is true for all $\tan\beta$.
 
The situation is similar for $\mu>0$, as seen in 
Fig.~\ref{fig:teardropp}. In
this case, the dominant experimental constraint comes from the
chargino limits, which are shown as dashed lines for both 1999 and the
`realistic' 2K scenario, whereas we plot only the 1999 contour for
$e^+ e^- \rightarrow \chi_i \chi_j$. The chargino constraints
alone exclude $\ohsq>0.025$ for a
Higgsino-like neutralino.  In all cases, an interesting amount
of cold dark matter is possible only if $p^2 \le 0.7$, i.e., 
the LSP is either predominantly a gaugino or a strongly mixed state.
We conclude that {\it a predominantly Higgsino state cannot
provide a substantial component of the dark matter}.

This conclusion is robust with respect
to variations in the other sparticle masses.
The sfermion masses have already been taken large enough for the
contribution to neutralino annihilation from sfermion exchange to be
negligible, and, as already noted, lighter sfermions yield tighter
constraints from associated neutralino production.  
The experimental chargino limits fall below the kinematic limit when
the sneutrino is closely degenerate with the chargino: however, again,
a light sneutrino enhances the associated neutralino production
limits, so this also provides no loophole.  {\it A cosmologically
interesting relic density can therefore only be achieved by either
heavily mixed or pure gaugino neutralino states.}

\section{Limits from the Tevatron Collider}

The limits on the mSUGRA (UHM)
parameter space from Run I of the Tevatron are summarized
in~\cite{sugrarept}, and we have already made use of their lower limits
on stop masses. Two other D0 limits are reported in~\cite{sugrarept}: one
is for squark and gluino jets and missing energy, and the other is
for dilepton events. The former analysis is for $\tan\beta = 2$, and hence
is not directly comparable with our plots. The analysis also assumes $A_0
= 0$ and $\mu < 0$, and yields a lower limit
\beq
m_{1/2} > 100~{\rm to}~50~{\rm GeV}
\eeq
for $m_0$ between 0 and 300 GeV. The dilepton analysis in
\cite{sugrarept} has
been performed for several values of
$\tan\beta$ including the values 3 and 5 studied in this paper. Again
for $A_0 = 0$ and $\mu < 0$, the lower limit
\beq
m_{1/2} > 60~{\rm to}~40~{\rm GeV}
\eeq
was found for $m_0$ between 0 and 300 GeV. These Tevatron Run~I limits
do not constrain the MSSM parameter space as strongly as do
the chargino and Higgs limits from LEP~II, unless scalar mass
universality between squarks and sleptons is relaxed.  We have
therefore not
displayed the Run I bounds in Figs.~\ref{fig:m0mhalfnegative} and
\ref{fig:m0mhalfpositive}. 

On the other hand,
Run II of the Tevatron, scheduled to begin
in 2001,
will impose strict new limits at low $\m12$.  The dominant
experimental constraint is expected to come from searches for
trilepton signatures of $\chi_1^\pm\chi_2^0$ production~\cite{trilep,sugrarept}.
We display in Fig.~\ref{fig:m0mhalfpositive} as long-dashed contours
the anticipated reach of the upgraded Tevatron in this channel, assuming 2
fb$^{-1}$
integrated luminosity. We have taken the published curves from
\cite{sugrarept}, which does not display results for $\tb=20$ or $\mu<0$
\footnote{The $\tb=3,10$ Run II curves and the $\tb=5$ Run II curve in
  \protect{Fig.~\ref{fig:m0mhalfpositive}} come from separate
  analyses and reflect slightly different confidence levels, 99\% and
  3$\sigma$, respectively.}.  The bounds are tightest at low $m_0$,
where they cut into the cosmological region as shown. 
At $\tb\ga3$,
the Run II curves bend over and intersect the LEP2 chargino bound
within the cosmological region. Their 
most significant impact is for low values of $\tan \beta$. 
However, since the LEP~II Higgs bounds dominate
at low $\tb$, the trilepton analysis at the Tevatron Run II will therefore not 
increase the absolute lower bound on the neutralino mass given by our
analysis in this region.  However, if the Tevatron is able to improve the LEP
Higgs bound, this could raise
substantially the neutralino limit at low $\tb$. 
The Higgs bound is no longer important
for $\tan \beta = 10$, but in this case the Tevatron limit bites
away less of the region favoured by cosmology.

\section{Summary and Conclusions}

\begin{figure}[htbp]
\begin{center}
\mbox{\epsfig{file=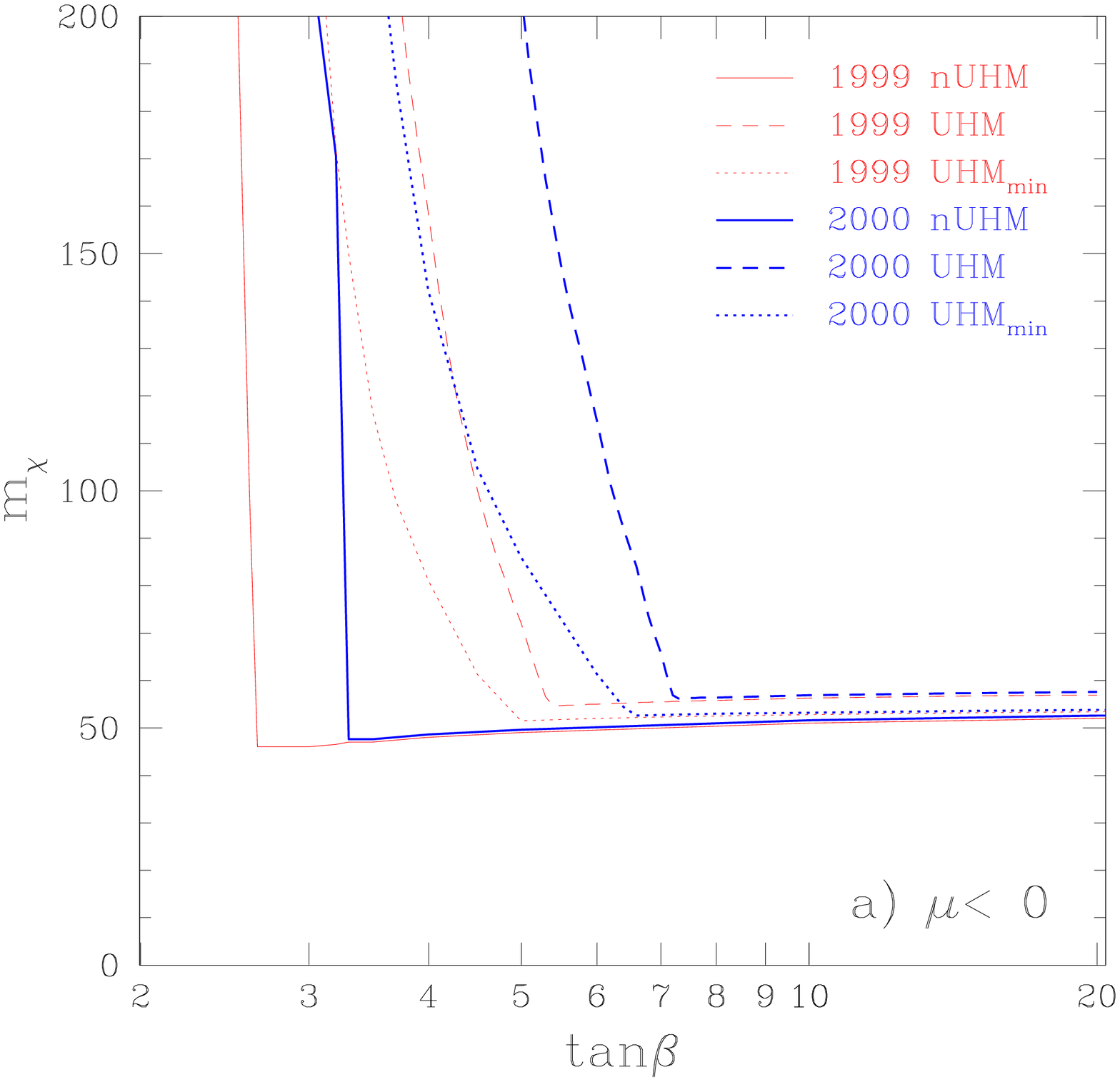,height=8.0cm}}
\mbox{\epsfig{file=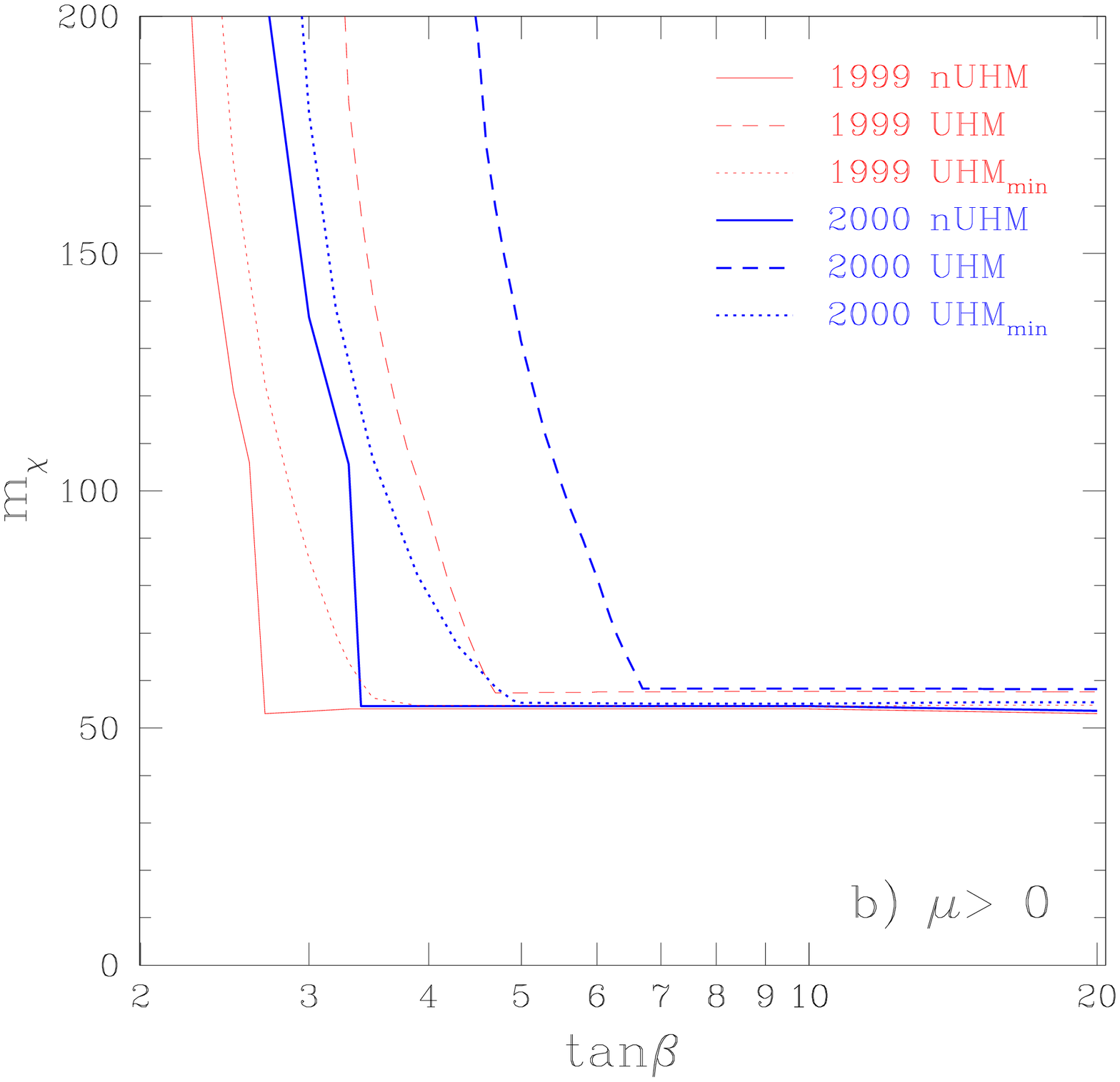,height=8.0cm}}
\end{center}
\caption[.]{\label{fig:mvtb}\it
 Lower limits on the neutralino mass $m_\chi$ as functions of
$\tan \beta$ for (a) $\mu< 0$ and (b) $\mu>0$. The curves correspond to
the final 1999 LEP results (thin lines) and our `realistic' expectations
for the 2K LEP run (thick lines).  We show the UHM case
with $A_0 = -m_{1/2}$ to avoid CCB minima (dashed curves): these are the
strongest constraints. We also show (dotted lines) the more general
UHM$_{\rm min}$ case
where $A_0$ is left free, and we do not require
the absence of CCB vacua. We also display additionally the nUHM case, which
is the most conservative and allows both $\mu$ and $m_A$ to be free in
addition to $A_0$. }
\end{figure}

One of our principal goals in this paper has been to obtain strengthened 
lower limits on the neutralino mass, combining the latest LEP data with
the cosmological dark matter requirement $0.1 < \Omega_\chi h^2 < 0.3$.
We summarize our limits in Figs.~\ref{fig:mvtb}, under various different
assumptions: universal (UHM) or non-universal (nUHM) soft
supersymmetry-breaking scalar
masses for Higgs bosons and (in the former case) whether one requires the 
present vacuum to be stable against transition to a charge- and
colour-breaking (CCB) vacuum or not (UHM$_{\rm min}$). Also, we give
limits both for the
available 1999 LEP data and with a `realistic' assessment of the likely
sensitivity of data to be taken in 2K. 

In all cases, for both positive
and negative $\mu$, the lower limits on $m_\chi$ are relatively
insensitive to
$\tan \beta$ at large $\tan \beta$. Here they are determined by the
LEP chargino bound, as the LEP Higgs mass bound
is weaker than the chargino bound at large $\tan \beta$.  
In fact, in the two UHM cases shown,
the points at which the limiting curves bend upward, as one decreases
$\tan \beta$, are precisely the points at which the Higgs mass bound
becomes more
stringent than the chargino bound. In the UHM cases, the
neutralino mass limits are strong at intermediate values of
$\tan \beta \simeq$ 4--7 because, as discussed earlier, the cosmological
bound on the relic density prohibits going to large values of $m_0$, and 
ensures that the Higgs bound places a strong constraint.
 Below this break point, the lower
limit on $m_\chi$ increases rapidly with decreasing $\tan \beta$. Above
this break point, the limit on $m_\chi$ is relatively insensitive to the
additional theoretical assumptions made, such as UHM vs. UHM$_{\rm min}$ 
or nUHM. However,
in the nUHM cases, because one can increase $m_0$ sufficiently to weaken
the Higgs mass bound, the break point occurs at a lower value of $\tan
\beta$. To go to lower values of $\tan \beta$ then requires a substantial
increase in $m_\chi$. 

\begin{figure}[htbp]
\begin{center}
\mbox{\epsfig{file=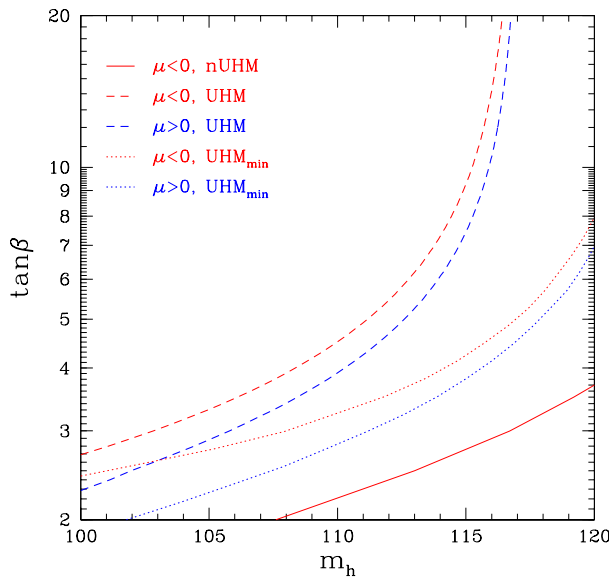,height=8cm}}
\end{center}
\caption[.]{\label{fig:tbvmh}\it Lower limit on $\tb$ imposed by the experimental
and cosmological constraints, as a function of the experimental Higgs mass limit.  The
UHM, UHM$_{min}$ and nUHM labels are as in \protect{Fig.~\ref{fig:mvtb}}.  The $\mu>0$
curve in the nUHM case is very similar to the $\mu<0$ curve.}
\end{figure}

In the UHM cases with and without the restriction forbidding CCB vacua,
the lower limit on the lightest MSSM Higgs mass, in particular, implies
lower limits on $\tan \beta$ which are plotted in Fig.~\ref{fig:tbvmh}.
 The limits are somewhat
stricter for $\mu < 0$ than for $\mu > 0$, whether (UHM) or not
(UHM$_{\rm min}$) one requires the absence of CCB vacua.
Indeed, Fig.~\ref{fig:tbvmh} will enable the appropriate conclusion to be
drawn from whatever lower limit on the Higgs mass LEP eventually provides.
We recall that the existing Higgs mass calculations in the MSSM are
believed to be accurate to about 3~GeV.  Therefore in computing the bounds on $\tb$ for 
Tables \ref{tab:tbvmh} and \ref{tab:tbvmhc}, for example, we have conservatively 
shifted the exclusion curves of Fig.~\ref{fig:Higgs} by 3 GeV to the left before 
reading the values of $\tb$ off of Fig.~\ref{fig:tbvmh}.
We also show in Fig.~\ref{fig:tbvmh}
the lower bound on $\tan \beta$ obtained in the nUHM, which is
significantly weaker than in the UHM cases, and essentially independent
of the sign of $\mu$.  

If LEP does achieve the `optimistic' 2K
energies and luminosities (\ref{optimistic}), the above constraints
will be somewhat tighter.  The horizontal segments
of Fig.~\ref{fig:mvtb}, corresponding to the chargino limits, increase
by a fraction of a GeV;  the vertical branches move to the right,
intersecting the horizontal segments at $\tb=8\, (7.5)$ for $\mu<0\, (\mu>0)$.
And lastly, the lower limits on $\tb$ improve to

\begin{table}[htbp]
  \begin{center}
    \begin{tabular}{|l|c|c|c|}
\hline
&UHM&UHM$_{\rm min}$&nUHM\\
\hline
$\mu<0\;\;\;\;\;$&4.7&3.4&2.3\\[.2em]
$\mu>0$&4.2&3.0&2.3\\
\hline
    \end{tabular}
    \caption{\label{tab:tbvmho}
    {\it Limits on $\tb$, assuming the 'optimistic' 2K energies and
      luminosities  (\protect{\ref{optimistic}})}.}
  \end{center}
\end{table}

In many respects, LEP has provided the most stringent constraints
on the parameters of the MSSM. This is true, in particular, for its
lower limits on the Higgs mass, and the chargino, neutralino and slepton
constraints from LEP compare favourably with the Tevatron bounds on
squark and gluino masses, once the different mass renormalizations
of electroweakly- and strongly-interacting sparticles are taken into
account. As we have shown, the present LEP data may be combined to set
interesting lower bounds on the lightest neutralino mass, in particular if
it is assumed to constitute the dark matter favoured by astrophysics and
cosmology, and on $\tan \beta$. It may well be that these lower limits
will be further strengthened by the LEP run during 2000, as we have
discussed in this paper. However, this would be the pessimistic
scenario. There is still a chance that sparticles or the Higgs boson
may turn up this year, in which case we would be delighted to see our
bounds superseded. LEP may not yet have discovered supersymmetry, but
it certainly deserves to!

\vskip 0.5in
\vbox{
\noindent{ {\bf Acknowledgments} } \\
\noindent  
We would like to thank P. Gambino and C. Kao and especially
M. Schmitt and M. Spiropulu for useful
discussions. 
This work was supported in part by DOE grant DE--FG02--94ER--40823.  The
work of T.F. was supported in part by DOE   
grant DE--FG02--95ER--40896 and in part by the University of Wisconsin  
Research Committee with funds granted by the Wisconsin Alumni Research  
Foundation.}

\end{document}